\documentclass[prb,aps,a4paper,notitlepage]{revtex4-1}
\pdfoutput=1
\usepackage{amsmath}
\usepackage{amssymb}
\usepackage{epsfig}
\usepackage{subfig}
\usepackage{color}
\usepackage{enumerate}
\usepackage{verbatim}
\usepackage{tikz}

\usepackage[english]{babel}

\usetikzlibrary{intersections}
\def \d {\mathrm{d}}

\def \OO {\mathcal{O}}
\def\fr#1{(\ref{#1})}
\def \qm {{Q_-}}
\def \qp {{Q_+}}
\def\eps{\varepsilon}
\def\figr#1{Fig.~\ref{#1}}
\def \am {{A_-}}
\def \ap {{A_+}}
\def \imp {{\rm imp}}

\def\pd{{\phantom\dagger}}
\def\nn{\nonumber\\}
\newcommand{\be}{\begin{equation}}
\newcommand{\ee}{\end{equation}}
\newcommand{\beA}{\begin{equation}\begin{aligned}}
\newcommand{\eeA}{\end{aligned}\end{equation}}
\newcommand{\bem}{\begin{multline}}
\newcommand{\eem}{\end{multline}}
\newcommand{\up}{\uparrow}
\newcommand{\down}{\downarrow}
\newcommand{\bea}{\begin{eqnarray}}
\newcommand{\eea}{\end{eqnarray}}

\begin{document}
\title{Mobile impurity approach to the optical 
conductivity in the Hubbard chain}
\author{Thomas Veness and Fabian H.~L.~Essler}
\affiliation{The Rudolf Peierls Centre for Theoretical Physics,
  University of Oxford, Oxford OX1 3NP, UK} 
\begin{abstract}
We consider the optical conductivity in the one dimensional Hubbard
model in the metallic phase close to half filling. In this regime most
of the spectral weight is located at frequencies above an energy scale
$E_{\rm opt}$ that tends towards the optical gap in the Mott
insulating phase for vanishing doping. Using the Bethe Ansatz we
relate $E_{\rm opt}$ to thresholds of particular kinds of excitations
in the Hubbard model. We then employ a mobile impurity models to
analyze the optical conductivity for frequencies slightly above these
thresholds. This entails generalizing mobile impurity models to excited
states that are not highest weight with regards to the SU(2)
symmetries of the Hubbard chain, and that occur at a maximum of the
impurity dispersion.
\end{abstract}
\maketitle
\section{Introduction}
Electron-electron interactions play a crucial r\^{o}le in determining the physical response to external probes of various quasi-one-dimensional materials e.g. organic semiconductors\cite{semicond}. 
In order to successfully describe the mechanisms and excitations
responsible for distinct physical phenomena, it is imperative to have
a microscopic model capturing the essence of the physics involved;
providing a framework within which realistic physical systems may be
interpreted. The one-dimensional Hubbard model\cite{book} offers an
excellent theoretical laboratory in which a comprehensive microscopic
understanding of the origin of various behaviours can be developed. 
The Hamiltonian for the Hubbard model is given by 
\beA
	H &= -t \sum_{i,\sigma} c^\dagger_{i+1,\sigma} c^{\pd}_{i,\sigma} + c^\dagger_{i,\sigma}c^{\pd}_{i+1,\sigma} + U\sum_i n^{\pd}_{i,\up} n^{\pd}_{i,\down}\\
	&-\mu \sum_i (n^{\pd}_{i,\up} + n^{\pd}_{i,\down} ) - B\sum_i (n^{\pd}_{i,\up} - n^{\pd}_{i,\down}).
\eeA
Here, $c^{\phantom\dagger}_{j,\sigma}$ annihilates a fermion with spin $\sigma=\, \up,\down$ at site $j$, $n^{\phantom\dagger}_{j,\sigma} = c^\dagger_{j,\sigma} c^{\phantom\dagger}_{j,\sigma}$ is the number operator, $t$ is the hopping parameter, $\mu$ is the chemical potential, $B$ is the magnetic field, and $U\geq 0$ is the strength of the on-site repulsion. 

The low-energy degrees of freedom in the metallic phase of the Hubbard
chain are described \cite{Woynarovich,FrahmKorepin,book} by a (perturbed)
spin-charge separated Luttinger liquid \cite{Haldane1,Haldane2,GNT,Giamarchi},  
with Hamiltonian
\be
H = \sum_{\alpha=c,s} \frac{v_\alpha}{16\pi} \int \d x\, \left[  \frac{1}{K_\alpha} \left( \frac{\partial \Phi_\alpha}{\partial x} \right)^2 + K_\alpha \left(  \frac{\partial \Theta_\alpha}{\partial x} \right)^2 \right] + {\rm irrelevant\ operators}. \label{eq:llHamiltonian}
\ee
The parameters $K_\alpha$, $v_\alpha$ can be calculated for the
Hubbard model by solving a system of linear integral equations (see
Appendix \ref{app:vK}). The Bose fields $\Phi_\alpha(x)$ and dual
fields $\Theta_\alpha(x)$ obey the commutation relation 
\be
[ \Phi_\alpha(x), \Theta_\beta (y) ] = 4\pi i \delta_{\alpha\beta} {\rm sgn}(x-y).
\ee
The spectrum of low-lying excitations relative to the ground state for
a large but finite system of length $L$ in zero magnetic field is
given by \cite{Woynarovich,FrahmKorepin,book}
\be
\Delta E = \frac{2\pi v_c}{L} \left[ \frac{\left( \Delta N_c \right)^2}{8K_c} +  2K_c \left( D_c + \frac{D_s}{2} \right)^2  + N_c^+ + N_c^- \right] +
\frac{2\pi v_s}{L} \left[ \frac{\left( \Delta N_s - \frac{\Delta N_c}{2} \right)^2}{2} +  \frac{D_s^2}{2}  + N_s^+ + N_s^- \right] ,
\ee
where $\Delta N_\alpha$, $D_\alpha$ and $N_\alpha^\pm$ are integers or half-odd integers subject to the ``selection rules''
\be
N_\alpha^\pm \in \mathbb{N}_0, \qquad \Delta N_\alpha \in \mathbb{Z}, \qquad D_c = \frac{\Delta N_c + \Delta N_s}{2}\,{\rm mod}\,1,\qquad D_s = \frac{\Delta N_c}{2}\,{\rm mod}\,1.
\ee
At low energies, correlation functions can be calculated from
(\ref{eq:llHamiltonian}) and generically exhibit singularities at the
thresholds of the allowed collective spin and charge degrees of
freedom, with power-law exponents given in terms of the quantities
$\Delta N_\alpha$, $D_\alpha$ and $N_\alpha^\pm$. However, when
working at a finite energy scale RG irrelevant terms have a non-zero
coupling and may (and in fact generically do) significantly alter the
predictions of the unperturbed Luttinger-liquid
\cite{Rozhkov,Rozhkov2,PereiraSirker,PerAffWhit,BettelheimAbanov,PerAffWhit2,PustilnikKhodas,BettelheimAbanov2,KhodasPustilnik,KhodasPustilnik2,ImambekovGlazman,ImambekovGlazman2,ImambekovGlazman3,AbanovBettelheim,Pereira2,SIGPRL,FHLE,SIGPRB,Seabra,PereiraPenc,Rozhkov3,Carmelo,Carmelo2,Carmelo3,Cheianov,PriceLamacraft,EsslerPereira}.  
Over the last decade or so a fairly general method for taking into
account the effects of certain irrelevant operators in the vicinities
of kinematic thresholds has been developed, which is reviewed in
Refs.~\onlinecite{SIG,Pereira}. The case of spin-charge separated
Luttinger liquids has very recently been revisited
\cite{EsslerPereira} in order to make it explicitly compatible with
exactly known properties of the Hubbard model. 
The essence of this approach is that, when considering a response
function, there are thresholds in the $(k,\omega)$-plane that
correspond to particular excitations. In integrable models, these
excitations hold privileged positions: they 
are stable (i.e. have infinite lifetimes) and  can be identified in
terms of the exact solution. If the kinematics near the threshold are
described by a case in which small number of high-energy excitations
carry most of the energy (in the precise sense of up to corrections of
$\mathcal{O}(L^{-1})$) then the problem becomes analogous to that of
the X-ray edge singularity problem for a mobile impurity \cite{Balents}.

In this work we employ mobile impurity methods to study the optical
conductivity 
\be
\sigma_1(\omega) = - \frac{ {\rm Im}\;\chi^J(\omega)}{\omega},\qquad
\chi^J(\omega)= -ie^2 \int_0^\infty dt\ e^{i\omega t} \sum_{l=-L/2}^{L/2-1} \langle GS | [J_{l}(t),J_0(0)]|GS\rangle, \label{eq:optCondDef}
\ee
where $J_j$ is the density of the current operator
\be
J_j = -it \sum_{\sigma} \left[ c^\dagger_{j,\sigma} c^{\pd}_{j+1,\sigma} - c^\dagger_{j+1,\sigma} c^{\pd}_{j,\sigma} \right]. \label{eq:currentOperator}
\ee
In the Mott insulating phase of the Hubbard model the optical
conductivity has been previously determined
\cite{JeckEssler,ControzziEsslerTsvelik,ControzziEsslerTsvelik2,Eric02,Eric03}: 
$\sigma_1(\omega)$ vanishes inside the optical gap $2\Delta$,
where $\Delta$ is the Mott gap. At frequencies $\omega>2\Delta$ there
is a sudden power-law onset
$\sigma_1(\omega)\sim\sqrt{\omega-2\Delta}$. Away from half-filling,
the system is a metal and therefore has a finite conductivity for all
$\omega$, specifically acquiring a Drude
peak\cite{schulzprl,ShastrySutherland} at $\omega=0$.
The low-frequency behaviour has been previously
studied\cite{Giamarchi1,Giamarchi2,Giamarchi3} in the framework of 
Luttinger liquid theory, predicting $\omega^3$ behaviour for $0<\omega\ll t$.
Close to
half-filling one expects most of the spectral weight in
$\sigma_1(\omega)$ to be located above an energy scale $E_{\rm opt}$
that tends to $2\Delta$ as we approach half-filling. 
The scale $E_{\rm opt}$ has been previously correctly identified in
Ref.~\onlinecite{PhysRevLett.84.4673}. In the same work it was
conjectured that the optical conductivity increases in a power-law
fashion above $E_{\rm opt}$ 
\begin{align}
\sigma_1(\omega) \sim \left( \omega - E_{\rm opt} \right)^\zeta \, 
   \Theta(\omega-E_{\rm opt})\ .
   \label{eq:carmeloPrediction}
\end{align}
As we will see in the following, the mobile impurity approach leads to
different results.

The outline of this paper is as follows. In Sec.~\ref{sec:specRep}, we
consider the spectral representation of the optical conductivity and
identify the quantum numbers of the states contributing non-zero
spectral weight. In Sec.~\ref{sec:baRev} we review the Bethe Ansatz
description of the ground state and construct the excited states
considered in Sec.~\ref{sec:specRep}, specifically identifying the
thresholds of these continua. In Sec.~\ref{sec:MIM} we  calculate the
threshold/edge behaviour for the associated excitations via the mobile
impurity approach, fixing the coupling constants using the Bethe
Ansatz to determine the finite-size corrections to the energy in the
presence of the high-energy excitation.  
\section{Spectral representation of the current-current correlator }
\label{sec:specRep}
In considering the optical conductivity as defined in
(\ref{eq:optCondDef}), the basic quantity of interest is  
\be
\langle GS| J_{j+\ell}(t) J_{j}(0)|GS\rangle = \sum_n \langle GS|J_{j+\ell}|n\rangle\langle n|J_{j}| GS\rangle e^{-i(E_n-E_{GS})t},\label{eq:simpleIdentRes}
\ee
where $\{|n\rangle\}$ constitute a complete set of energy
eigenstates. To understand threshold behaviours, we wish to identify
the states contributing to this sum. A crucial insight to this
end are global continuous symmetries and their relation to the energy
eigenstates provided by the exact Bethe Ansatz
solution\cite{EsslerKorepin,EKS,EK1,EK2,EF}. In the case of zero magnetic field
and chemical potential, the Hubbard model possesses two independent
SU(2) symmetries\cite{Heilmann,Yang,book}:
\beA
S^z &= \frac{1}{2} \sum_{i=1}^L (c_{i,\up}^\dagger c_{i,\up}^{\pd} - c_{i,\down}^\dagger c_{i,\down}^{\pd}),&&
S^+ = \sum_{i=1}^L c_{i,\up}^\dagger c_{i,\down}^{\pd},&&
S^- = \sum_{i=1}^L c_{i,\down}^\dagger c_{i,\up}^{\pd}\ ,\\
\eta^z &= \frac{1}{2} \sum_{i=1}^L (c_{i,\up}^\dagger c_{i,\up}^{\pd} + c_{i,\down}^\dagger c_{i,\down}^{\pd} - 1),&&
\eta^+ = \sum_{i=1}^L (-1)^i c_{i,\down}^\dagger c_{i,\up}^{\dagger},&&
\eta^- = \sum_{i=1}^L (-1)^i c_{i,\up}^{\pd} c_{i,\down}^{\pd}. \label{eq:su2eta}
\eeA
The $S^\alpha$ generate the well known spin rotational SU(2)
symmetry, while the $\eta^\alpha$ are known as $\eta$-pairing
generators. The Bethe Ansatz provides us with the \emph{lowest weight
  states}\cite{EsslerKorepin}, which we denote by $|\mathit{LWS};{\bf
  m}\rangle$. Here ${\bf m}$ is a multi-index which labels all
distinct regular Bethe Ansatz states in the sense of
Ref. \onlinecite{EsslerKorepin}. The states are lowest-weight with
respect to the two SU(2) algebras in the sense that
\be
\eta^-|\mathit{LWS};{\bf m}\rangle = 0=S^+|\mathit{LWS};{\bf m}\rangle\ .
\ee 
Each state
$|\mathit{LWS};{\bf m}\rangle$ is defined on a system of length $L$ and has a
well-defined number of electrons $N$ and $z$-component of spin
$S_z$. A complete basis of states is given by 
$\left\{(\eta^+)^{k} (S^-)^{l} |\mathit{LWS};{\bf m}\rangle\;|\; k=0,\dots, L-N;\; l = 0,\dots,2S^z\right\}$. 
For the repulsive Hubbard model below half-filling, the ground state in zero magnetic field and finite chemical potential is a spin singlet and a lowest-weight $\eta$-pairing state i.e.
\be
S^-|GS\rangle = S^+|GS\rangle = \eta^-|GS\rangle =  0.
\ee
Using the algebra defined in (\ref{eq:su2eta}) it is readily verified that $[\eta^-,[\eta^-,J_j]] = 0$ and therefore for integer $m\geq 0$
\be
\langle \mathit{LWS}; {\bf n}|(\eta^-)^{m+1} J_j |GS\rangle = \langle \mathit{LWS};  {\bf n}| (\eta^-)^{m} [\eta^-,J_j] |GS\rangle =  \delta_{m,0}  \langle \mathit{LWS}; {\bf n}| [\eta^-,J_j] |GS\rangle.
\ee
This shows that the only states that may have a non-zero overlap with
$J_j|GS\rangle$ are lowest weight states $|\mathit{LWS};{\bf m}\rangle$ or
$\eta$-pairing descendant states of the form $\eta^+|\mathit{LWS};{\bf
  m}\rangle$, which implies the expansion
\be
J_j|GS\rangle = \sum_{\bf m}\left( a_{\bf m}|\mathit{LWS};{\bf m}\rangle + b_{\bf m} \eta^+|\mathit{LWS};{\bf m}\rangle \right),
\ee
where $a_{\bf m}$, $b_{\bf m}$ are complex coefficients.
Substituting this into (\ref{eq:simpleIdentRes}) provides further
constraints on the subset of energy eigenstates that may make
non-vanishing contributions to the correlator. The subset consists of
\begin{enumerate}
	\item Lowest-weight states with $N_{GS}$ electrons with ${\bf S}^2 = S^z = 0$;
	\item  States of the form $\eta^+|\mathit{LWS};{\bf m}\rangle$, with $|\mathit{LWS};{\bf m}\rangle$ having $N_{GS}-2$ electrons and ${\bf S}^2 = S^z = 0$.
\end{enumerate}
Using that $[H,\eta^+] = -2\mu \eta^+$ we can thus express the
current-current correlator in the form 
\bea
C_{\rm JJ}(\ell,t)&=&\langle GS| J_{j+\ell}(t) J_{j}(0)|GS\rangle\nn
&=& \sum_{\bf m} 
\langle GS|J_{j+\ell} |\mathit{LWS};{\bf m}\rangle\langle \mathit{LWS}; {\bf
  m}|J_{j}|GS\rangle e^{-i(E_{\bf m}-E_{GS})t}\nn
  &+& \sum_{\bf m} \frac{1}{2\eta^z_{\bf m}}\langle GS|J_{j+\ell} \eta^+ |\mathit{LWS}; {\bf m}\rangle
\langle \mathit{LWS}; {\bf  m}|\eta^- J_{j}|GS\rangle e^{-i(E_{\bf m}-E_{GS}-
  2\mu)t}. \label{eq:specAll} 
\eea
The factor of $(2\eta^z_{\bf m})^{-1}$ arises from the normalization
of the state $\eta^+|\mathit{LWS},{\bf m}\rangle$. We note that
$\mu<0$ and hence $-2\mu$ is a positive energy shift. It is not
obvious how to understand the second term in the framework of a mobile
impurity model. However, using the lowest-weight property
$\eta^-|GS\rangle=0$, we can rewrite (\ref{eq:specAll}) in the form  
\bea
C_{\rm JJ}(\ell,t)&=&
\sum_{\bf m} 
\langle GS|J_{j+\ell} |\mathit{LWS};{\bf m}\rangle\langle \mathit{LWS}; {\bf
  m}|J_{j}|GS\rangle e^{-i(E_{\bf m}-E_{GS})t} \nn
  &+&\sum_{\bf m}  \frac{1}{2\eta^z_{\bf m}}
\langle GS|[J_{j+\ell}, \eta^+] |\mathit{LWS}; {\bf m}\rangle\langle \mathit{LWS}; {\bf
  m}|[\eta^-, J_{j}]|GS\rangle e^{-i(E_{\bf m}-E_{GS}-
  2\mu)t}. \label{eq:simplifiedSum0} 
\eea
The main advantage of the representation (\ref{eq:simplifiedSum0}) is
that it only involves regular Bethe Ansatz states, which can be
constructed by standard methods. As we concern ourselves only with the
threshold behaviours of the optical conductivity, we need only focus
on the lower edges of the various excitation continua. As a
consequence of kinematic constraints and matrix-element effects,
processes with a small number of excitations above the ground state
give the dominant contributions to response functions. Defining 
\be
\mathcal{O}_j = [\eta^-, J_j] = 2it(-1)^j\left( c_{j,\down} c_{j+1,\up} + c_{j+1,\down} c_{j,\up}\right), \label{eq:ojDef}
\ee
we can recast \fr{eq:simplifiedSum0} in the form
\bea
C_{\rm JJ}(\ell,t)&=&\sum_{\bf m} |\langle GS|J_j |\mathit{LWS};{\bf m}\rangle|^2
e^{-i(E_{\bf m}-E_{GS})t+iP_{\bf m}\ell}\nn
&+&\sum_{\bf m}  \frac{1}{2\eta^z_{\bf m}}
|\langle GS|{\cal O}_j^\dagger |\mathit{LWS}; {\bf m}\rangle|^2
e^{-i(E_{\bf m}-E_{GS}-  2\mu)t+i(P_{\bf m}-\pi)\ell}\equiv
C^{(1)}_{\rm JJ}(\ell,t)+C^{(2)}_{\rm JJ}(\ell,t).
\label{eq:simplifiedSum} 
\eea
Here the additional contribution to the momentum arises because
acting with $\eta^+$ shifts the momentum by $\pi$. 
If the ground state contains $N$ fermions, the contribution $C_{\rm
  JJ}^{(2)}(\ell,t)$ is proportional to $1/(L-N+2)$, and can therefore
be dropped in the thermodynamic limit away from half filling. However,
as we are interested in densities close to one fermion per site it is
useful to retain it in view of potential comparisons to numerical
results for finite-size systems. The optical conductivity can
then be written as
\be
\sigma_1(\omega)=\frac{e^2}{2\omega}\left[\sum_{a=1}^2\sum_\ell\int_{-\infty}^\infty
dt\ e^{i\omega t}\ C_{\rm
JJ}^{(a)}(\ell,t)-\big\{\omega\rightarrow-\omega\}\right]
\equiv\sum_{a=1}^2\sigma_1^{(a)}(\omega).
\label{sigma_master}
\ee
\section{Bethe Ansatz for the Hubbard model}\label{sec:baRev}
To gain further insight into the representation (\ref{eq:simplifiedSum}) we now
construct the ground state and low-lying excitations above it. We
first  calculate the energy of such excitations in the thermodynamic
limit. This will allow us to identify, on kinematic grounds, which
states within the manifold identified earlier are important with
respect to the threshold behaviours we aim to describe. 
We therefore recapitulate some results from Ref. \onlinecite{book} to allow a self-contained discussion.

For large system sizes, the eigenstates of the repulsive Hubbard model can be expressed in terms of solutions of the Takahashi equations, expressed 
in terms of so-called \emph{counting functions}. In the case of $N$ electrons, $M$ of which are spin-down, these are defined by
\begin{align}
	z_c(k_j) &= k_j +  \frac{1}{L} \sum_{n=1}^\infty \sum_{\alpha=1}^{M_n} \theta \left( \frac{\sin k_j - \Lambda^n_\alpha}{nu}\right) + \frac{1}{L} \sum_{n=1}^\infty \sum_{\alpha=1}^{M_n'}\theta\left( \frac{\sin k_j - {\Lambda'}_\alpha^n}{nu}\right)  ,&&\qquad j=1,\dots, N-2M', \\
z_n(\Lambda^n_\alpha) &= \frac{1}{L}\sum_{j=1}^{N-2M'} \theta\left( \frac{\Lambda^n_\alpha - \sin k_j}{nu}\right) - \frac{1}{L}\sum_{m=1}^\infty \sum_{\beta=1}^{M_m} \Theta_{nm} \left( \frac{\Lambda^n_\alpha - \Lambda^m_\beta}{u} \right),&&\qquad \alpha=1,\dots,M_n,\\
z'_n({\Lambda'}^n_\alpha) &= -\frac{1}{L}\sum_{j=1}^{N-2M'} \theta\left( \frac{ {\Lambda'}^n_\alpha - \sin k_j}{nu}\right) - \frac{1}{L}\sum_{m=1}^\infty \sum_{\beta=1}^{M'_m} \Theta_{nm} \left( \frac{ {\Lambda'}^n_\alpha - {\Lambda'}^m_\beta}{u} \right) \nn
&+ 2{\rm Re}[\arcsin( {\Lambda'}^n_\alpha + niu)],&&\qquad \alpha=1,\dots,M_n',
\end{align}
where $\theta(x) = 2 \arctan(x)$, $u=U/4t$, 
\be
\Theta_{nm}(x) = \begin{cases} \theta\left( \frac{x}{|n-m|} \right) + 2\theta\left( \frac{x}{|n-m|+2} \right)+ \dots + 2\theta\left( \frac{x}{n+m-2} \right) + \theta\left( \frac{x}{n+m} \right),& n\neq m\\
2\theta\left( \frac{x}{2} \right) + 2\theta\left( \frac{x}{4} \right) + \dots + 2\theta\left( \frac{x}{2n-2} \right) + \theta\left( \frac{x}{2n} \right),& n=m \end{cases},
\ee
and
\be
M = \sum_{n=1}^\infty n(M_n + M_n'), \qquad M' = \sum_{n=1}^\infty n M_n'.
\ee
Takahashi's equations are 
\be
z_c(k_j) = \frac{2 \pi I_j}{L},\qquad z_n(\Lambda^n_\alpha) = \frac{2\pi J^n_\alpha}{L},\qquad z'_n({\Lambda'}^n_\alpha) = \frac{2\pi {J'}^n_\alpha}{L}.
\ee
Here the sets $\{ I_j\}$, $\{ J^n_\alpha\}$, $\{ {J'}^n_\alpha \}$ consist of integers or half-odd integers depending on the particular state under consideration, obeying the ``selection rules''
\be
\begin{aligned}
I_j &\in \begin{cases} \mathbb{Z} + \frac{1}{2} & {\rm if}\ \sum_m (M_m +M'_m)\ {\rm odd}\\
	\mathbb{Z} & {\rm if}\ \sum_m (M_m +M'_m)\ {\rm even}
\end{cases},&& \qquad -\frac{L}{2} < I_j \leq \frac{L}{2},\\
J^n_\alpha &\in \begin{cases} \mathbb{Z} & {\rm if}\ N-M_n\ {\rm odd}\\
	\mathbb{Z}+\frac{1}{2} & {\rm if}\ N-M_n\ {\rm even}
\end{cases},&& \qquad |J^n_\alpha| \leq \frac{1}{2} (N-2M' - \sum_{m=1}^\infty t_{nm} M_m -1),\\
{J'}^n_\alpha &\in \begin{cases} \mathbb{Z} & {\rm if}\ L-N+M'_n\ {\rm odd}\\
	\mathbb{Z}+\frac{1}{2} & {\rm if}\ L-N+M'_n\ {\rm even}
\end{cases},&&
\qquad |J'^n_\alpha| \leq \frac{1}{2}\left( L - N + 2M' - \sum_{m=1}^\infty t_{nm} M_m' - 1 \right),
\label{eq:integerParity}
\end{aligned}
\ee
where $t_{nm}= 2\, { \rm min} (m,n) - \delta_{mn}$.
The energy and momentum, measured in units of $t$, of an eigenstate characterised by the set of roots $\{ k_j, \Lambda_\alpha^n, {\Lambda'}_\beta^m\}$ are given by
\be
E = - \sum_{j=1}^{N-2M'} \left( 2\cos k_j + \mu + 2u + B\right) + 2BM + 4 \sum_{n=1}^\infty \sum_{\beta=1}^{M_n'} {\rm Re} \sqrt{ 1 - ( {\Lambda'}^n_\beta + niu)^2} + Lu, \label{eq:energyFull}
\ee
\be
	P = \left[  \sum_{j=1}^{N-2M'} k_j - \sum_{n=1}^\infty \sum_{\beta=1}^{M_n'} \left( 2\,{\rm Re} \arcsin\left( {\Lambda'}^n_\beta + niu \right) - (n+1)\pi \right) \right] {\rm mod}\, 2\pi.
\ee
The monotonicity of the counting functions ensures that specifying a set of integers/half-odd integers in accordance with the ``selection rules'' uniquely determines a solution of the Takahashi equations.
\subsection{Ground state}
\begin{figure}[ht!]
\begin{center}
	\includegraphics{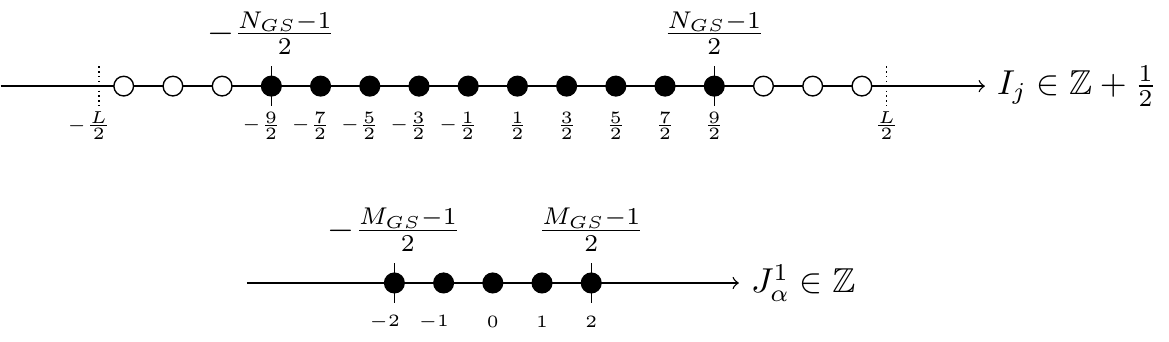}
\end{center}
\caption{Configuration of the integers for the ground state, explicit numbers given are for $L=16$, $N_{GS}=10$, $M_{GS}=5$} \label{fig:gsInts}
\end{figure}

We consider the case where $L$ is even, the total number of electrons
$N_{GS}$ is even and the number of down spins $M_{GS}$ is odd. The
ground state is then obtained by choosing the set $\{ I_j, J_\alpha^n,
{J'}_\beta^m \}$ to be\cite{book} 
\begin{align}
I_j &= - \frac{N_{GS}}{2} - \frac{1}{2} + j,&&  j=1,\ldots,N_{GS} ,\\
J^1_\alpha &= - \frac{M_{GS}}{2} - \frac{1}{2} + \alpha,&&  \alpha = 1,\ldots, M_{GS}.
\end{align}
This configuration is shown for the example $L=16$, $N_{GS}=2M_{GS}=10$ in
\figr{fig:gsInts}. We denote the ground 
state, in the previously established notation, by 
\be
|GS\rangle = |\mathit{LWS};\{I_j\},\{J_\alpha^1\}\rangle.
\ee

\subsubsection{Thermodynamic limit}
On taking the thermodynamic limit at fixed density $n_{GS}$ and
magnetisation $m_{GS}$ the roots become dense and we can describe the
ground state in terms of root densities $\rho_{c,0}$, $\rho_{s,0}$,
which satisfy linear integral 
equations\cite{book} 
\begin{align}
\rho_{c,0}(k) &= \frac{1}{2\pi} + \cos k \int_{-A}^A \d \Lambda\,
a_1(\sin k - \Lambda) \rho_{s,0}(\Lambda), \label{eq:gsrhoc} \\
\rho_{s,0}(\Lambda) &= \int_{-Q}^Q \d k\, a_1(\Lambda - \sin k)
\rho_{c,0}(k) - \int_{-A}^A \d \Lambda'\, a_2(\Lambda-\Lambda')
\rho_{s,0}(\Lambda'). \label{eq:gsrhos} 
\end{align}
Here $a_n(x) = \frac{2nu}{2\pi} \frac{1}{(nu)^2 + x^2}$ and the
integration boundaries $Q$ and $A$ are determined by 
\be
\int_{-Q}^Q \d k\, \rho_{c,0}(k) = n_{GS},\qquad \int_{-A}^A \d
\Lambda\, \rho_{s,0}(\Lambda) = \frac{1}{2} \left( n_{GS}-2m_{GS} \right). 
\ee
The energy density of the system is given to $o(1)$ by\cite{book}
\be
e_{GS} = \int_{-Q}^Q \frac{\d k}{2\pi} \varepsilon_c(k) + u, \label{eq:egs}
\ee
where
\begin{align}
\varepsilon_c(k) &= -2 \cos k -\mu - 2u - B + \int_{-A}^A \d \Lambda \,a_1(\sin k - \Lambda) \varepsilon_s(\Lambda), \label{eq:ec}\\
\varepsilon_s(\Lambda) &=2B + \int_{-Q}^Q \d k\, \cos k\, a_1(\Lambda - \sin k) \varepsilon_c(k) - \int_{-A}^A \d \Lambda'\, a_2(\Lambda - \Lambda')\varepsilon_s(\Lambda'). \label{eq:es}
\end{align}
The dressed energies $\varepsilon_c(k)$ and $\varepsilon_s(\Lambda)$ satisfy $\varepsilon_c(\pm Q) = \varepsilon_s(\pm A)=0$. 
The dressed momenta are given by\cite{book}
\begin{align}
p_c(k) &=  k + \int_{-A}^A \d\Lambda\, \rho_{s,0}(\Lambda) \theta\left( \frac{\sin k - \Lambda}{u} \right),\\
p_s(\Lambda) &= \int_{-Q}^Q \d k\, \rho_{c,0}(k) \theta\left( \frac{\Lambda - \sin k}{u} \right) - \int_{-A}^A \d\Lambda'\,\rho_{s,0}(\Lambda') \theta\left( \frac{\Lambda-\Lambda'}{2u} \right).
\end{align}
\subsection{Excitations contributing to $C^{(1)}_{\rm JJ}(\ell,t)$.}
We now turn to excited states that contribute to the spectral
representation \fr{eq:simplifiedSum} of 
$C^{(1)}_{\rm JJ}(\ell,t)$. These are lowest weight states
of the spin and $\eta$-pairing SU(2) algebras with quantum numbers
$N=N_{GS}$, $M=M_{GS}$.
\subsubsection{``Particle-hole'' excitation with $N=N_{GS}$, $M=M_{GS}$. }
\label{PHexc}
\begin{figure}[ht!]
\begin{center}
	\includegraphics{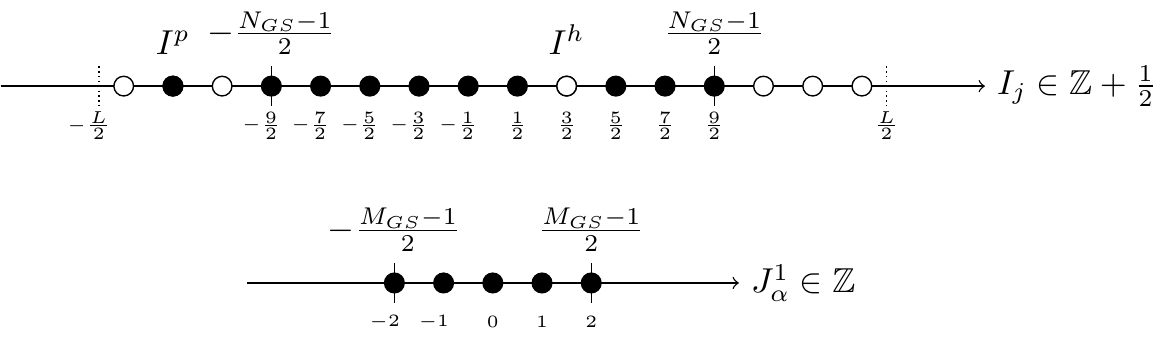}
\end{center}
\caption{Configuration of the integers for the particle-hole excitation above the ground state, explicit numbers given are for $L=16$, $N_{GS}=10$, $M_{GS}=5$} \label{fig:phIntegerConfiguration}
\end{figure}

\begin{figure}[ht!]
	\includegraphics[width=0.6\columnwidth]{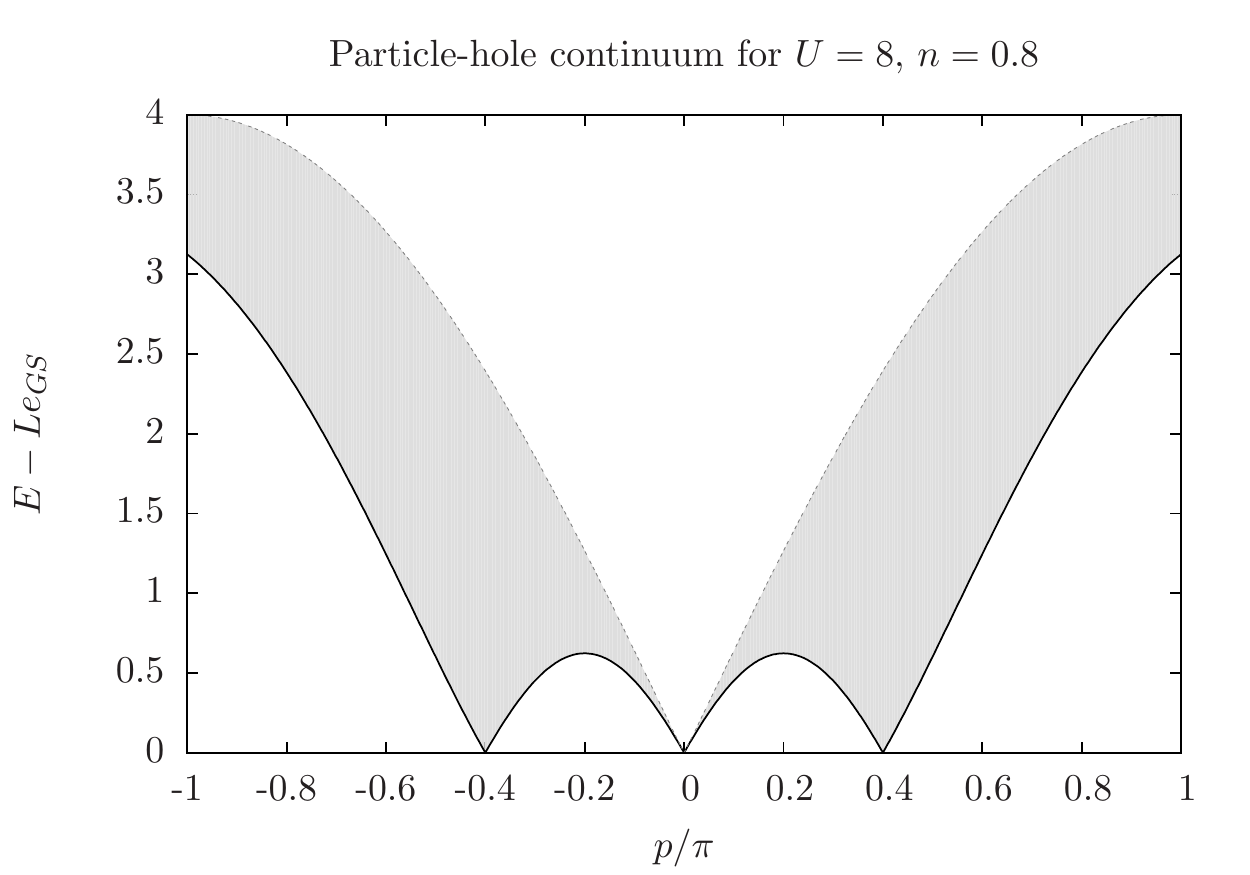}
	\caption{Particle-hole excitation continuum above the ground state} \label{fig:particleHoleContinuum}
\end{figure}

Creating a particle-hole excitation in the charge degrees of freedom yields a
state with the same charge and spin quantum numbers as the ground
state, but with a finite momentum and energy difference. The
(half-odd) integers for this type of excitation are given by 
\be
I_j =  \begin{cases} -\frac{N_{GS}+1}{2} + j + \Theta\left( -\frac{N_{GS} +1}{2} +j-I^h\right), & j=1,\dots,N_{GS}-1\\
		I^p, &				j=N_{GS}\\
	\end{cases},
	\ee
	\be
\begin{aligned}
J_\alpha &= -\frac{M_{GS}+1}{2} + \alpha, &\qquad& \alpha=1,\dots,M_{GS},
	\end{aligned}
\ee
where $\Theta(x) = 1$ for $x\geq0$ and $0$ otherwise.
The arrangement for these integers is shown in \figr{fig:phIntegerConfiguration}.
This excitation is two-parametric and has an energy and momentum of the form
\beA
E&= e_{GS}L + \varepsilon_c(k^p) - \varepsilon_c(k^h) + o(1),\\
P&= p_c(k^p) - p_c(k^h) + o(1),
\eeA
where the rapidities are determined by  $z_c(k^h) = \frac{2\pi I^h}{L}$, $z_c(k^p) = \frac{2\pi I^p}{L}$. This forms a continuum of excitations above the ground state, shown in \figr{fig:particleHoleContinuum}.

\subsubsection{``$k$-$\Lambda$ string'' excitation}
\label{sec:klStr}
\begin{figure}[ht!]
\begin{center}
	\includegraphics{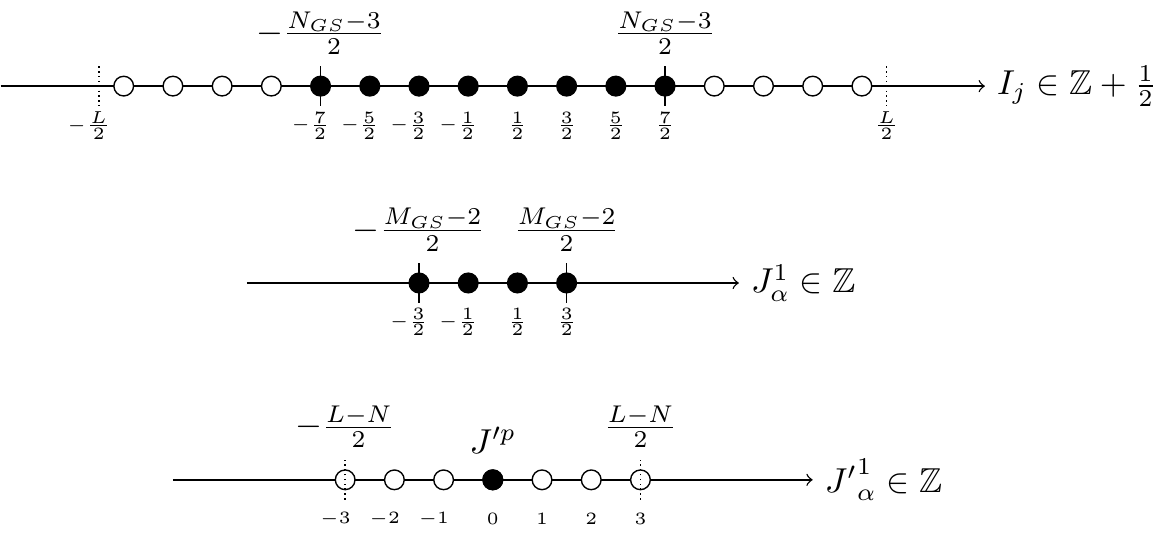}
\end{center}
	\caption{Configuration of the integers for the $k$-$\Lambda$ string excited state} \label{fig:exInts}
\end{figure}

\begin{figure}[ht!]
	\subfloat[ {Dispersion relation
		$\varepsilon_{k\Lambda}(\Lambda(p))$ for the
	$k$-$\Lambda$ string for various $U$ and $n$. Each curve has been
shifted down by $-2\mu$.}]{
		\includegraphics[width=0.45\columnwidth]{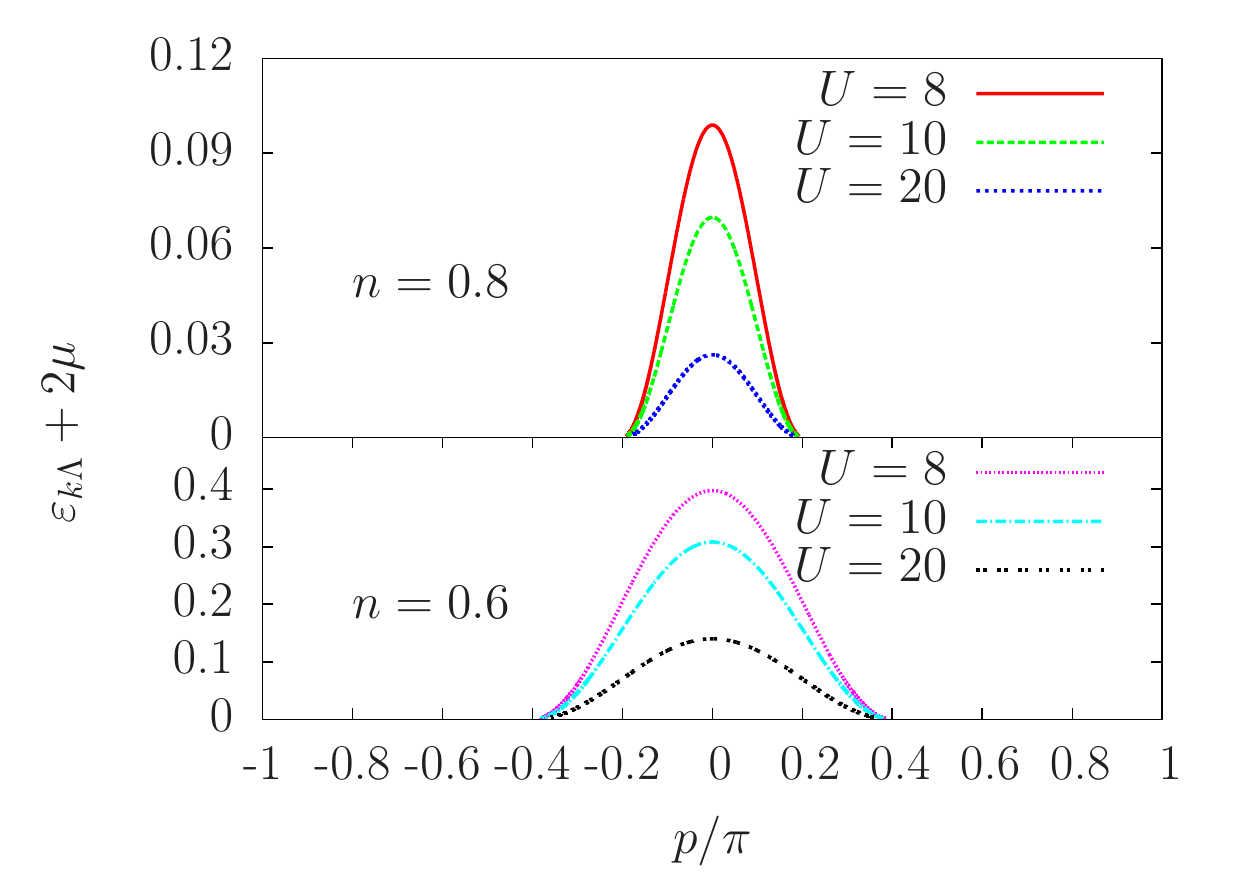}
		\label{subfig:kld}
	}
	\subfloat[$k$-$\Lambda$ and charge particle-hole excitation continuum. ]{
		\includegraphics[width=0.45\columnwidth]{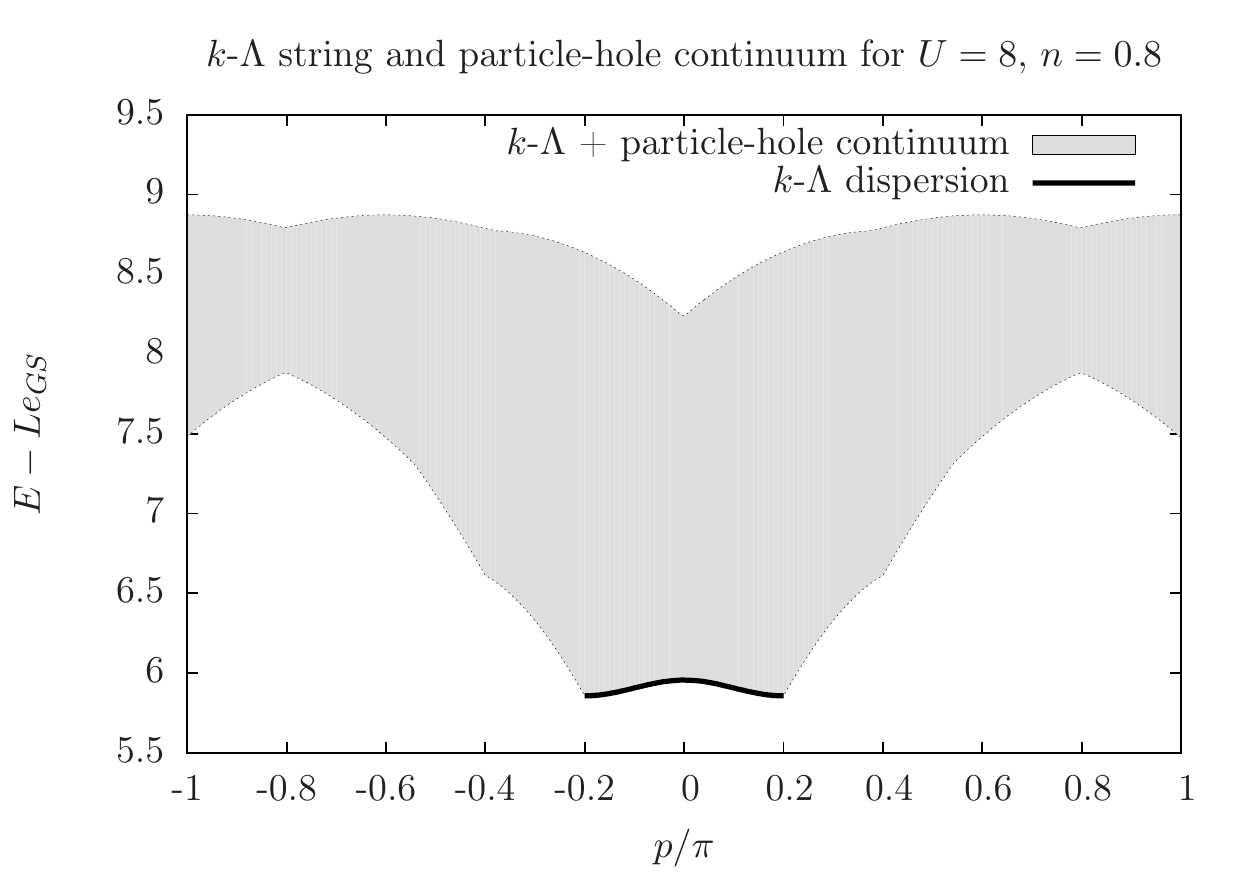} \label{subfig:klco}
	}
	\caption{$k$-$\Lambda$ string dispersion for various $U$ and $n$, and particle-hole excitation continuum above this for $U=8$, $n=0.8$. 
	For small momenta, the $k$-$\Lambda$ string dispersion marks the lower edge of a continuum described by additional excitations e.g. particle-hole excitations in the charge sector.	
	}\label{fig:kldisp}
\end{figure}

We start by considering excitations with $N=N_{GS}$, $M=M_{GS}$
involving a single (``$k$-$\Lambda$ string'') bound state. This
excitation has been considered previously e.g. in Section 7.7.2 of
Ref. \onlinecite{book}. It involves having a 
single (half-odd) integer in the sector corresponding to the set $\{
{J'}^1_\alpha \}$.  
The lowest-energy bound state which can be created comprises of two
$k$s and one $\Lambda$ forming a string pattern in the complex plane.
The Takahashi equations describe the real centres of these and other 
root patterns. 
The case we consider is realised by the integer configuration
\begin{align}
I_j &= - \frac{N_{GS}-2}{2} - \frac{1}{2} + j, \qquad &j=1,\ldots,N_{GS}-2,\\
J^1_\alpha &= - \frac{M_{GS}-1}{2} - \frac{1}{2} + \alpha, \qquad &\alpha = 1,\ldots, M_{GS}-1,\\
{J'}^1_\beta &= {J'}^p, \qquad &\beta=1,
\end{align}
which is displayed in \figr{fig:exInts}. In the notations used above,
we can denote this excited state by $|\mathit{LWS}; \{I_j\},\{J_\alpha^1\}, \{ {J'}_\beta^1 \} \rangle$.
We can again take the thermodynamic limit and compare the energy of this excited state with that of the ground state. Following similar manipulations to the case of the ground state energy, the $\mathcal{O}(1)$ corrections can be calculated\cite{book}.
The energy is given by
\be
E = L e_{\rm GS} + \varepsilon_{k\Lambda}(\Lambda^p) + o(1),
\ee
where 
\be
\varepsilon_{k\Lambda}(\Lambda) = 4{\rm Re}\sqrt{ 1- (\Lambda - iu)^2} - 2\mu - 4u + \int_{-Q}^Q dk \cos k\; a_1(\sin k -\Lambda)\; \varepsilon_c(k).
\ee
The momentum is given by $P=p_{k\Lambda}(\Lambda^p)$, where
\be
p_{k\Lambda}(\Lambda') = -2{\rm Re} \arcsin\left( \Lambda' + iu \right) + \int_{-Q}^Q \d k\, \rho_{c,0}(k) \theta\left( \frac{\Lambda'-\sin k}{u} \right),
\ee
and $\Lambda^p$ is determined by $z'_1(\Lambda^p) = \frac{2 \pi J'^p}{L}$. 
This form can be readily interpreted physically as a particle-like
excitation above the ground state. 
The $k$-$\Lambda$ string dispersion describes the threshold of an
excitation continuum obtained by adding e.g. particle-hole
excitations in the charge sector. 
The dispersion relation for this excitation and the 
particle-hole continuum is shown in \figr{fig:kldisp}. 
The existence of such a continuum at $p=0$ is necessary to understand 
the problem within the mobile impurity approach to threshold 
singularities.

\subsection{Excitations contributing to $C^{(2)}_{\rm JJ}(\ell,t)$.}
We now turn to excited states that contribute to the spectral
representation \fr{eq:simplifiedSum} of 
$C^{(2)}_{\rm JJ}(\ell,t)$. {As we have re-expressed $C^{(2)}_{\rm
  JJ}(\ell,t)$ in terms of matrix elements of the operator ${\cal
  O}_j^\dagger$ defined in \fr{eq:ojDef}, we will focus on excited states
$|\mathit{LWS}; {\bf m}\rangle$ that have non-vanishing matrix elements
$\langle GS|{\cal O}_j^\dagger |\mathit{LWS}; {\bf m}\rangle\neq 0$ 
These are lowest weight states of the spin and $\eta$-pairing SU(2)
algebras and their quantum numbers are $N=N_{GS}-2$, $M=M_{GS}-1$. It
is of course straightforward to translate back to excitations
contributing to the original spectral representation \fr{eq:specAll}:
all that is required is to act with $\eta^\dagger$ on the states we
discuss in the following. 
}
\subsubsection{``Particle-hole'' excitation with $N=N_{GS}-2$, $M=M_{GS}-1$. }
\label{sec:shiftedAntiHolon}
\begin{figure}[ht!]
\begin{center}
	\includegraphics{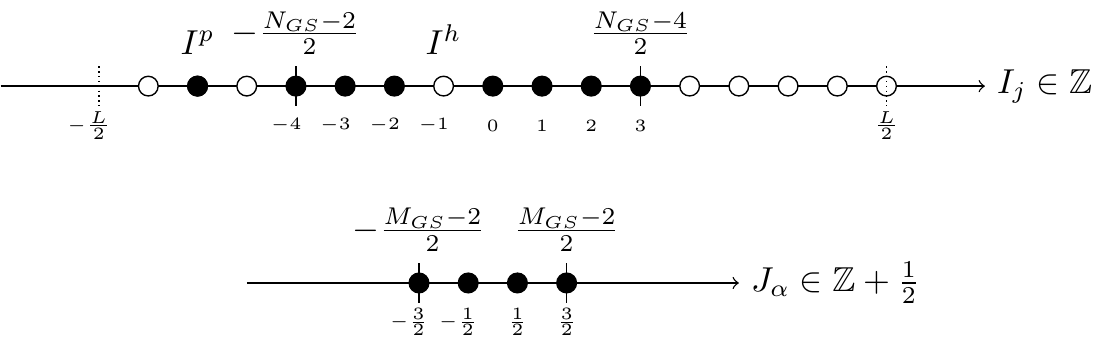}
\end{center}
\caption{Integer configuration for the particle-hole excitation, explicit numbers for $L=16$, $N_{GS}=10$, $M_{GS}=5$. 
}
\label{fig:heIntEx}
\end{figure}

The integer configuration for this type of excitation is given by
\be
I_j =  \begin{cases} -\frac{N_{GS}}{2} + j + \Theta\left( \frac{-N_{GS}}{2} + j - I^h \right), & j=1,\dots,N_{GS}-3\\
		I^p,&				j=N_{GS}-2
	\end{cases},
	\ee
	\be \label{eq:heInts}
\begin{aligned}
J_\alpha &= -\frac{M_{GS}}{2} + \alpha, &\qquad& \alpha=1,\dots,M_{GS}-1
	\end{aligned}.
\ee
This is shown graphically in \figr{fig:heIntEx}. 

\begin{figure}[ht!]
	\includegraphics[width=0.6\columnwidth]{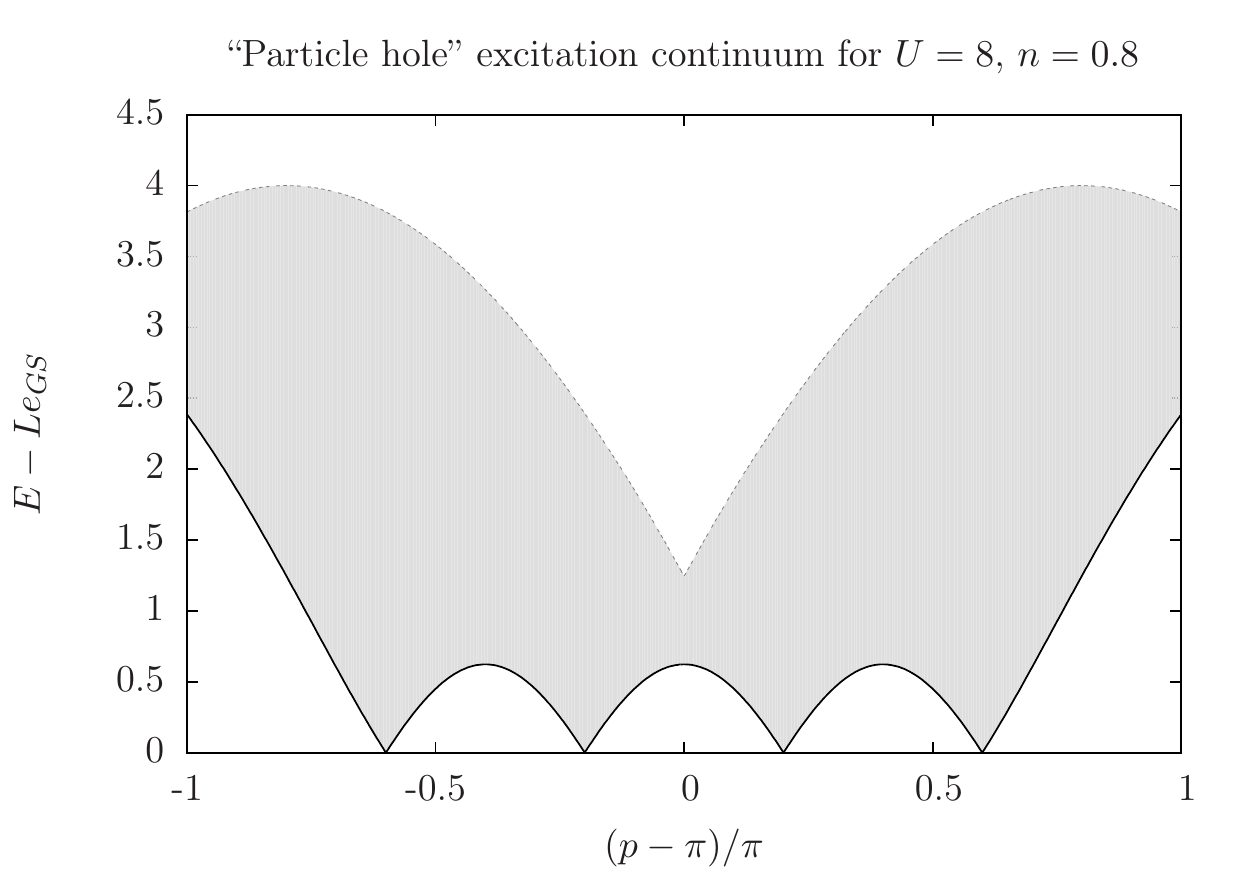}
	\caption{Continuum for particle-hole excitation with momentum shifted for clarity.}\label{fig:shiftedParticleHoleContinuum}
\end{figure}

In complete analogy to the previous case, the energy and momentum of
this state are given by 
\bea
E &=& L e_{\rm GS}  + \varepsilon_c(k^p) - \varepsilon_c(k^h) + o(1),\nn
P &=& p_c(k^p) - p_c(k^h) \pm 2k_F  + o(1),
\label{EPdesc1}
\eea
where $k^p$ and $k^h$ are determined by $z_c(k^p) = \frac{2\pi
  I^p}{L}$, $z_c(k^h) = \frac{2\pi I^h}{L}$. The contributions $\pm
2k_F$ arise from the asymmetry of the charge ``Fermi sea'', leaving
a choice of two parity-related states. The continuum of excitations
given by \fr{EPdesc1} is shown in
\figr{fig:shiftedParticleHoleContinuum}, and consists of the union of
two copies of  the continuum depicted in
\figr{fig:particleHoleContinuum} shifted by $\pm 2k_F$
respectively. We note that in order to make closer contact with the
spectral representation \fr{eq:simplifiedSum} we have shifted the
momentum by $\pi$.
\subsubsection{``Two particle'' excitation with $N=N_{GS}-2$, $M=M_{GS}-1$.}
\label{descpp}
\begin{figure}[ht!]
\begin{center}
	\includegraphics{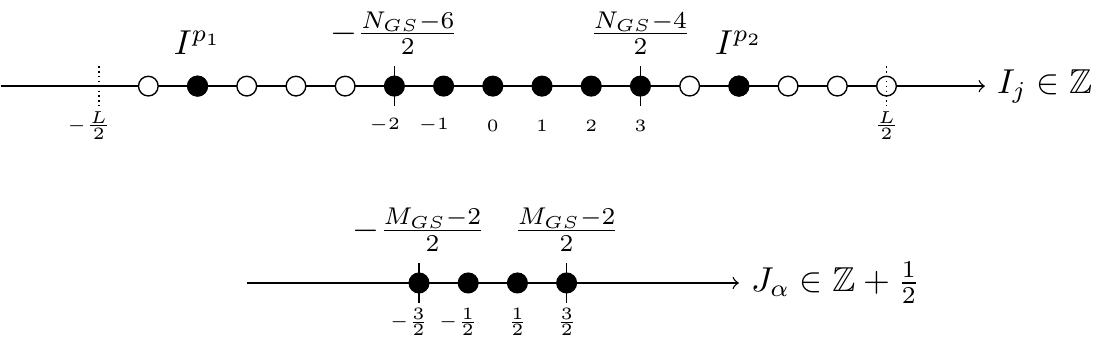}
\end{center}
\caption{Integer configuration for the particle-particle excitation, explicit numbers for $L=16$, $N_{GS}=10$, $M_{GS}=5$. 
}
\label{fig:ppIntEx}
\end{figure}

\begin{figure}[ht!]
\begin{center}
	\includegraphics[width=0.6\columnwidth]{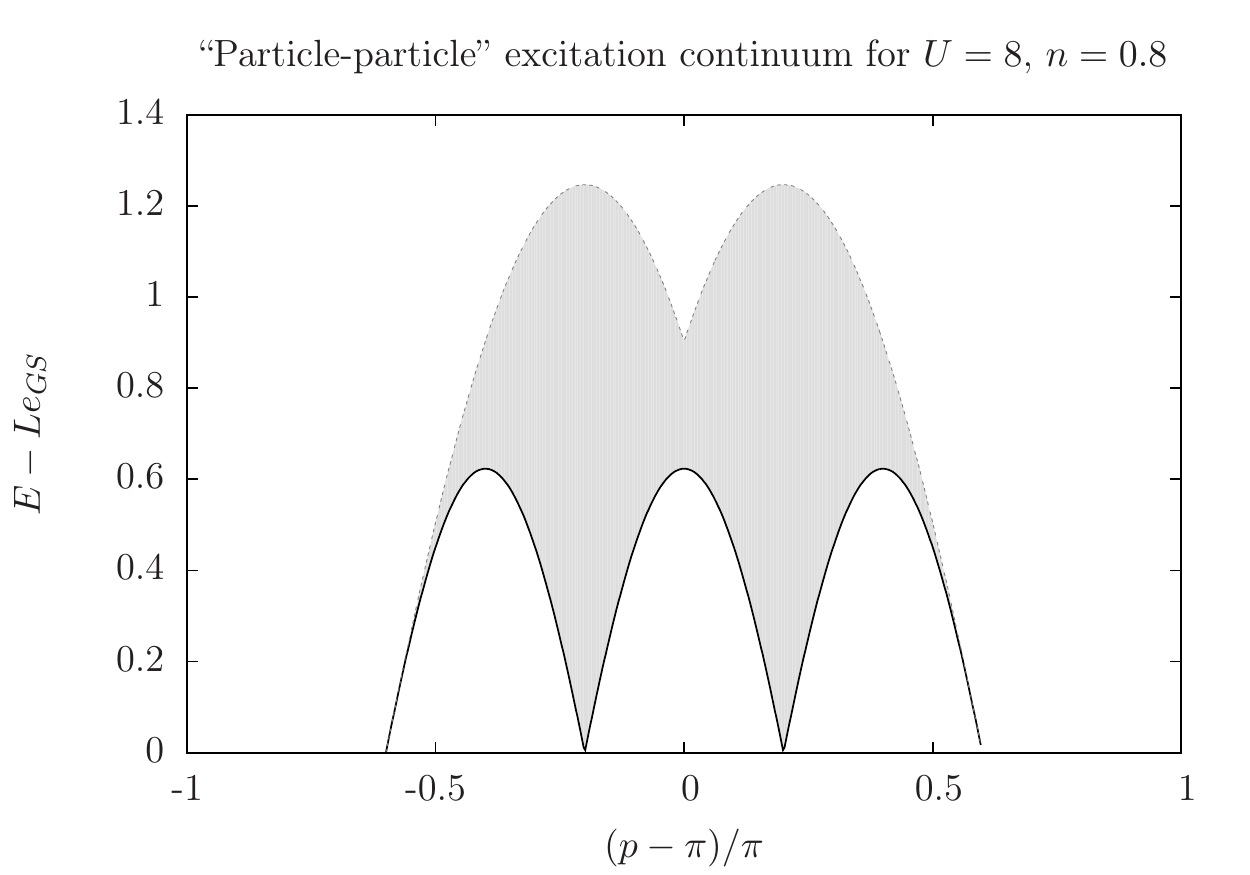}
	\caption{Continuum for particle-particle excitation with momentum shifted for clarity.} \label{fig:twoPartContinuum}
\end{center}
\end{figure}

A closely related type of excitation corresponds to the choice of
(half-odd) integers
\be
I_j =  \begin{cases} -\frac{N_{GS}-4}{2} + j,  & j=1,\dots,N_{GS}-4\\
	I^{p_1},&				j=N_{GS}-3,\\
	I^{p_2},&				j=N_{GS}-2\\
	\end{cases},
	\ee
	\be
\begin{aligned}
J_\alpha &= -\frac{M_{GS}}{2} + \alpha, &\qquad& \alpha=1,\dots,M_{GS}-1.
\end{aligned}
\ee
Such a configuration is shown in \figr{fig:ppIntEx} and can be thought
of as involving two particles associated with $I^{p_1}$ and $I^{p_2}$
respectively. 
The energy and momentum of this excitation are
\bea
E &=& L e_{\rm GS} + \varepsilon_c(k^{p_1}) + \varepsilon_c(k^{p_2})  + o(1),\nn
P &=& p_c(k^{p_1}) + p_c(k^{p_2}) \pm 2k_F + o(1), \label{eq:ppMtm}
\label{EPdesc2}
\eea
with $z_c(k^{p_i}) = \frac{2\pi I^{p_i}}{L}$. The continua
corresponding to \fr{EPdesc2} are shown in
\figr{fig:twoPartContinuum}. We note that both possible choices $\pm
2k_F$ have been taken into account, and we have again shifted the
total momentum by $\pi$ in order to make closer contact with
the spectral representation \fr{eq:simplifiedSum} of our correlator.

\subsubsection{``Two hole'' excitation with $N=N_{GS}-2$, $M=M_{GS}-1$.}
\label{sec:shiftedTwoHolon}
\begin{figure}[ht!]
\begin{center}
	\includegraphics{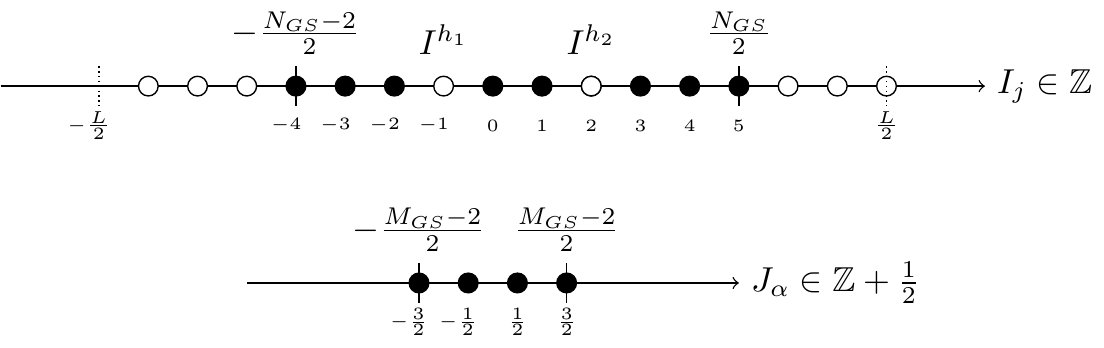}
\end{center}
\caption{Integer configuration for two hole excited state, explicit numbers for $L=16$, $N_{GS}=10$, $M_{GS}=5$.
}\label{fig:ex3Ints}
\end{figure}

\begin{figure}[ht!]
	\includegraphics[width=0.6\columnwidth]{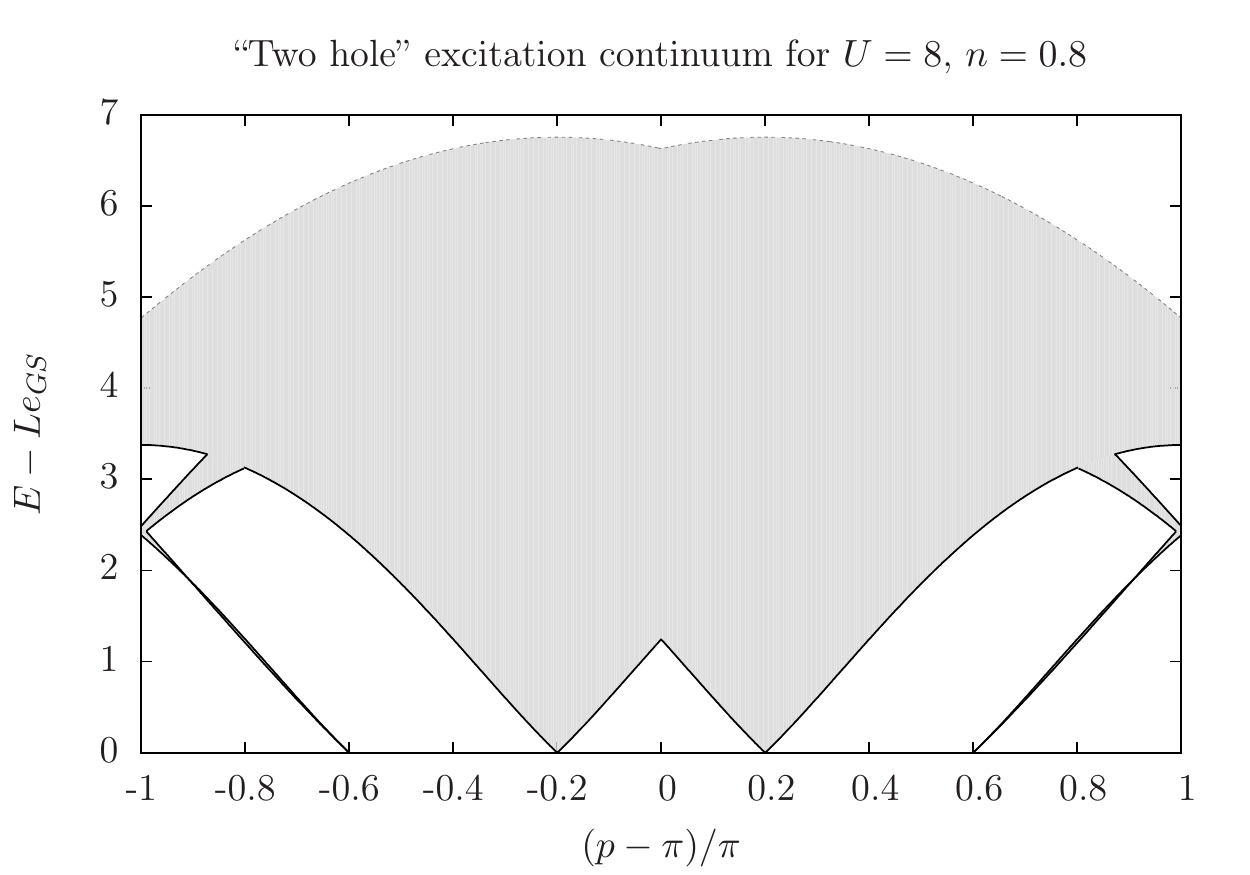}
	\caption{Continuum for two hole excitation with momentum shifted for clarity.}\label{fig:twoHoleContinuum}
\end{figure}

Finally, we consider excitations characterised by the distribution of
(half-odd) integers
\begin{align}
	I_j &= - \frac{N_{GS}}{2} + j + \Theta\left(- \frac{N_{GS}}{2} + j -I^{h_1}\right)+  \Theta\left(- \frac{N_{GS}}{2} + j -I^{h_2}\right), \qquad &j=1,\ldots,N_{GS}-2,\\
J_\alpha &= - \frac{M_{GS}}{2} +  \alpha, \qquad &\alpha = 1,\ldots,M_{GS}-1,
\end{align}
which is displayed in \figr{fig:ex3Ints}.  
We see that these states can be viewed as involving two holes
associated with $I^{h_1}$ and $I^{h_2}$ respectively. The energy and
momentum of this excitation are given by
\bea
E &=& L e_{\rm GS} - \varepsilon_c(k^{h_1}) - \varepsilon_c(k^{h_2}) + o(1),\nn
P &=& -p_c(k^{h_1}) - p_c(k^{h_2}) \pm 2k_F  + o(1),
\eea
with $z_c(k^{h_i}) = \frac{2\pi I^{h_i}}{L}$. The continua for these
excitations are shown in \figr{fig:twoHoleContinuum}, where we have taken both possible choices of $\pm 2k_F$ into account and we again have shifted the 
total momentum by $\pi$ in order to make closer contact with
the spectral representation \fr{eq:simplifiedSum} of our correlator.

\subsection{Excitation thresholds at commensurate fillings}
\begin{figure}[ht!]
\includegraphics[width=0.6\columnwidth]{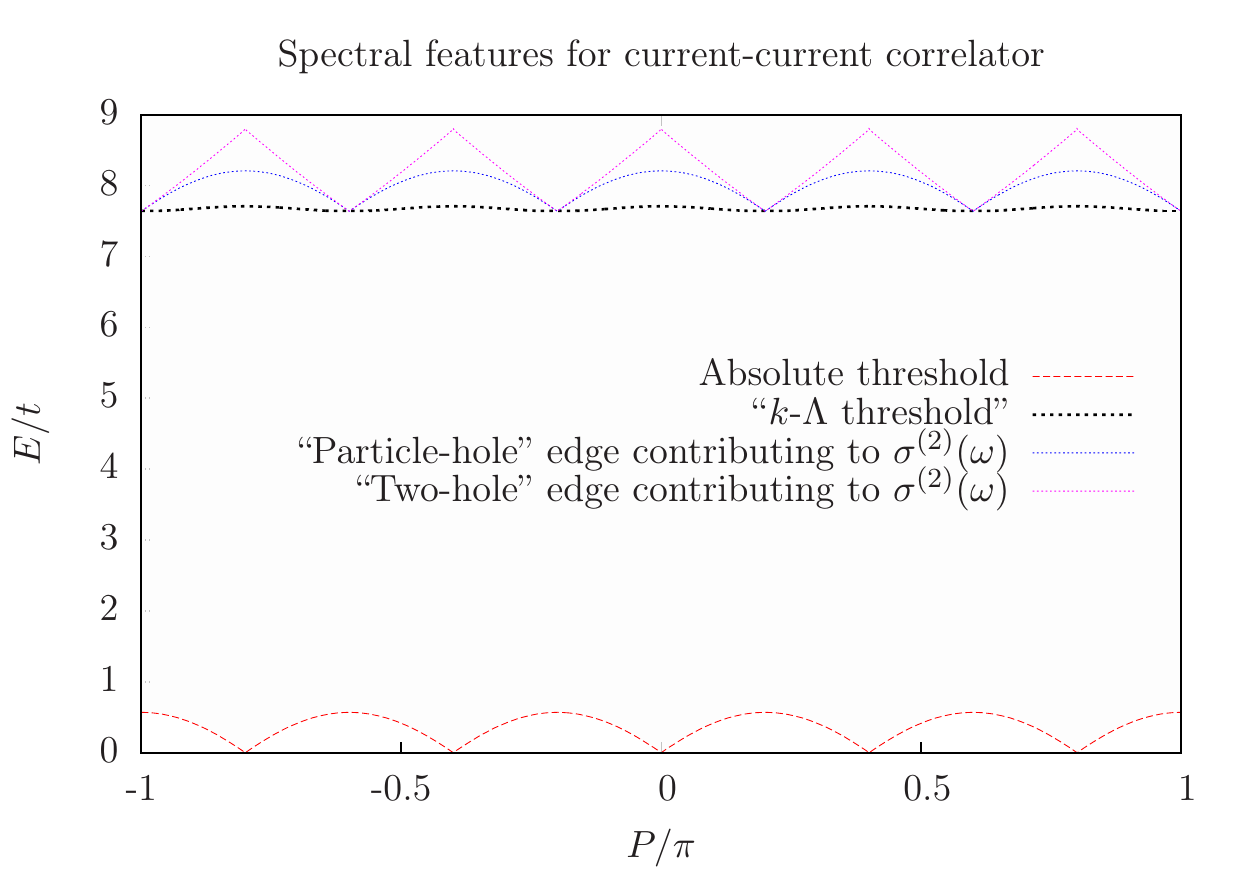}
\caption{The continuum of lowest-lying excitations of the Hubbard
model involving only the charge sector for $n=0.8$, $U=10$. As the
optical conductivity is defined at zero momentum, only the features
encountered at $P=0$ are relevant. The $k$-$\Lambda$ string
dispersion defines the lower edge of a continuum of excitations
involving the $k$-$\Lambda$ string. The contributions to
$\sigma^{(2)}(\omega)$ are shifted by $-2\mu$, in accordance with the
spectral representation (\ref{eq:simplifiedSum}).}\label{fig:spectralPlot} 
\end{figure}

By considering additional excitations around the ``Fermi points'' in
the charge sector we can construct other excitations that are
degenerate in energy (to $o(1)$), but differ in their momenta by
integer multiples of $4 k_F$. As we consider the case of zero
magnetic field, there is no freedom to rearrange the integers in the
spin sector that leads to a lower energy for a given momentum. 
In this way we can determine the thresholds for a given
class of excited states. 
\begin{enumerate}
\item{} The absolute threshold is obtained by combining the particle-hole
excitation of Sec. \ref{PHexc} with zero-energy particle-hole
excitations at the ``Fermi points'' in the charge sector, which shift
the momentum by multiples of $4k_F$. It is depicted by a dashed red line in
Fig.~\ref{fig:spectralPlot}. At zero momentum, the relevant value for
the optical conductivity, the absolute threshold occurs at zero
energy. At low energies the optical conductivity is dominated by
particle-hole excitations. Close to half-filling, the spectral weight
of this contribution is small and tends to zero for $n\to
1$. In the vicinity of half-filling most of the spectral weight
concomitantly occurs above a ``pseudo-gap'' that is close in value to
the Mott gap of the half-filled system.
\item{} Above an energy scale that tends to the Mott gap as the band
filling approaches one from below, excitations involving a single
$k$-$\Lambda$ string of length two exist. Their threshold is shown as
a dashed black line in Fig.~\ref{fig:spectralPlot}. Precisely at half
filling these excitations do not contribute to the optical
conductivity \cite{JeckEssler,ControzziEsslerTsvelik,book} as a
result of the enhanced symmetry: at half filling this excitation
describes a singlet of the $\eta$-pairing SU(2) algebra and does not
contribute to $\sigma_1(\omega)$.
\item{} For band fillings close to $n=1$ there are other excitations
of the form $\eta^+|\mathit{LWS},{\bf m}\rangle$ that contribute to the
optical conductivity. At half-filling these are the only states
contributing to $\sigma_1(\omega)$ in the frequency regime
$2\Delta\leq\omega\leq 4\Delta$, where $\Delta$ is the Mott gap. Below
half-filling, their contribution to $\sigma_1(\omega)$ can be cast in
the form of $C_{\rm JJ}^{(2)}(\ell,t)$ in \fr{eq:simplifiedSum}, and
the states to be considered are then given by Sec.
\ref{sec:shiftedAntiHolon}, \ref{descpp} and \ref{sec:shiftedTwoHolon}.
\end{enumerate}

The thresholds shown in \figr{fig:spectralPlot} are at a high 
commensurability: $5(4k_F)=8\pi$. We note that thresholds involving
high-order umklapp processes are suppressed in $\sigma_1(\omega)$,
cf. Refs.~\onlinecite{PustilnikKhodas,KhodasPustilnik,KhodasPustilnik2}. 
Analytic forms for the ``high-energy'' thresholds are given below in 
eqns (\ref{klthres}), (\ref{eq:fixMomentum}) and (\ref{eq:halfFillThresh}). 
These results show that the contributions from $\sigma^{(2)}(\omega)$  
do not constitute the ``pseudo-gap threshold'', and moreover are
suppressed by a factor $1/L$ as we have pointed out before. Hence
they do not play an important role in the initial growth of
$\sigma_1(\omega)$, even for finite-size systems, and we therefore 
relegate their discussion to Appendix \ref{app:sigma2MIM}.

\section{Mobile impurity approach to threshold singularities}
\label{sec:MIM}
Our goal is to determine the behaviour of the optical conductivity
in the metallic phase of the Hubbard model close to half-filling
above the excitation thresholds occurring in the vicinity of the Mott
gap at $n=1$. This can be achieved by following the mobile impurity
approach to the Hubbard chain set out in
Ref.~\onlinecite{EsslerPereira}. In the main cases of interest to us 
here, the mobile impurity model describes low-energy degrees of
freedom in the presence of a single high-energy excitation with
momentum $q$ and takes the general form 
\begin{align}
	H &= H_{\rm LL}+H_{\rm imp}+H_{\rm int}, \label{eq:MIMtot}\\
H_{\rm LL}&=\int
dx\Big\lbrack\sum_{\alpha=c,s}\frac{v_\alpha}{16\pi}\left(\frac{1}{2K_\alpha}
\big(\partial_x\Phi_\alpha^*\big)^2
+2K_\alpha\big(\partial_x\Theta_\alpha^*\big)^2\right)\Big\rbrack, \label{eq:MIM1}\\
H_{\rm
  imp}&= \int \d x \,B^\dagger(x)\left[\epsilon(q)-i\epsilon'(q)\partial_x
-\frac{1}{2}\epsilon''(q)\partial_x^2\right]B(x)\ ,\\
H_{\rm int}&=\int
dx\ B^\dagger(x)B(x)\left[f_\alpha(q)\partial_x\varphi^*_\alpha(x)+
\bar{f}_\alpha(q)\partial_x\bar{\varphi}^*_\alpha(x)\right]+\ldots\ .
\label{MIM}
\end{align}
Here $v_{c,s}$ and $K_{c,s}$ are respectively the velocities and
Luttinger parameters of low-energy collective spin and charge degrees
of freedom, $\varphi^*_{c,s}$, $\bar\varphi^*_{c,s}$ are chiral charge
and spin Bose fields, and
\be
\Phi^*_\alpha=\varphi_{\alpha}^*+\bar\varphi_{\alpha}^*\ ,\quad
\Theta^*_\alpha=\varphi_{\alpha}^*-\bar\varphi_{\alpha}^*\ ,\qquad \alpha=c,s.
\ee
The high-energy excitation under consideration has a ``bare'' dispersion
$\epsilon(q)$ and is described in terms of the field $B(x)$. Finally, 
the functions $f_{c,s}(q)$ and $\bar{f}_{c,s}(q)$ parametrise the
interactions between the high-energy excitation and the low-energy
degrees of freedom. Our Bose fields are related to the usual
spin and charge bosons\cite{Giamarchi,GNT} by a canonical transformation
\be
\Phi_\alpha=\frac{\Phi_\alpha^*}{\sqrt{2}}\ ,\quad
\Theta_\alpha=\sqrt{2}\Theta_\alpha^*\ ,
\label{old_new}
\ee
and were introduced in Ref.~\onlinecite{EsslerPereira} by bosonizing the
physical fermionic spin and charge excitations in the Hubbard
model. The form of $H_{\rm int}$ is fixed by symmetry considerations
and assuming the high-energy excitation to be a point-like
object. Within the mobile impurity model the current operator is
represented as
\be
J_j\rightarrow B^\dagger(x) {\cal O}_{\rm LL}(x),
\ee
where ${\cal O}_{\rm LL}(x)$ is an operator acting in the Luttinger
liquid sector of the model \fr{MIM} only. In order to fully specify
our problem we proceed as follows:
\begin{enumerate}
\item{} The spin and charge velocities and Luttinger parameters are
determined directly from the exact solution of the Hubbard model, see
Appendix~\ref{app:vK} for a brief summary.
\item{} The relevant (``dressed'') dispersion relations for the
various excitations we need to consider have already been determined
above in section \ref{sec:baRev}.
\item{} For a given threshold, the projection ${\cal O}_{\rm LL}$ of
the current operator onto the Luttinger liquid sector is 
determined by bosonisation/refermionisation techniques. This is done
in sections \ref{sec:proj1}, \ref{sec:proj2} and \ref{sec:proj3} below.
\item{} Finally, the interaction parameters $f_{c,s}(q)$, $\bar{f}_{c,s}(q)$
are determined in sections \ref{sec:fsComp1}, \ref{sec:fsComp2} and
\ref{sec:fsComp3} by comparing finite-size corrections to excitation
energies in the Hubbard model and the mobile impurity model \fr{eq:MIMtot}.
\end{enumerate}
\subsection{$k$-$\Lambda$ threshold in $\sigma^{(1)}(\omega)$}
\label{sec:klSection}
\begin{figure}[ht!]
	\includegraphics[width=0.6\columnwidth]{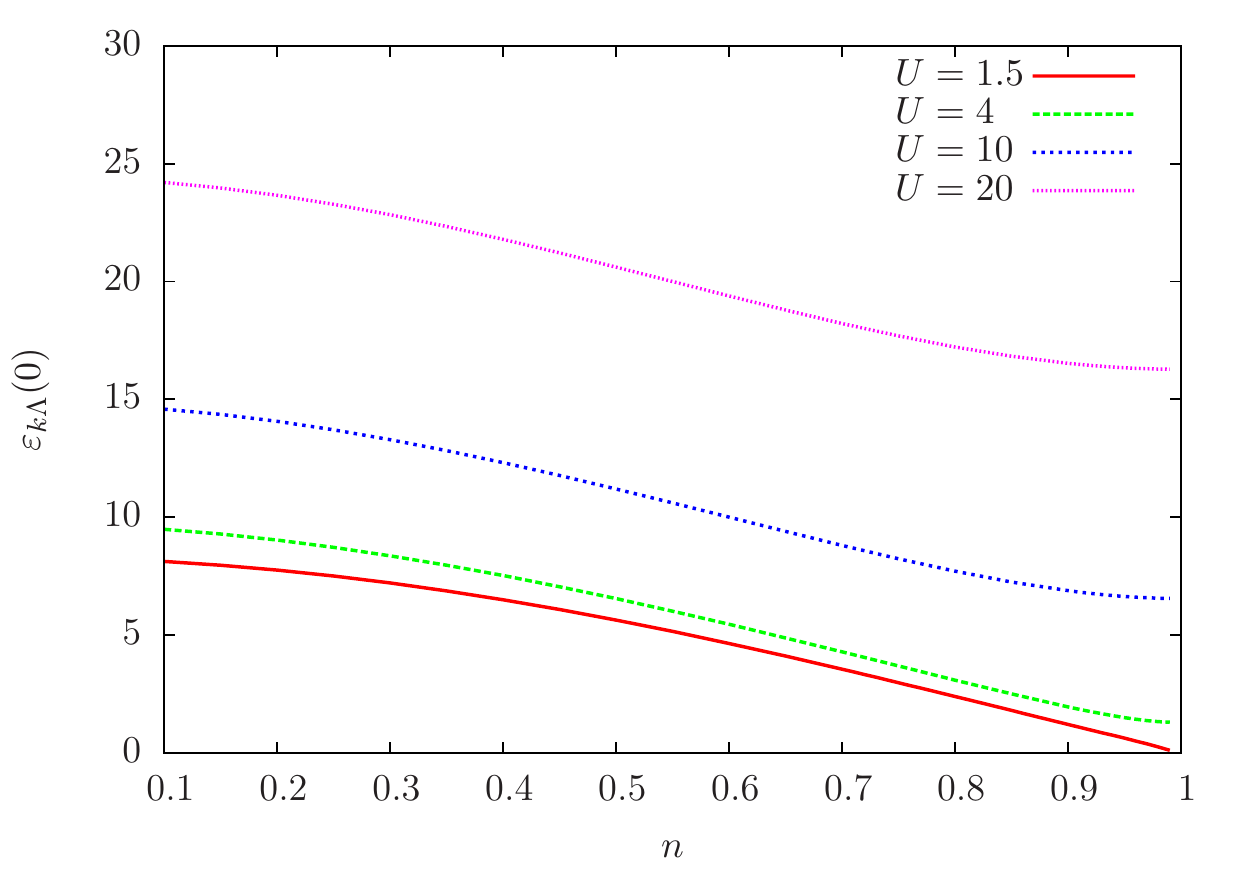}
	\caption{The threshold of the $k$-$\Lambda$ string $\varepsilon_{k\Lambda}(0)$ is shown for various $U$ and $n$}\label{fig:eoptPlot}
\end{figure}

This threshold is obtained when the entire ${\cal O}(1)$
contribution to the excitation energy and momentum are carried by the
$k$-$\Lambda$ string. The functional form of the threshold is
\bea
E_{\rm thres}^{k\textrm{-}\Lambda}(q)&=&\varepsilon_{k\Lambda}\big(\Lambda(q)\big),\nn
q&=&-2\, {\rm Re}\ {\rm arcsin}(\Lambda+iu)+\int_{-Q}^Qdk\, \theta\left(\frac{\Lambda-\sin k}{u}\right)\rho_{c,0}(k),
\label{klthres}
\eea
where $\rho_{c,0}(k)$ is the ground state root density
(\ref{eq:gsrhoc}). It important to note that in the case relevant for
the optical conductivity the $k$-$\Lambda$ string sits at $q=0$, which
corresponds to a \emph{maximum} of $\varepsilon_{k\Lambda}(\Lambda)$.
The mobile impurity Hamiltonian appropriate for the description of
this case is therefore of the form
\be
H_{\rm imp} = \int \d x\, B^\dagger(x) \left( \epsilon(0) -  \frac{1}{2} \epsilon''(0) \partial_x^2 \right) B(x),
\ee
where $\epsilon''(0) < 0$. We note that by virtue of the
interactions between the mobile impurity and the Luttinger liquid
degrees of freedom, the bare dispersion $\epsilon(q)$ is differs
from the actual threshold
$\varepsilon_{k\Lambda}\big(\Lambda(q)\big)$. The relationship between
the two quantities is established below.
The threshold $\varepsilon_{k\Lambda}(0)$ is shown in Fig.~\ref{fig:eoptPlot} for various $U$ and $n$
\subsubsection{Projection of the current operator}
\label{sec:proj1}
Having identified the state involving the $k$-$\Lambda$ string as
contributing to $\sigma_1^{(1)}(\omega)$, we wish to project the
current operator (\ref{eq:currentOperator}) onto the operators
involved in the mobile impurity model. To this end we
introduce the Hubbard projection operators\cite{book}, defined on site $j$ as 
\be
X_{j}^{ab}:= |a\rangle_j { }_j\langle b|, \qquad a,b=0,\up,\down,2\, (\up\down).
\ee
The current operator is expressed in terms of the $X_j^{ab}$ as
\be
J_{j} = -it \sum_{\sigma} \left(  \sigma X_j^{2\bar{\sigma}} + X_j^{\sigma 0} \right)\left( \sigma X_{j+1}^{\bar{\sigma}2} + X_{j+1}^{0\sigma} \right) - 
\left(  \sigma X_{j+1}^{2\bar{\sigma}} + X_{j+1}^{\sigma 0} \right)\left( \sigma X_{j}^{\bar{\sigma}2} + X_{j}^{0\sigma} \right).
\ee
In order to proceed further, we now consider the large-$U$ limit, in
which the $k$-$\Lambda$ string corresponds to a doubly-occupied site.
The current operator $J$ can be decomposed into three terms: a piece
which increases the double occupancy by one ($J^+$), a piece which
decreases it by one ($J^-$) and a piece that leaves the double
occupancy unchanged ($J^0$) i.e. 
\be
J_{j} = J^+_{j} + J^-_{j} + J^0_{j}\ .
\ee
As we are concerned with creating an excitation involving
double-occupation, we are interested in $J^+_{j}$ only. This is given by
\be 
J^+_{j} = -it \sum_\sigma \sigma X_j^{2\bar{\sigma}} X_{j+1}^{0\sigma} - \sigma X_{j+1}^{2\bar{\sigma}} X_j^{0\sigma},
\ee
and can be suggestively rewritten as
\be
J^+_{j} = -it \left[  X_j^{20}\left( X_j^{0\down} X_{j+1}^{0\up} - X_j^{0\up} X_{j+1}^{0\down} \right) - X_{j+1}^{20} \left(  X_{j+1}^{0\down} X_j^{0\up} - X_{j+1}^{0\up} X_j^{0\down} \right)\right].
\ee
As, in the large-$U$ limit, a $k$-$\Lambda$ string corresponds to a
doubly occupied site, while the ground state has zero double occupancy,
we can identify the operator creating the $k$-$\Lambda$ string as
$B^\dagger(x) \sim X^{20}_j$. This allows us to recast $J^+$ in the form
\be
J^+_{j} \sim -it \left[  B_j^\dagger - B_{j+1}^\dagger \right] \left( c^{\pd}_{j,\down} c^{\pd}_{j+1,\up} \left( 1-n^{\pd}_{j,\up} \right) \left(  1-n^{\pd}_{j+1,\down} \right)
-
c^{\pd}_{j,\up} c^{\pd}_{j+1,\down} \left( 1-n^{\pd}_{j,\down} \right) \left(  1-n^{\pd}_{j+1,\up} \right)
\right). \label{eq:Jpluslatt}
\ee
In order to complete the projection of the current operator onto the
mobile impurity model we simply bosonize all remaining electron operators.
The final result is
\be
J_{k\Lambda}(x) \sim \left(\partial_x B^\dagger(x)\right) e^{-i
  \Theta^*_c(x)/\sqrt{2}} \sin
\left(\frac{\Phi_s^*}{2\sqrt{2}}\right)+\ldots 
\label{currentproj}
\ee
\subsubsection{Finite-size corrections to excitation energies in the
  mobile impurity model}
\label{sec:MIMfinSize1}
Energies of excited states in the mobile impurity model in a large,
finite volume can be calculated following Refs.~\onlinecite{PerAffWhit,EsslerPereira}.
The chiral spin and charge Bose fields have mode expansions
\be
\varphi_\alpha^*(x) = \varphi_{\alpha,0}^* + \frac{x}{L} Q_\alpha^* + \sum_{n=1}^\infty \sqrt{\frac{2}{n}} \left[ e^{i \frac{2\pi n}{L}x} a_{\alpha,R,n} + e^{-i \frac{2\pi n}{L} x} a^\dagger_{\alpha,R,n} \right],
\ee
\be
\bar{\varphi}_\alpha^*(x) = \bar{\varphi}_{\alpha,0}^* + \frac{x}{L} \bar{Q}_\alpha^* + \sum_{n=1}^\infty \sqrt{\frac{2}{n}} \left[ e^{-i \frac{2\pi n}{L}x} a_{\alpha,L,n} + e^{i \frac{2\pi n}{L} x} a^\dagger_{\alpha,L,n} \right].
\ee
Here $Q^*_\alpha$, $\bar{Q}^*_\alpha$, $\varphi_{\alpha,0}$,
$\bar{\varphi}_{\alpha,0}$ are zero-mode operators, obeying the
commutation relations 
\be
[\varphi_{\alpha,0}^*,Q^*_\alpha] = - [\bar{\varphi}_{\alpha,0}^*, \bar{Q}^*_\alpha] = -4\pi i.
\ee
The eigenvalues $q_\alpha$, $\bar{q}_\alpha$ of the operators
$Q_\alpha^*$, $\bar{Q}^*_\alpha$ depend on the boundary conditions of
the fields $\varphi_\alpha^*(x)$, $\bar{\varphi}_\alpha^*(x)$. These
boundary conditions are, crucially, influenced by the presence of a
mobile impurity: coupling the impurity to the Luttinger liquid will
change the boundary conditions and therefore modify the eigenvalue
spectrum, causing a shift in the $\mathcal{O}(L^{-1})$ spectrum. It is
precisely this relationship that will allow us to determine the
coupling constants by examining the finite-size spectrum of the
Hubbard model in the presence of a high-energy excitation. 
An important distinction from previous calculations is that the
dispersion of the mobile impurity is quadratic in our case and has
negative curvature. 

The interactions between the impurity and the LL degrees of freedom
in (\ref{eq:MIMtot}) can be removed by a unitary transformation of 
the form \cite{SIG,EsslerPereira} 
\be
U = e^{-i \int_{-\infty}^\infty dx\ \sum_\alpha \left(\gamma^{\phantom
    *}_\alpha \varphi_\alpha^*(x) + \bar{\gamma}^{\phantom *}_\alpha
  \bar{\varphi}_\alpha^*(x) \right) B^\dagger(x) B(x)}. 
\label{eq:unitaryDef}
\ee
The transformed fields are given by
\bea
\varphi_\alpha^\circ &=& U\varphi_\alpha^*U^\dagger = \varphi_\alpha^*(x) - 2\pi \gamma^{\phantom *}_\alpha C(x),\nn
\bar{\varphi}_\alpha^\circ &=& U\bar{\varphi}_\alpha^*U^\dagger =
\bar{\varphi}_\alpha^*(x) + 2\pi \bar{\gamma}^{\phantom *}_\alpha
C(x), \nn
\widetilde{B}(x) &=& UB(x)U^\dagger = B(x) e^{i\sum_\alpha
  (\gamma_\alpha\varphi^*_\alpha(x) + \bar{\gamma}_\alpha
  \bar{\varphi}_\alpha^*(x))} e^{-i\pi\sum_\alpha (\gamma_\alpha^2 -
  \bar{\gamma}_\alpha^2)C(x)},
\label{eq:transformedFields}
\eea
where
\be
C(x) = \int_{-\infty}^\infty \d y\, \mathrm{sgn}(x-y) B^\dagger(y) B(y).
\ee
By choosing the parameters $\gamma_\alpha$, $\bar{\gamma}_\alpha$ to fulfil
\be
\begin{pmatrix} f_\alpha\\ \bar{f}_\alpha \end{pmatrix} = \begin{pmatrix} -v_\alpha^+ & -v_\alpha^- \\ v_\alpha^- & v_\alpha^+ \end{pmatrix} \begin{pmatrix} \gamma_\alpha \\ \bar{\gamma_\alpha} \end{pmatrix}, 
\qquad
v_\alpha^\pm = \frac{v_\alpha}{2} \left( 2K_\alpha \pm \frac{1}{2K_\alpha} \right),
\label{eq:unitTrans}
\ee
we find that, retaining only the most relevant terms, the impurity decouples 
in the new basisi.e.
\be
H = \int \d x\left[  \sum_{\alpha=c,s} \frac{v_\alpha}{16\pi} \left(
\frac{1}{2K_\alpha}\left( \partial_x \Phi_\alpha^\circ \right)^2 +
2K_\alpha \left( \partial_x \Theta^\circ_\alpha \right)^2 \right)\right] 
+ \int \d x\, \widetilde{B}^\dagger(x)
\left[\widetilde{\epsilon}(q) - \frac{1}{2}
  \widetilde{\epsilon}''(q) \partial_x^2 \right]\widetilde{B}(x) +
\dots 
. \label{eq:transformedHamiltonian}
\ee
We note that the ``dressed'' impurity dispersion for momenta $k\approx q$ is
$\widetilde{\epsilon}(q) - \frac{1}{2}\widetilde{\epsilon}''(q)
(k-q)^2$ and differs from its ``bare'' value
$\epsilon(k)$ by a constant\cite{endNote}. 
Importantly, it is the \emph{dressed} dispersion
that relates directly to the Bethe Ansatz result for $E_{\rm
thres}^{k\textrm{-}\Lambda}(k)$ in \fr{klthres}. 
In the decoupled theory of (\ref{eq:transformedHamiltonian}) it is a 
straightforward matter to calculate the
spectrum of \emph{low-energy} excitations above the ground state in
the presence of an impurity. The result is\cite{EsslerPereira} 
\be
\Delta E_{LL} = \sum_{\alpha=c,s} \frac{2\pi v_\alpha}{L} \left[
  \frac{1}{4K_\alpha} \left(  \frac{q_\alpha + \bar{q}_\alpha}{4\pi }
  - \gamma_\alpha + \bar{\gamma}_\alpha \right)^2 + K_\alpha \left(
  \frac{q_\alpha - \bar{q}_\alpha}{4\pi} - \gamma_\alpha -
  \bar{\gamma}_\alpha \right)^2 + \sum_{n>0} n\left[ M_{n,\alpha}^+ +
    M_{n,\alpha}^- \right]\right]. 
\label{eq:generalImpModelFSS} 
\ee
Here $M_{n,\alpha}^\pm$ are non-negative integers corresponding to 
particle-hole excitations at the edge of the ``Fermi seas''.
Any operator acting on the ground state will, in general, 
produce a superposition of energy eigenstates. 
Noting that the ground state is annihilated by 
$Q^*_\alpha$, $\bar{Q}^*_\alpha$, the state $\mathcal{O}(x)|GS\rangle$ 
has well-defined quantum numbers $q_\alpha^{(0)}$, $\bar{q}_\alpha^{(0)}$ 
if $\mathcal{O}(x)$ satisfies the relations
\be
[Q_\alpha^*,\mathcal{O}(x)] = q_{\alpha}^{(0)} \mathcal{O}(x), \qquad 	
[\bar{Q}_\alpha^*,\mathcal{O}(x)] = \bar{q}_{\alpha}^{(0)} \mathcal{O}(x).
\label{q0s}
\ee
If the operator satisfies such a property then all states in the 
superposition defined by $\mathcal{O}(x)|GS\rangle$ must have the 
\emph{same} $q_\alpha$, $\bar{q}_\alpha$, namely $q_\alpha^{(0)}$, $\bar{q}_\alpha^{(0)}$.
The only difference in the energies comes from having different $M_{n,\alpha}^\pm$.
We can therefore identify the ``minimal'' excitation\cite{EsslerPereira}: this is the state with all $M_{n,\alpha}^\pm = 0$ 
i.e. no particle-hole excitations. For the specific case of interest
here, namely acting with the projected current operator
$J_{k\Lambda}(x)$ on the ground state, this can be represented pictorially as
\be
J_{k\Lambda}(x)
\Bigg\vert
\begin{tikzpicture}[scale=0.5,baseline=-.5ex]
	\draw (-1,1) node[anchor=east]{\tiny $c$} -- (1,1);
	\draw plot [smooth] coordinates {(-1,1.2) (-.8,1.15) (-.3,.8) (0,.7) (.3,.8) (.8,1.15) (1,1.2)};
	\begin{scope}
		\clip 	(-1,1) rectangle (1,-1);
		\draw[very thick] plot [smooth] coordinates {(-1,1.2) (-.8,1.15) (-.3,.8) (0,.7) (.3,.8) (.8,1.15) (1,1.2)};
	\end{scope}
	\draw (-1,0) node[anchor=east]{\tiny $s$} -- (1,0);
	\draw plot [smooth] coordinates {(-1,0.2) (-.5,-.2) (0,-.3) (.5,-.2) (1,0.2)};
	\begin{scope}
		\clip 	(-1,0) rectangle (1,-2);
		\draw[very thick] plot [smooth] coordinates {(-1,0.2) (-.5,-.2) (0,-.3) (.5,-.2) (1,0.2)};
	\end{scope}
	\node at (-1.2,-1) {\tiny$k\Lambda$};
	\draw plot [smooth] coordinates { (-1,-1.4) (-.85,-1.35) (-.6,-1.2) (-.2,-0.95) (0,-0.9) (.2,-0.95) (.6,-1.2) (.85,-1.35) (1,-1.4) };
\end{tikzpicture}
\Bigg\rangle
\sim 
A
\underbrace{
\Bigg\vert
\begin{tikzpicture}[scale=0.5,baseline=-.5ex]
	\draw (-1,1) node[anchor=east]{\tiny $c$} -- (1,1);
	\draw plot [smooth] coordinates {(-1,1.2) (-.8,1.15) (-.3,.8) (0,.7) (.3,.8) (.8,1.15) (1,1.2)};
	\begin{scope}
		\clip 	(-1,1) rectangle (1,-1);
		\draw[very thick] plot [smooth] coordinates {(-1,1.2) (-.8,1.15) (-.3,.8) (0,.7) (.3,.8) (.8,1.15) (1,1.2)};
	\end{scope}
	\draw (-1,0) node[anchor=east]{\tiny $s$} -- (1,0);
	\draw plot [smooth] coordinates {(-1,0.2) (-.5,-.2) (0,-.3) (.5,-.2) (1,0.2)};
	\begin{scope}
		\clip 	(-1,0) rectangle (1,-2);
		\draw[very thick] plot [smooth] coordinates {(-1,0.2) (-.5,-.2) (0,-.3) (.5,-.2) (1,0.2)};
	\end{scope}
	\node at (-1.2,-1) {\tiny$k\Lambda$};
	\draw plot [smooth] coordinates { (-1,-1.4) (-.85,-1.35) (-.6,-1.2) (-.2,-0.95) (0,-0.9) (.2,-0.95) (.6,-1.2) (.85,-1.35) (1,-1.4) };
	\fill (0,-.9) circle (0.05);
\end{tikzpicture}
\Bigg\rangle}_{\substack{q_\alpha = q_\alpha^{(0)},\,\bar{q}_\alpha = \bar{q}_\alpha^{(0)}\\\textrm{``minimal''} }}
+ 
B
\underbrace{
\Bigg\vert
\begin{tikzpicture}[scale=0.5,baseline=-.5ex]
	\draw (-1,1) node[anchor=east]{\tiny $c$} -- (1,1);
	\draw plot [smooth] coordinates {(-1,1.2) (-.8,1.15) (-.3,.8) (0,.7) (.3,.8) (.8,1.15) (1,1.2)};
	\begin{scope}
		\clip 	(-1,1) rectangle (1,-1);
		\draw[very thick] plot [smooth] coordinates {(-1,1.2) (-.8,1.15) (-.3,.8) (0,.7) (.3,.8) (.8,1.15) (1,1.2)};
	\end{scope}
	\draw (-1,0) node[anchor=east]{\tiny $s$} -- (1,0);
	\draw[name path=spin] plot [smooth] coordinates {(-1,0.2) (-.5,-.2) (0,-.3) (.5,-.2) (1,0.2)};
	\begin{scope}
		\clip 	(-1,0) rectangle (1,-2);
		\draw[very thick] plot [smooth] coordinates {(-1,0.2) (-.5,-.2) (0,-.3) (.5,-.2) (1,0.2)};
	\end{scope}
	\path[name path=ph1] (.5,1) -- (.5,-1) -- (.9,-1) -- (.9,1);
	\draw[->,>=latex, name intersections={of=spin and ph1}] (intersection-1) to[out=135,in=135,looseness=3] (intersection-2);
	\fill[white,draw=black, name intersections={of=spin and ph1}] (intersection-1) circle (0.05);
	\fill[black,draw=none, name intersections={of=spin and ph1}] (intersection-2) circle (0.05);
	\node at (-1.2,-1) {\tiny$k\Lambda$};
	\draw plot [smooth] coordinates { (-1,-1.4) (-.85,-1.35) (-.6,-1.2) (-.2,-0.95) (0,-0.9) (.2,-0.95) (.6,-1.2) (.85,-1.35) (1,-1.4) };
	\fill (0,-.9) circle (0.05);
\end{tikzpicture}
\Bigg\rangle}_{\substack{q_\alpha = q_\alpha^{(0)},\,\bar{q}_\alpha = \bar{q}_\alpha^{(0)}\\M_{n,s}^+\neq0}}
+ 
C
\underbrace{
\Bigg\vert
\begin{tikzpicture}[scale=0.5,baseline=-.5ex]
	\draw (-1,1) node[anchor=east]{\tiny $c$} -- (1,1);
	\draw[name path=charge] plot [smooth] coordinates {(-1,1.2) (-.8,1.15) (-.3,.8) (0,.7) (.3,.8) (.8,1.15) (1,1.2)};
	\begin{scope}
		\clip 	(-1,1) rectangle (1,-1);
		\draw[very thick] plot [smooth] coordinates {(-1,1.2) (-.8,1.15) (-.3,.8) (0,.7) (.3,.8) (.8,1.15) (1,1.2)};
	\end{scope}
	\path[name path=ph2] (.6,2) -- (.6,-1) -- (.9,-1) -- (.9,2);
	\draw[->,>=latex, name intersections={of=charge and ph2}] (intersection-1) to[out=135,in=135,looseness=3] (intersection-2);
	\fill[white,draw=black, name intersections={of=charge and ph2}] (intersection-1) circle (0.05);
	\fill[black,draw=none, name intersections={of=charge and ph2}] (intersection-2) circle (0.05);
	\draw (-1,0) node[anchor=east]{\tiny $s$} -- (1,0);
	\draw[name path=spin] plot [smooth] coordinates {(-1,0.2) (-.5,-.2) (0,-.3) (.5,-.2) (1,0.2)};
	\begin{scope}
		\clip 	(-1,0) rectangle (1,-2);
		\draw[very thick] plot [smooth] coordinates {(-1,0.2) (-.5,-.2) (0,-.3) (.5,-.2) (1,0.2)};
	\end{scope}
	\node at (-1.2,-1) {\tiny$k\Lambda$};
	\draw plot [smooth] coordinates { (-1,-1.4) (-.85,-1.35) (-.6,-1.2) (-.2,-0.95) (0,-0.9) (.2,-0.95) (.6,-1.2) (.85,-1.35) (1,-1.4) };
	\fill (0,-.9) circle (0.05);
\end{tikzpicture}
\Bigg\rangle}_{\substack{q_\alpha = q_\alpha^{(0)},\,\bar{q}_\alpha = \bar{q}_\alpha^{(0)}\\M_{n,c}^+\neq0}}
+ 
D
\underbrace{
\Bigg\vert
\begin{tikzpicture}[scale=0.5,baseline=-.5ex]
	\draw (-1,1) node[anchor=east]{\tiny $c$} -- (1,1);
	\draw[name path=charge] plot [smooth] coordinates {(-1,1.2) (-.8,1.15) (-.3,.8) (0,.7) (.3,.8) (.8,1.15) (1,1.2)};
	\begin{scope}
		\clip 	(-1,1) rectangle (1,-1);
		\draw[very thick] plot [smooth] coordinates {(-1,1.2) (-.8,1.15) (-.3,.8) (0,.7) (.3,.8) (.8,1.15) (1,1.2)};
	\end{scope}
	\path[name path=ph2] (.6,2) -- (.6,-1) -- (.9,-1) -- (.9,2);
	\draw[->,>=latex, name intersections={of=charge and ph2}] (intersection-1) to[out=135,in=135,looseness=3] (intersection-2);
	\fill[white,draw=black, name intersections={of=charge and ph2}] (intersection-1) circle (0.05);
	\fill[black,draw=none, name intersections={of=charge and ph2}] (intersection-2) circle (0.05);
	\draw (-1,0) node[anchor=east]{\tiny $s$} -- (1,0);
	\draw[name path=spin] plot [smooth] coordinates {(-1,0.2) (-.5,-.2) (0,-.3) (.5,-.2) (1,0.2)};
	\begin{scope}
		\clip 	(-1,0) rectangle (1,-2);
		\draw[very thick] plot [smooth] coordinates {(-1,0.2) (-.5,-.2) (0,-.3) (.5,-.2) (1,0.2)};
	\end{scope}
	\path[name path=ph1] (.5,1) -- (.5,-1) -- (.9,-1) -- (.9,1);
	\draw[->,>=latex, name intersections={of=spin and ph1}] (intersection-1) to[out=135,in=135,looseness=3] (intersection-2);
	\fill[white,draw=black, name intersections={of=spin and ph1}] (intersection-1) circle (0.05);
	\fill[black,draw=none, name intersections={of=spin and ph1}] (intersection-2) circle (0.05);
	\node at (-1.2,-1) {\tiny$k\Lambda$};
	\draw plot [smooth] coordinates { (-1,-1.4) (-.85,-1.35) (-.6,-1.2) (-.2,-0.95) (0,-0.9) (.2,-0.95) (.6,-1.2) (.85,-1.35) (1,-1.4) };
	\fill (0,-.9) circle (0.05);
\end{tikzpicture}
\Bigg\rangle}_{\substack{q_\alpha = q_\alpha^{(0)},\,\bar{q}_\alpha = \bar{q}_\alpha^{(0)}\\M_{n,s}^+\neq0,\,M_{n,c}^+\neq0}}
+
\dots \label{eq:schematic}
\ee
From the bosonised expression for $J_{k\Lambda}(x)$, and focussing only on the $e^{i\Phi_s^*/2\sqrt{2}}$ term with the other following from parity, it follows that
\be
q_c^{(0)} = \bar{q}_c^{(0)} = 2\pi \sqrt{2};\qquad q_s^{(0)} = -\bar{q}_s^{(0)} = -\pi\sqrt{2}.
\ee
The total momentum can also be calculated using the mode expansion,
and is found to be of the form
\bea
P &=&\frac{k_F}{\pi\sqrt{2}} \left(  \bar{q}_c - q_c \right) +P_{\rm
  imp}(k^p_L) 
+ \frac{2\pi}{L}\sum_{\alpha=c,s} \left[ \left( \frac{q_\alpha +
    \bar{q}_\alpha}{4\pi} - \gamma_\alpha +  \bar{\gamma}_\alpha\right)
\left(\frac{q_\alpha - \bar{q}_\alpha}{4\pi} - \gamma_\alpha - \bar{\gamma}_\alpha \right)
+\left( N_\alpha^+ - N_\alpha^- \right) \right],
\label{eq:mimFSS}
\eea
where $k^p_L$ includes finite-size shifts to the rapidity $k^p$. 
We can identify the ``minimally excited'' state with the Bethe Ansatz
excitation at the relevant threshold. By matching the expressions for
the finite-size energies, we will be able to constrain the parameters
$\gamma_\alpha$, $\bar{\gamma}_\alpha$.  

\subsubsection{Finite-size corrections to excitation energies from Bethe Ansatz}
\label{sec:zeroField}
Finite-size corrections to the energies of states involving both high-
and low-energy excitations can be determined from the Bethe Ansatz
solution of the Hubbard model following Ref.~\onlinecite{FHLE}. The
details for the excitations of interest here involving a $k$-$\Lambda$
string are given in Appendix \ref{app:klBAfull}. The final result
for zero magnetic field and total momentum $P={\cal O}(L^{-1})$ is
\bea
E =  e_{GS} L +\varepsilon_{k\Lambda} (0) 
	- \frac{\pi}{6L} (v_c + v_s) &+&
\frac{2\pi v_c}{L} \left[ \frac{ (\Delta N_c - N_c^\imp)^2}{8K_c} +
  2K_c \left( D_c - D_c^\imp  + \frac{D_s - D_s^\imp}{2} \right)^2
  \right]\nn
&+&\frac{2\pi v_s}{L} \left[ \frac{1}{2} 
\left( \Delta N_s - \frac{ \Delta N_c}{2}
\right)^2 + \frac{(D_s - D_s^\imp)^2}{2} \right]. 
\label{eq:BAfinitespecKL}
\eea
Here $e_{GS}$ is the ground state energy per site in the thermodynamic
limit, while $\varepsilon_{k\Lambda}(0)$ is the contribution due to
the (high-energy) $k$-$\Lambda$ string excitation and is obtained from
the solution of the integral equations 
\bea
\varepsilon_{k\Lambda}(\Lambda) &=& 
4{\rm Re}\sqrt{ 1- (\Lambda - iu)^2} - 2\mu - 4u + \int_{-Q}^Q \d k\, \cos
k\; a_1(\sin k -\Lambda)\; \varepsilon_c(k),\nn
\varepsilon_c(k) &=&
-2\cos k - \mu - 2u + \int_{-Q}^Q \d k'\ \cos k'\; R(\sin k - \sin k')
\varepsilon_c(k'), \label{eq:dressedEnergyZeroField}
\eea
where the function $R(x)$ is given by
\be
R(x)=\int_{-\infty}^\infty \frac{d\omega}{2\pi}\frac{e^{i\omega x}}
{1+\exp(2u|\omega|)}.
\label{rofx}
\ee
The spin and charge velocities $v_{s,c}$ and the Luttinger parameter
$K_c$ are given in Appendix~\ref{app:vK}, while the quantities
$N_c^\imp$ and $D_{c,s}^\imp$ are given by 
\bea
D_c^\imp &=& 0, \quad D_s^\mathrm{imp} = 0\ ,\quad
N_c^\imp = \int_{-Q}^Q \d k\, \rho_{c,1}(k),
\eea
where
\be
\rho_{c,1}(k) = \cos k\; a_1(\sin k - \Lambda'^p) + \cos k \int_{-Q}^Q
dk'\ \rho_{c,1}(k') R(\sin k - \sin k'). 
\ee
Finally, the quantities $\Delta N_c$, $\Delta N_s$, $D_c$ and $D_s$
characterise low-energy excitations of the spin and charge degrees of
freedom and for the ``minimal'' excitation of interest are given by
\be
\Delta N_c =-2\ , \quad D_c = 0\ ,\quad
\Delta N_s =-1\ , \quad D_s = 0.
\ee
We note that in order to fully specify the mobile impurity model we
require the value of the curvature of the impurity dispersion at its
maximum. This is given by
\be
\frac{1}{m} =\left. \frac{\partial^2 \varepsilon_{k\Lambda}(\Lambda)}{\partial p^2}\right|_{\Lambda = 0} = \frac{\varepsilon''_{k\Lambda}(0)}{(2\pi \sigma'^h_1 (0))^2},
\ee
where
\bea
\varepsilon_{k\Lambda}''(0) &=& -\frac{4}{(1+u^2)^{3/2}} - \int_{-Q}^Q
\d k\, a_1'(\sin k)\varepsilon_c'(k), \nn
2\pi \sigma'^h_1(0) &=& \frac{2}{\sqrt{1+u^2}} - 2\pi \int_{-Q}^Q \d k\, a_1(\sin k)\rho_{c,0}(k).
\eea
The total momentum of the state of interest can also be calculated
from the Bethe Ansatz and for the case of interest results in
\be
P = q_L + 2 k_F (2D_c + D_s) +  \frac{2\pi}{L}\sum_{\alpha=c,s}\left[ 
(\Delta N_\alpha-N_\alpha^{\rm imp})(D_\alpha-D_\alpha^{\rm imp})
+ (N_\alpha^+ - N_\alpha^- )\right] , 
\label{PBA}
\ee
with $q_L$ the contribution, including finite-size shifts of the rapidities, from the high-energy impurity and $N_\alpha^\pm$ are integers corresponding to particle-hole pairs at the edge of the ``Fermi seas''. 
The method used for deriving this result is summarised in Appendix~\ref{app:fsm}.
\subsubsection{Fixing the parameters $\gamma_\alpha$, $\bar\gamma_\alpha$}
\label{sec:fsComp1}
By equating the Bethe Ansatz results
(\ref{eq:BAfinitespecKL}) and \fr{PBA} for energy and momentum with
the ones obtained in the framework of the mobile impurity model 
(\ref{eq:generalImpModelFSS}), \fr{eq:mimFSS} we can fix the parameters
$\gamma_\alpha$, $\bar\gamma_\alpha$ to be
\be
\gamma_c = -\bar{\gamma}_c =\frac{1}{\sqrt{2}}  + \frac{(\Delta N_c -
  N_c^\imp)}{2\sqrt{2}}  ;\qquad \gamma_s = \bar{\gamma}_s =
-\frac{1}{2\sqrt{2}}. \label{eq:signchoice} 
\ee
\subsubsection{Current-current correlator in the mobile impurity
model}
\begin{figure}[ht!]
\includegraphics[width=0.6\columnwidth]{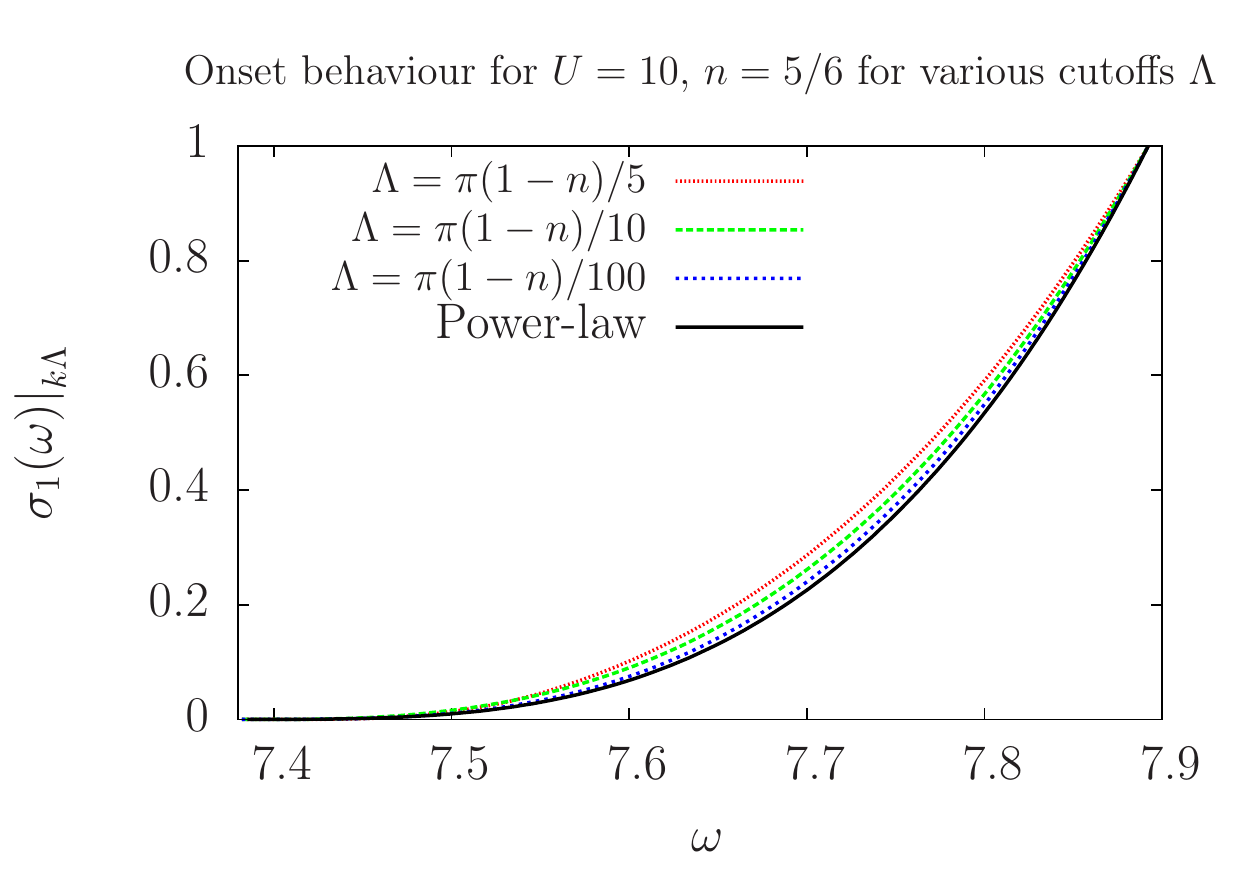}
\caption{Optical conductivity (\ref{eq:klOnsetForm}) for $U=10$,
 $n=5/6$ and several values of the cutoff $\Lambda$. The different
curves have been normalized such that $\sigma_1(\varepsilon_{k\Lambda}
+ 0.5)\Big\vert_{k\Lambda}=1$. For comparison we show
the power-law behaviour \fr{powerlaw}, valid for
$\omega\to\varepsilon_{k\Lambda}$. We see that the power law in fact
provides a good approximation over the entire range of comparison.
}\label{fig:klComparison}
\end{figure}

\begin{figure}[ht!]
\includegraphics[width=0.6\columnwidth]{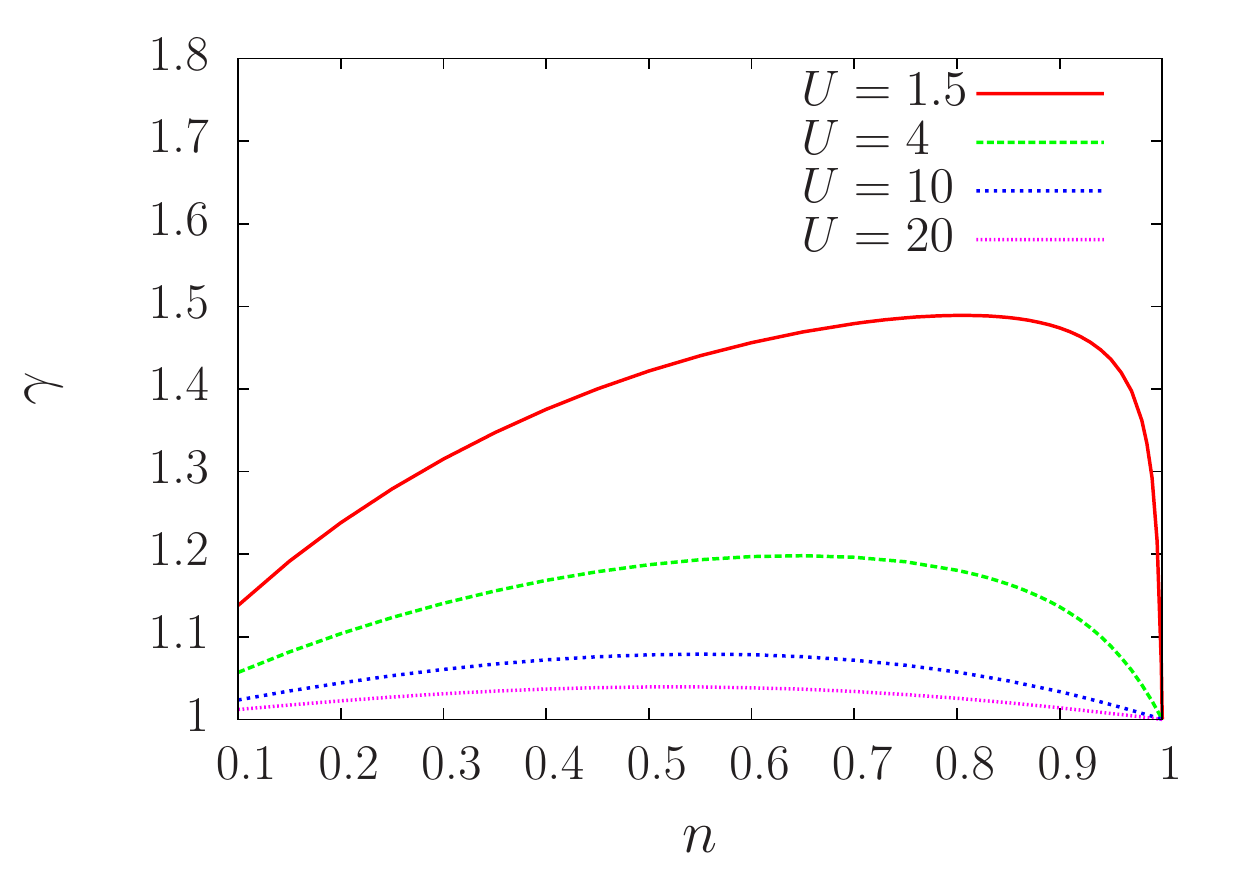}
\caption{Value of the exponent $\gamma$ in \fr{powerlaw}
characterizing the power-law behaviour of $\sigma_1(\omega)$ just
above the pseudo-gap, for several values of $U$ and $n$}\label{fig:gammaPlot}
\end{figure}

We are now in a position to work out the current-current correlation
function \fr{eq:specAll} in the mobile impurity model framework. Given
the expression \fr{currentproj} for the projection of the current
operator, we have 
\be
C^{(1)}_{\rm JJ}(\ell,t)\sim G(x,t)=\langle J^\dagger_{k\Lambda}(x,t)
J_{k\Lambda}(0,0)\rangle.
\label{jdaggerj}
\ee
In order to evaluate $G(x,t)$ we go over to the transformed basis, in
which the impurity decouples from the LL degrees of freedom. 
Given \fr{eq:signchoice}, the leading contribution takes the form
\bea
J_{k\Lambda}(x)&\sim&\partial_x\widetilde{B}^\dagger(x)
e^{i \Theta_c^\circ(x) \frac{\Delta N_c -
    N_c^\imp}{2\sqrt{2}}}
+i\gamma_c\widetilde{B}^\dagger(x)\partial_x\Theta_c^\circ(x)
e^{i \Theta_c^\circ(x) \frac{\Delta N_c -
    N_c^\imp}{2\sqrt{2}}}\nn
&-&\frac{i}{2\sqrt{2}}
\widetilde{B}^\dagger(x)\partial_x\Phi_s^\circ(x)e^{i \Theta_c^\circ(x) \frac{\Delta N_c -
    N_c^\imp}{2\sqrt{2}}}.
\eea
Substituting this back into \fr{jdaggerj} leads to three kinds of
contributions to the correlator
\bea
G(x,t) &=& G_1(x,t) \langle \partial_x \widetilde{B}(x,t) \partial_x
\widetilde{B}^\dagger(0,0)\rangle + 
G_2(x,t) \langle \widetilde{B}(x,t)
\widetilde{B}^\dagger(0,0)\rangle\nn
& +& G_3(x,t)i\left[   \langle \partial_x \widetilde{B}(x,t)  
\widetilde{B}^\dagger(0,0) -\langle \widetilde{B}(x,t) 
\partial_x \widetilde{B}^\dagger(0,0) \right].
\label{gxt}
\eea
Here $G_j(x,t)$ are correlation functions in the LL sector of the
theory and can be evaluated by standard methods. The results are
\bea
G_1(x,t) &=& \frac{1}{(x^2 - v_c^2t^2)^{\gamma}}\ ,\nn
\frac{G_2(x,t)}{G_1(x,t)} &=&  
-\frac{2\gamma_c^2}{K_c} \left[ \frac{x^2+v_c^2t^2}{(x^2-v_c^2t^2)^2}
+\frac{2\gamma x^2}{(x^2-v_c^2t^2)^2}\right] 
- \frac{1}{2}\frac{x^2+v_s^2t^2}{(x^2-v_s^2t^2)^2}\ ,\nn
\frac{G_3(x,t)}{G_1(x,t)} &=& i\gamma_c \sqrt{\frac{\gamma}{K_c}}
{\rm sgn}(N_c^{\rm imp}+2)\ \frac{2x}{x^2-v_c^2t^2}\ ,
\label{g123}
\eea
where we have defined
\be
\gamma = \frac{1}{2K_c} 
\left( 1 + \frac{N_c^\imp}{2}\right)^2. \label{eq:gammakl}
\ee
The free impurity correlator is given by
\be
\langle \widetilde{B}(x,t) \widetilde{B}^\dagger(0,0)\rangle = \int_{-\Lambda}^\Lambda \frac{\d p}{2\pi} e^{-ipx} e^{-i\varepsilon(p)t},
\label{impcorr}
\ee
where $\varepsilon(p)$ is the dispersion relation for the
$k$-$\Lambda$ string and $\Lambda$ is a  momentum cutoff for the
impurity excitation. Using \fr{jdaggerj}, \fr{gxt}, \fr{g123} and
\fr{impcorr} we may now determine the contribution from the
$k$-$\Lambda$ string excitation to the retarded correlator
(\ref{eq:optCondDef}). The result can be written in the form
\bea
\sigma_1(\omega)\Big\vert_{k\Lambda} \sim
\frac{1}{\omega}\int_{-\Lambda}^\Lambda \d p  &\Bigg\{&
\frac{\gamma_c^2}{K_c} \left(
\left( 1 + \gamma\right) 
\left[\widetilde{G}^{c}_{\gamma+2,\gamma}\big(\omega - \varepsilon(p),p\big)  
+ \widetilde{G}^{c}_{\gamma,\gamma+2}\big(\omega - \varepsilon(p),p\big)\right] 
- 2\gamma\widetilde{G}^{c}_{\gamma+1,\gamma+1}\big(\omega - \varepsilon(p),p\big) 
\right) \nn
&+&\sqrt{\frac{4\gamma}{K_c}} \gamma_c p \left[ 
\widetilde{G}^c_{\gamma+1,\gamma}\big(\omega-\varepsilon(p),p\big) 
-\widetilde{G}^c_{\gamma,\gamma+1}\big(\omega-\varepsilon(p),p\big) \right]
+ p^2 \widetilde{G}^{c}_{\gamma,\gamma}\big(\omega -
\varepsilon(p),p\big)\nn
&+&\gamma_s^2\left[
\widetilde{G}^{s}_{\gamma}\big(\omega - \varepsilon(p),p\big) 
+ \widetilde{G}^{s}_{\gamma}\big(\omega - \varepsilon(p),-p\big)
\right] \Bigg\}, \label{eq:klOnsetForm}
\eea
where we have defined
\bea
\widetilde{G}^{c}_{\gamma_+,\gamma_-}(\omega,p) &=& 
\frac{(2\pi)^2}{\Gamma(\gamma_+)\Gamma(\gamma_-)(2v_c)^{\gamma_++\gamma_--1}}
 (\omega+v_c p)^{\gamma_+-1}(\omega-v_cp)^{\gamma_--1}
\Theta(\omega-v_c|p|)\ ,
\label{eq:Gtildec}\\
\widetilde{G}^{s}_{\gamma}(\omega,p) &=& \int_0^1 \d s \left[ \frac{2\pi}{\Gamma\left( \gamma \right)} \right]^2 \frac{(\omega-v_sp)^{2\gamma-1}}{(v_c^2-v_s^2)^{\gamma}} \Theta(\omega-v_sp) s^{\gamma-1}(1-s)^{\gamma-1}\nn
&&\times\left[
\frac{2v_c(\omega-v_sp)}{v_c^2-v_s^2} s- \frac{\omega-v_cp}{v_c-v_s}  \right]
\Theta \Big(
\frac{2v_c(\omega-v_sp)}{v_c^2-v_s^2}s
 - \frac{\omega-v_cp}{v_c-v_s}  \Big) .
\eea
The dependence of \fr{eq:klOnsetForm} on the momentum  cutoff
$\Lambda$ is shown in \figr{fig:klComparison}. We see that over a wide
range the result is only weakly cutoff-dependent. 
Unfortunately, the mobile impurity method provides no simple way of predicting how
large the cutoff should be. The only obvious constraint is that it
should fulfil $\Lambda \ll
\pi(1-n)$. If we approach the threshold from above, i.e. consider the
limit $\omega \to \varepsilon_{k\Lambda}(0)$, the remaining integral
in \fr{eq:klOnsetForm} can be carried out and yields a power-law
behaviour of the form
\be
\lim_{\omega \to \varepsilon_{k\Lambda}(0)} \sigma_1(\omega)\Big\vert_{k\Lambda} \sim \frac{1}{\omega} \left[ \omega-\varepsilon_{k\Lambda}(0) \right]^{\gamma-1} \Theta\left(\omega - \varepsilon_{k\Lambda}(0)\right).
\label{powerlaw}
\ee
This is shown in \figr{fig:klComparison} together with numerical
evaluations of \fr{eq:klOnsetForm} for several values of the cutoff
$\Lambda$, and is seen to provide a good approximation
across the entire frequency range examined. Importantly, the exponent
$\gamma-1$ is always larger than one, which is in disagreement with
that of Ref.~\onlinecite{PhysRevLett.84.4673} .
Additionally, the exponent $\gamma$ can be calculated for a variety of parameters, as presented in Fig.~\ref{fig:gammaPlot}.

\section{Comparison with numerical results}
\label{sec:results}
In Ref.~\onlinecite{Tiegel15} the optical conductivity of the one
dimensional Hubbard model has been computed by matrix product
methods. The approach requires introduction of a damping parameter
$\eta>0$ and provides ${\rm Im}\ \chi^J(\omega+i\eta)$ for a chain of
finite length. In order to facilitate a comparison with the results
obtained here it is necessary to remove this broadening. In order to
do this approximately we proceed as follows. For positive frequencies
the zero temperature optical conductivity can be expressed as
\bea
\sigma_1(\omega>0)&=&-\frac{{\rm Im}\; \chi_+(\omega) }{\omega}\ ,\nn
\chi_+(\omega)&=&\frac{e^2}{L}\sum_n
\frac{|\langle {\rm GS}|J|n\rangle|^2}{\omega+i0-E_n+E_{\rm GS}}\ .
\eea
Ref.~\onlinecite{Tiegel15} provides results for the quantity
\be
\chi_+(\omega;L,\eta)=
\int_{-\infty+i0}^{\infty+i0}\frac{d\omega'}{2\pi}\
\frac{2\eta}{\eta^2+(\omega-\omega')^2}
{\rm Im}\; \chi_+(\omega')\ ,
\ee
where $L$ is the chain length and $\omega$ takes values on a regular
grid of frequencies. We first use rational function interpolation to
extract a continuous function from the numerical data, which we then
deconvolve using the Richardson-Lucy algorithm \cite{RL1,RL2}. 

\begin{figure}[ht!]
		\includegraphics[width=0.45\columnwidth]{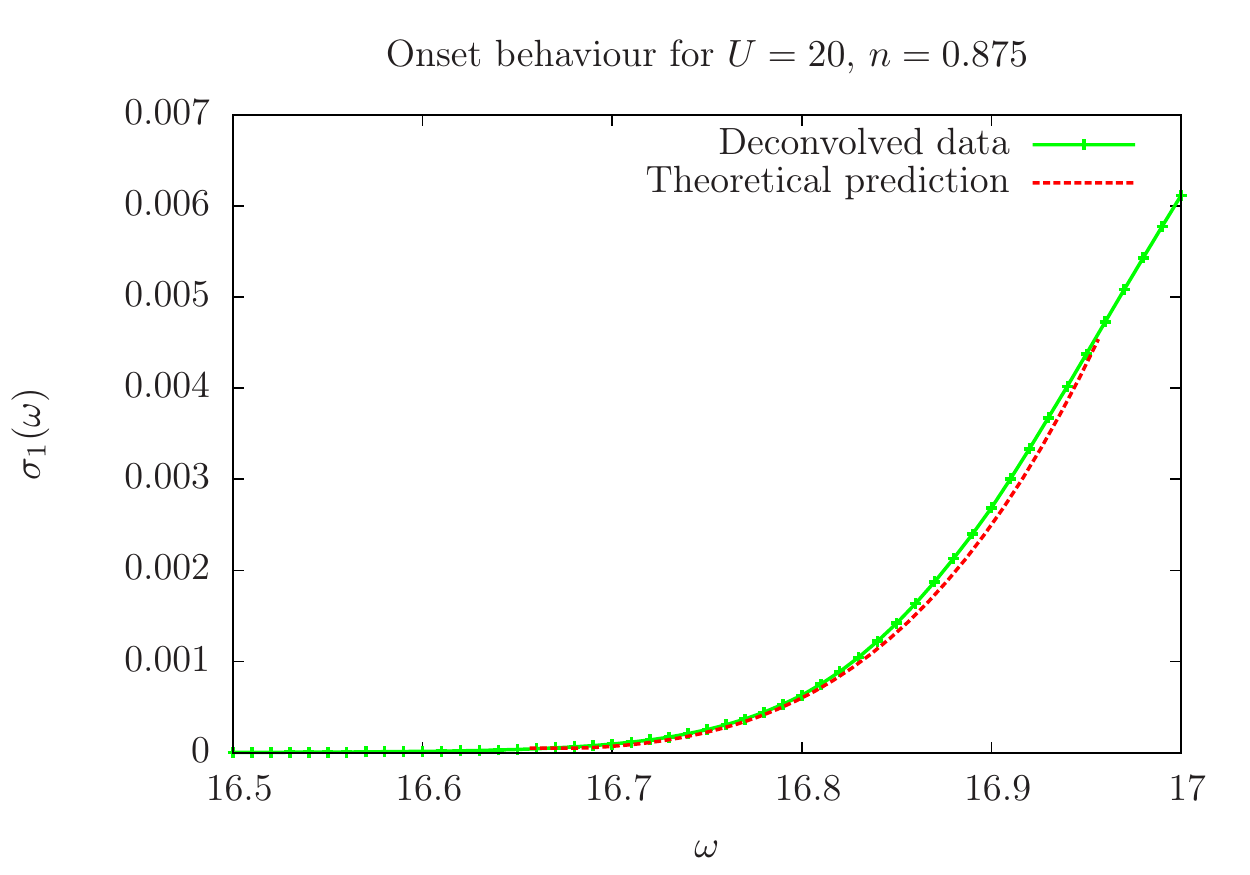}
		\label{fig:numCom1}
		\includegraphics[width=0.45\columnwidth]{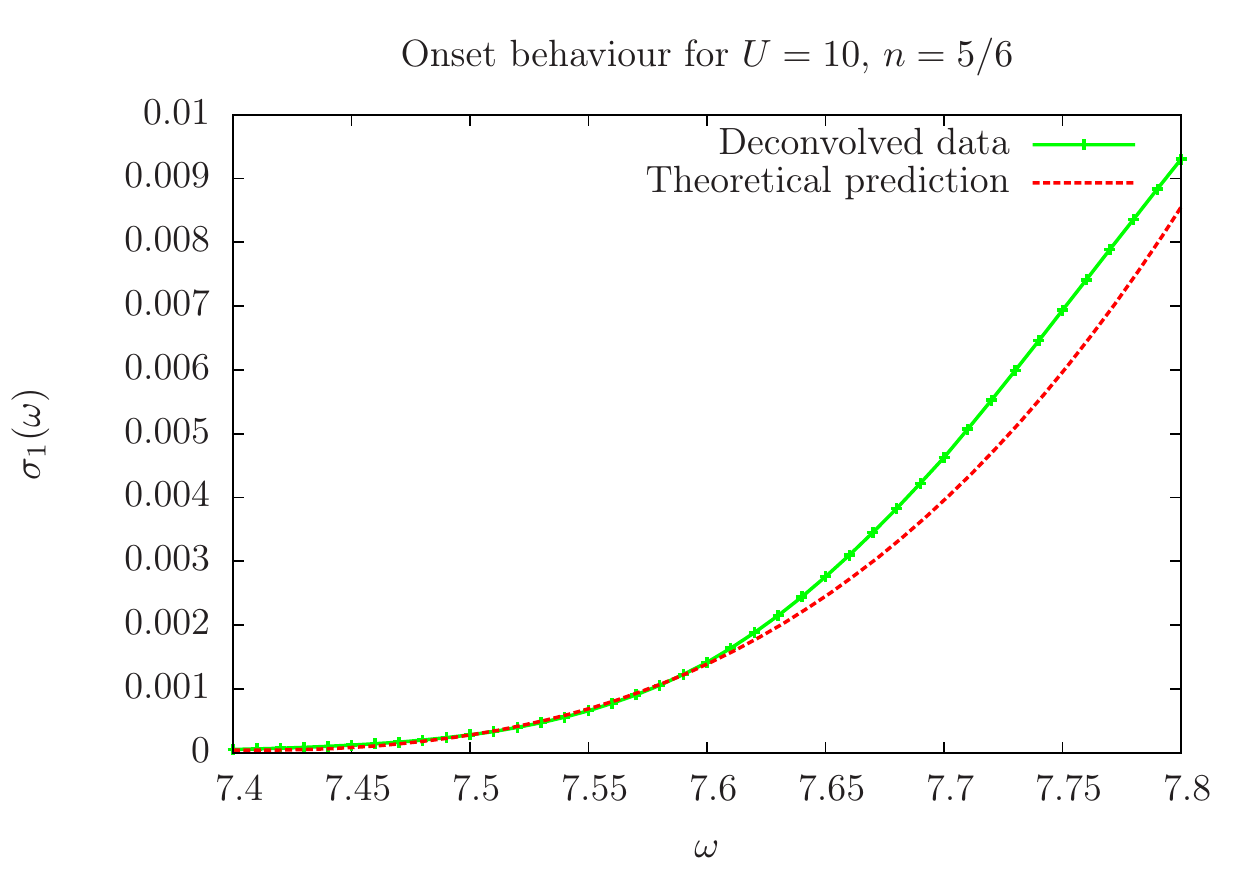}
		\label{fig:numCom2}
	\caption{Comparison of deconvolved DMRG data with the onset as predicted in (\ref{eq:klOnsetForm}), 
	varying the offset and overall scale factor and choosing $\Lambda = (1-n)\pi/10.$} \label{fig:onset}
\end{figure}

The deconvolved numerical results obtained in this way can then be
compared to the onset predicted at the lowest threshold, as given by
(\ref{eq:klOnsetForm}). We allow for an unknown scale factor in the
calculation, as well as a small constant contribution from the
particle-hole excitations. We choose a specific value of the cutoff
$\Lambda$, but as noted earlier, the results do not depend strongly on
the precise choice. Due to the soft nature of the onset predicted, it
is only realistic to compare the initial onset due to the
$k$-$\Lambda$ string, as when moving away from this point less
relevant operators will begin to contribute. Comparisons between the
prediction of the MIM and numerical results are shown in
Fig.~\ref{fig:onset}. The agreement is not perfect, but the results
are seen to be compatible. As usual the size of the frequency window
in which the MIM prediction applies is not known.
The theoretical and numerical results of the onset being convex 
are in stark contrast to the results of Ref.~\onlinecite{PhysRevLett.84.4673}, 
which for the parameters we consider predicts concave power-law behaviour.

\section{Away from $q=0$}
One can easily generalise the results here to examine the conductivity at \emph{finite momentum} i.e. consider
\be
\chi^J(\omega,q)= -ie^2 \int_0^\infty dt\ \sum_{l=-L/2}^{L/2-1} e^{i(\omega t -  q l a_0)} \langle GS | [J_{l}(t),J_0(0)]|GS\rangle, \label{eq:condDef}
\ee
for $q\neq0$. The analysis proceeds in an identical manner to identify the quantities $N_{c,s}^\imp$, $D_{c,s}^\imp$, as the high-energy impurity simply shifts its momentum.
However, in order to find the threshold away from $q\in [(n-1)\pi,(1-n)\pi]$, umklapp processes must be involved. For an arbitrary filling, these will generically be of a very large order, which will suppress the contributions. 
The dispersion relation is generically no longer a saddle point and therefore in the region where one impurity is involved and one can linearise, power-law behaviour for the onset will be obtained. 
For the $k$-$\Lambda$ string at momentum $q$, one must first solve for $\Lambda^p$ such that $p_{k\Lambda}(\Lambda^p) = q$, and then $N_c^\imp(\Lambda^p)$, $D_c^\imp(\Lambda^p)$ can be calculated. $\gamma$ is still given by (\ref{eq:gammakl}). The most relevant contribution will then have the form
\be
{\rm Im}\; \chi(\omega, q\neq 0) \sim \frac{1}{(\omega - \omega_{\rm th}(q))^{\gamma(\Lambda^p)}} ,
\ee
where $\gamma(\Lambda^p)$ is a function of the quantities $N_\alpha^\imp$, $D_\alpha^\imp$ and $K_\alpha$.

\section{Summary and conclusions}\label{sec:conc}
We have studied the optical conductivity $\sigma_1(\omega)$ in the one
dimensional Hubbard at zero temperature and close to half
filling. Recent DMRG computations \cite{Tiegel15} have shown that
in this regime $\sigma_1(\omega)$ is very small within a ``pseudo-gap''
and exhibits a rapid increase above an energy scale $E_{\rm opt}$ that
depends on doping as well as the interaction strength $U$. Using the
Bethe Ansatz we have identified the relevant excitations that
contribute to $\sigma_1(\omega)$ for $\omega>E_{\rm opt}$. One of
these, the $k$-$\Lambda$ string excitation, had been previously
proposed to describe the scale $E_{\rm opt}$
\cite{PhysRevLett.84.4673}. We then followed
Ref.~\onlinecite{EsslerPereira} to construct a mobile impurity model
describing the behaviour of $\sigma_1(\omega)$ above $E_{\rm
  opt}$. The analysis of this model entailed several generalizations
relating to the projection of lattice operators to local fields in the
MIM, the treatment of excitations that are not highest weight states
with respect to the $\eta$-pairing algebra of the Hubbard model, and
considering the mobile impurity to be located at a maximum of its
dispersion. We also derived an explicit expression for the finite-size
momentum of the relevant Bethe Ansatz states, which is useful in
determining the various unknown parameters in the MIM. Our main
result is to show that the MIM approach predicts a smooth, slow
increase in $\sigma_1(\omega)$ for frequencies above $E_{\rm
  opt}$. This is in contrast to the half-filled case \cite{JeckEssler}
and previous predictions\cite{PhysRevLett.84.4673}, but consistent
with recent dynamical DMRG computations\cite{Tiegel15}. 
The results presented in this work are by construction specific to the
Hubbard model. However, we expect the gross features seen in the
optical conductivity to be quite general for weakly doped Mott insulators.
In particular, we expect that such systems to exhibit a rapid increase of
$\sigma_1(\omega)$ above a pseudo-gap. As the functional form of the
increase is non-universal, it is conceivable that for other models it
could be considerably steeper than in the case considered here.

\section{Acknowledgements}
We thank Imke Schneider and Alexander Tiegel for helpful discussions.
This work was supported by the EPSRC under grants EP/I032487/1 and
EP/J014885/1. 
\appendix
\section{Velocities and Luttinger parameters in zero magnetic field}
\label{app:vK}
In zero magnetic field the charge and spin velocities are given in
terms of the solutions to the linear integral equations (\ref{eq:gsrhoc}), (\ref{eq:gsrhos}), (\ref{eq:ec}), (\ref{eq:es}) for the
dressed energies and root densities as 
\be
v_c=\frac{\eps'_c(Q)}{2\pi\rho_{c,0}(Q)}\ ,\qquad v_s=\frac{\eps'_s(\infty)}{2\pi\rho_{s,0}(\infty)}.
\ee
The spin Luttinger parameter is fixed by spin rotational symmetry to
be
\be
K_s=1.
\label{Kseq1}
\ee
We stress that all spin excitations relevant to our mobile impurity
model description occur at (approximately) zero energy, so that
corrections to \fr{Kseq1} are negligible. The charge Luttinger parameter is
\be
K_c=\frac{\xi^2(Q)}{2},
\ee
where $\xi(k)$ is the solution of the linear integral equation
\be
\xi(k)=1+ \int_{-Q}^Q\d k'\, \cos k' \ R(\sin k- \sin k')\ \xi(k'). \label{eq:dressedChargeZeroField}
\ee
Here $R(x)$ is defined in \fr{rofx}.

\section{Bethe Ansatz results for $k$-$\Lambda$ string} 
\label{app:klBAfull}
Having established that the threshold above the low-energy continuum can be explained by a $k$-$\Lambda$ string excitation, the simplest equations to consider  are the Takahashi equations\cite{book} in the presence of a single $k$-$\Lambda$ string of length $2$ i.e. consisting of 1 $\Lambda$ and 2 $k$s. 
It is also clear that as the correlator is a zero momentum quantity, $k$-$\Lambda$ string is pinned to zero momentum.
The Takahashi equations can be analysed for large $L$, keeping terms to $\mathcal{O}(L^{-2})$ in order to calculate the finite-size corrections to the energy.
The counting functions in this specific case are given by
\beA
L z_c(k_j) &=k_j L + \sum_{\alpha = 1}^{M-1} \theta\left( \frac{\sin k_j - \Lambda_\alpha}{u} \right) + \theta\left( \frac{\sin k_j - \Lambda'^p}{u} \right), && j=1,\dots,N-2, \\
L z_s(\Lambda_\alpha) &= \sum_{j=1}^{N-2} \theta\left( \frac{\Lambda - \sin k_j}{u} \right) - \sum_{\beta=1}^{M-1} \theta \left( \frac{ \Lambda_\alpha - \Lambda_\beta}{2u} \right), && \alpha=1,\dots,M-1.
\eeA
Employing the Euler-Maclaurin summation formula
\be
\frac{1}{L} \sum_{n=n_1}^{n_2} f\left( \frac{n}{L} \right) = \int_{\frac{n_-}{L}}^{\frac{n_+}{L}} dx\ f(x) + \frac{1}{24L^2} \left( f'\left( \frac{n_-}{L}\right) - f'\left(\frac{n_+}{L} \right) \right) + \dots, \label{eq:eulerMaclaurin}
\ee
where $n_+ = n_2 + \frac{1}{2}$ and $n_- = n_1 - \frac{1}{2}$, it can be seen that
\begin{align}
z_c(k) &= k + \int_{\am}^\ap \d\Lambda\ \theta\left( \frac{\sin k - \Lambda}{u} \right) \rho_s(\Lambda)+ \frac{1}{L} \theta\left( \frac{\sin k - \Lambda'^p}{u} \right)\nn
&+ \frac{2\pi}{24L^2} \left[ \frac{a_1(\sin k - \ap)}{\rho_s(\ap)} - \frac{a_1(\sin k - \am)}{\rho_s(\am)} \right],\\
z_s(\Lambda) &= \int_\qm^\qp dk\ \theta\left( \frac{\Lambda-\sin k}{u} \right) \rho_c(k) - \int_\am^\ap d\Lambda'\ \theta\left( \frac{\Lambda - \Lambda'}{2u} \right) \rho_s(\Lambda') \nn
&+ \frac{2\pi}{24L^2} \left[ \frac{a_1(\Lambda-\sin\qp)\cos \qp}{\rho_c(\qp)} - \frac{a_1(\Lambda-\sin\qm) \cos\qm}{\rho_c(\qp)} - \frac{a_2(\Lambda - \ap)}{\rho_s(\ap)} + \frac{a_2(\Lambda - \am)}{\rho_s(\am)} \right].
\end{align}
Taking derivatives, equations for the root densities can be found
\beA
\rho_c(k) &= \frac{1}{2\pi} + \int_\am^\ap d\Lambda\ \cos k\, a_1(\sin k - \Lambda) \rho_s(\Lambda) + \frac{1}{L} \cos k\, a_1(\sin k - \Lambda'^p) \\
&+\frac{1}{24L^2} \left[ \frac{\cos k\, a_1'(\sin k - \ap)}{\rho_s(\ap)} - \frac{\cos k\, a_1'(\sin k - \am)}{\rho_s(\am)} \right] ,
\eeA
\beA
\rho_s(\Lambda) &= \int_\qm^\qp dk\ a_1(\Lambda - \sin k) \rho_c(k) - \int_\am^\ap d\Lambda'\ a_2(\Lambda-\Lambda') \rho_s(\Lambda')\\
&+\frac{1}{24L^2} \left[ \frac{a_1'(\Lambda - \sin\qp)\cos\qp}{\rho_c(\qp)} - \frac{a_1'(\Lambda - \sin\qm)\cos\qm}{\rho_c(\qm)} - \frac{a_2'(\Lambda - \ap)}{\rho_s(\ap)} + \frac{a_2'(\Lambda - \am)}{\rho_s(\am)} \right] .
\eeA
As the integral equations are linear, we can write
\be
\rho_\alpha(z_\alpha) := \rho_{\alpha,0}(z_\alpha) + \frac{1}{L} \rho_{\alpha,1}(z_\alpha)  + \frac{1}{24L^2} \sum_{\beta,\sigma} \frac{ f_{\alpha\beta}^{(\sigma)}(z)}{\rho_\beta(X^\beta_\sigma)}, \label{eq:split1}
\ee
here $X_\sigma^c=Q_\sigma$, $X_\sigma^s=A_\sigma$.  The integral equations satisfied by the first two terms in (\ref{eq:split1}) are
\be
\rho_{\alpha,a}(z) = \rho_{\alpha,a}^{(0)}(z) + K_{\alpha\beta}*\rho_{\beta,a},\qquad a=0,1 , \label{eq:split2}
\ee
with $K_{\alpha\beta}*f_\beta$ denoting the convolution $\sum_{\beta} \int_{X^\beta_-}^{X^\beta_+} \d z_\beta\, K_{\alpha\beta}(z_\alpha,z_\beta) f_{\beta}(z_\beta)$,
the kernels defined by
\beA
K_{cc}(k,k') = 0,&\qquad K_{cs}(k,\Lambda) = \cos k\, a_1(\sin k - \Lambda),\\
K_{sc}(\Lambda,k) = a_1(\Lambda - \sin k), &\qquad K_{ss}(\Lambda,\Lambda') = - a_2(\Lambda-\Lambda') , \label{eq:kernels}
\eeA
and the driving terms given by
\begin{align}
\rho_{\alpha,0}^{(0)} &= \frac{\delta_{\alpha,c}}{2\pi},\\
\rho_{\alpha,1}^{(0)} &= \delta_{\alpha,c} \cos k\; a_1(\sin k - \Lambda'^p).
\end{align}
The final integral equation is determined by
\be
f_{\alpha\beta}^{(\sigma)} = d_{\alpha\beta}^{(\sigma)} + K_{\alpha\gamma}*f_{\gamma\beta}^{(\sigma)}, \label{eq:split3}
\ee
where
\be
d_{\alpha\beta}^{(\sigma)} = -\sigma \frac{\partial}{\partial z'} K_{\alpha\beta}(z,z') \bigg\vert_{z'=X^\beta_\sigma}.
\ee
The exact finite-size energy of the system is given by (\ref{eq:energyFull}). 
Using the Euler-Maclaurin summation formula (\ref{eq:eulerMaclaurin}) again, corrections can be kept to $\mathcal{O}(L^{-1})$, yielding
\be
E = Lu + L \sum_\alpha \int_{X_-^\alpha}^{X_+^\alpha} dz\ \varepsilon_{\alpha}^{(0)}(z)\rho_\alpha(z) + \varepsilon_{k\Lambda}^{(0)}(\Lambda'^p).
\ee
Expanding in powers of $L$ and exploiting the identical kernels of the integral equations for dressed charge and root density equations, if the energy is now considered as a functional of the integration boundaries, performing an expansion about $\sigma X^\alpha$ to second order (the first order term vanishes)\cite{FHLE}, it can be shown that
\be
E = Le_{GS}(\{ X^\alpha \} ) + \varepsilon_{k\Lambda}(\Lambda'^p) + L\pi \sum_\alpha v_\alpha \left\{ \left(\rho_{\alpha,0}(X^\alpha) (X_+^\alpha - X^\alpha )\right)^2 + \left(\rho_{\alpha,0}(X^\alpha) (X_-^\alpha + X^\alpha )\right)^2 \right\}
\ee
\subsection{Impurity densities}
The following are taken as definitions
\begin{align}
n_\alpha &= \int_{X_-^\alpha}^{X_+^\alpha} dz\ \rho_\alpha(z),\\
2D_c &= I_++I_- = \frac{L}{2\pi} \left[ z_c(\qp) + z_c(\qm) \right], \\
2D_s &= J_++J_- = \frac{L}{2\pi} \left[ z_s(\ap) + z_s(\am) \right].
\end{align}
The corrections from adding the ``impurity'' i.e. the high-energy excitation can be identified and separated off from the terms that would be present without it. This is achieved by using that
\begin{align}
\lim_{\Lambda \to \infty} z_s(\Lambda) &= - \lim_{\Lambda \to -\infty}z_s(\Lambda),\\
\lim_{k\to\pm\pi}z_c(k) &= - \int_\am^\ap d\Lambda\ \theta\left( \frac{\Lambda}{u} \right) \rho_s(\Lambda) - \theta\left( \frac{\Lambda'^p}{u} \right).
\end{align}
This allows the ``quantum numbers'' to be expressed in terms of integrals of the root densities, which can then be split off order-by-order in $1/L$. More explicitly, one finds that
\beA
2D_s     &= L \left( \int_{-\infty}^\am d\Lambda\ \rho_s(\Lambda) - \int_\ap^\infty d\Lambda\ \rho_s(\Lambda) \right),\\
     &= L\left( \int_{-\infty}^\am d\Lambda\ \rho_{s,0}(\Lambda) - \int_\ap^\infty d\Lambda\ \rho_{s,0}(\Lambda) \right) + 2D_s^\mathrm{imp} ,
\eeA
where
\be
2D_s^\mathrm{imp} = \int_{-\infty}^{-A} d\Lambda\ \rho_{s,1}(\Lambda) - \int_A^\infty d\Lambda\ \rho_{s,1}(\Lambda) .
\ee
Similarly for the charge sector
\begin{align}
	2D_c &= \frac{L}{2\pi} \left( z_c(\qp) + z_c(\qm) - z_c(\pi) - z_c(-\pi) - 2\int_\am^\ap d\Lambda\ \rho_{s}(\Lambda) \theta\left( \frac{\Lambda}{u} \right) - \frac{2}{L}\theta \left( \frac{\Lambda'^p}{u} \right)\right),\nn
	&= L \left( \int_{-\pi}^\qm dk\ \rho_{c,0}(k) -\int_\qp^\pi dk\ \rho_{c,0}(k) - \frac{1}{\pi}\int_\am^\ap d\Lambda\ \rho_{s,0}(\Lambda)\theta\left( \frac{\Lambda}{u} \right)\right) +2D_c^\mathrm{imp},\\
2D_c^\mathrm{imp}&= \int_{-\pi}^{-Q} dk\ \rho_{c,1}(k) -\int_Q^\pi dk\ \rho_{c,1}(k) - \frac{1}{\pi}\int_{-A}^A d\Lambda\ \rho_{s,1}(\Lambda)\theta\left( \frac{\Lambda}{u} \right) - \frac{1}{\pi} \theta \left( \frac{\Lambda'^p}{u} \right) .
\end{align}
Similarly,
\be
N_\alpha^\mathrm{imp} = \int_{-X^\alpha}^{X^\alpha} dz\ \rho_{\alpha,1}(z).
\ee
\subsection{Relation between $X_\sigma^\alpha-\sigma X^\alpha$ and the impurity densities}
Following Ref.~\onlinecite{FHLE}, considering the variation of the integration bounds 
$X_\sigma^\alpha$ with respect to $n_\beta$, it can be seen that, in terms of  
the dressed charge matrix\cite{book} $Z_{\alpha\beta}$, defined by 
\beA
Z_{\alpha\beta} &= \xi_{\alpha\beta}(X^\beta),\label{eq:dressedChargeDef}\\
\xi_{\alpha\beta}(z_\beta) &= \delta_{\alpha\beta} + \xi_{\alpha\gamma} * K_{\gamma\beta},
\eeA
with $K_{\alpha\beta}$ given by (\ref{eq:kernels}), 
one finds 
\be
X_\sigma^\alpha - \sigma X^\alpha = \sigma \frac{1}{2} \frac{ Z^{-1}_{\alpha\beta}}{\rho_{\alpha,0}(X^\alpha)} \left( \Delta n_\beta - \frac{1}{L} N^\imp_\beta \right) + \frac{Z^\top_{\alpha\beta}}{\rho_{\alpha,0}(X^\alpha)} \left( d_\beta - \frac{1}{L} D^\imp_\beta \right).
\ee
These results can be inserted into the finite-size energy, which now reads as
\be
E = L e_{GS}(\{X^\alpha\}) + \varepsilon_{k\Lambda}(0) + \frac{1}{L} \left( -\frac{\pi}{6}(v_s + v_c) + 2\pi\left[ \frac{1}{4} \Delta\widetilde{N}_\alpha (Z^\top)^{-1}_{\alpha\gamma} v_\gamma Z^{-1}_{\gamma\beta} \Delta \widetilde{N}_\beta + \widetilde{D}_\alpha Z_{\alpha\gamma} v_\gamma Z^\top_{\gamma\beta} \widetilde{D}_\beta \right] \right),
\ee
where
\begin{align}
\widetilde{D}_\alpha &= D_\alpha - D_\alpha^\imp, \label{eq:delDt}\\
\Delta \widetilde{N}_\alpha &= \Delta N_\alpha - N_\alpha^\imp. \label{eq:delNt}
\end{align}
\subsection{Simplifications for zero magnetic field}
 In the $B\to 0$ limit, the integration boundary $A \to \infty$ and many results simplify by use of Fourier transforms. Useful identities used can be found in Ch. 17 of Ref.~\onlinecite{book}.
First, the dressed charge matrix adopts the simple form
\be
Z = \begin{pmatrix} \xi & 0 \\ \frac{\xi}{2} & \frac{1}{\sqrt{2}} \end{pmatrix},
\ee
where $\xi = \xi(Q)$ and $\xi(k)$ obeys (\ref{eq:dressedChargeZeroField}).
Following a similar method to Ref. \onlinecite{FHLE}, the root densities can be shown to simplify as
\begin{align}
\rho_{c,1}(k) &= \cos k\; a_1(\sin k - \Lambda'^p) + \cos k \int_{-Q}^Q dk'\ \rho_{c,1}(k') R(\sin k - \sin k'),\\
\rho_{s,1}(\Lambda) &= \int_{-Q}^Q dk\ \rho_{c,1}(k)\; s(\Lambda - \sin k ), \label{eq:rhos1Bzero}
\end{align}
where
\be
s(x) = \frac{1}{4u\cosh\left( \frac{\pi x}{2u} \right)}.
\ee
Considering the Fourier transform of (\ref{eq:rhos1Bzero}), it can be shown that
\be
N_s^\mathrm{imp} =  \frac{1}{2} N_c^\mathrm{imp}.
\ee
In the $\Lambda'^p \to 0$ limit, both $\rho_{c,1}(k)$ and $\rho_{s,1}(\Lambda)$ are even functions and therefore
\be
D_c^\mathrm{imp} = D_s^\mathrm{imp} = 0.
\ee
It is useful to note that the dressed energies take the form
\begin{align}
\varepsilon_c(k) &= -2\cos k - \mu - 2u + \int_{-Q}^Q \d k'\, \cos k'\; R(\sin k - \sin k') \varepsilon_c(k'),\\
\varepsilon_s(\Lambda) &= \int_{-Q}^Q dk\ \cos k\; s\left[ \Lambda - \sin k \right]\;\varepsilon_c(k), \\
\varepsilon_{k\Lambda}(\Lambda) &= 4{\rm Re}\sqrt{ 1- (\Lambda - iu)^2} - 2\mu - 4u + \int_{-Q}^Q dk \cos k\; a_1(\sin k -\Lambda)\; \varepsilon_c(k).
\end{align}
The value of $\varepsilon_{k\Lambda}(0)$ provides the location of the threshold at zero momentum in this sector.
The finite-size energy can therefore be simply written as
\be
E = L e_{GS}(\{ X^\alpha \}) + \varepsilon_{k\Lambda}(0) - \frac{\pi v_c}{6L} +
\frac{2\pi v_c}{L} \left[ \frac{ (\Delta N_c - N_c^\imp)^2}{4\xi^2} + \xi^2 \left( D_c - D_c^\imp  + \frac{D_s - D_s^\imp}{2} \right)^2 \right] .
\ee

\section{Bethe Ansatz results for high-energy charge particle}
\label{app:chBAfull}
\subsection{Bethe Ansatz calculation}
Starting from the Takahashi equations
\beA
L z_c(k_j) &= k_j L + \sum_{\alpha=1}^M \theta\left( \frac{\sin k_j - \Lambda_\alpha}{u} \right), && j=1,\dots,N,\\
L z_s(\Lambda_\alpha) &= \sum_{j=1}^{N} \theta \left( \frac{ \Lambda_\alpha - \sin k_j}{u} \right) - \sum_{\beta=1}^M \theta \left( \frac{\Lambda_\alpha-\Lambda_\beta}{2u} \right), && \alpha = 1,\dots,M. \label{eq:takahashiStartingPoint}
\eeA
We can use the Euler-Maclaurin formula (\ref{eq:eulerMaclaurin}) to recast this as
\begin{align}
z_c(k) &= k + \int_{A_-}^{A^+} \d \Lambda\, \theta\left( \frac{\sin k-\Lambda}{u} \right) \rho_{s}(\Lambda) + \frac{2\pi}{24 L^2} \left[ \frac{a_1(\sin k - \ap)}{\rho_s(\ap)} - \frac{a_1(\sin k - \am)}{\rho_s(\am)} \right],\\
z_s(\Lambda) &= \int_{Q_-}^{Q_+} \d k\, \theta\left( \frac{\Lambda-\sin k}{u} \right) \rho_c(k) - \int_{A_-}^{A_+} \d \Lambda' \theta\left( \frac{\Lambda-\Lambda'}{2u} \right) \rho_s(\Lambda') + \frac{1}{L} \theta\left( \frac{\Lambda - \sin k^p}{u} \right) \nn
&+\frac{2\pi}{24L^2} \left[ \frac{a_1(\Lambda - \sin \qp)\cos\qp}{\rho_c(\qp)} - \frac{a_1(\Lambda - \sin \qm)\cos\qm}{\rho_c(\qm)} - \frac{ a_2(\Lambda - \ap)}{\rho_s(\ap)} + \frac{a_2(\Lambda - \am)}{\rho_s(\am)} \right].
\end{align}
Taking derivatives gives the root densities
\begin{align}
\rho_c(k) &= \frac{1}{2\pi} + \int_{A_-}^{A_+} \d \Lambda a_1(\sin k - \Lambda) \rho_s(\Lambda) \cos k + \frac{1}{24L^2} \cos k \left[ \frac{a_1'(\sin k -\ap)}{\rho_s(\ap)} - \frac{a_1'(\sin k - \am)}{\rho_s(\am)} \right],\\
\rho_s(\Lambda) &= \int_{Q_-}^{Q_+} \d k a_1(\Lambda - \sin k) \rho_c(k) - \int_{A_-}^{A_+} \d \Lambda' a_2(\Lambda - \Lambda') \rho_s(\Lambda') + \frac{1}{L} a_1(\Lambda - \sin k^p) \nn
&+\frac{1}{24L^2} \left[ \frac{a_1'(\Lambda - \sin \qp)\cos\qp}{\rho_c(\qp)} - \frac{a_1'(\Lambda - \sin \qm)\cos\qm}{\rho_c(\qm)} + \frac{a_2'(\Lambda - \am)}{\rho_s(\am)} - \frac{a_2'(\Lambda - \ap)}{\rho_s(\ap)} \right].
\end{align}
We can again split these linear integral equations into the form
(\ref{eq:split1}), (\ref{eq:split2}), (\ref{eq:split3}) 
where in this case 
\be
\rho_{\alpha,0}^{(0)} = \frac{\delta_{\alpha,c}}{2\pi},\qquad \rho_{\alpha,1}^{(0)}(z_\alpha) = \delta_{\alpha,s} a_1(z_\alpha - \sin k^p). \label{eq:rhoImp3}
\ee
and the integral kernels are again given by (\ref{eq:kernels}). 
We can then construct the impurity densities
\begin{align}
N_\alpha^\imp &= \int_{-X_\alpha}^{X^\alpha} \d z_\alpha \rho_{\alpha,1}(z_\alpha), \label{eq:naimpdef1}\\
2D_c^\imp &= \int_{Q}^\pi \d k \left[ \rho_{c,1}(-k) -\rho_{c,1}(k) \right] - \frac{1}{\pi} \int_{-A}^A \d \Lambda\, \rho_{s,1}(\Lambda)\, \theta\left( \frac{\Lambda}{u} \right), \label{eq:dcimpdef1}\\
2D_s^\imp &= \int_{A}^\infty\d \Lambda\, \left[ \rho_{s,1}(-\Lambda) - \rho_{s,1}(\Lambda) \right].
\end{align}
To determine the thermodynamic rapidity $k^p$ and the finite-size correction $\delta k^p$, we can examine the requirements that
\begin{align}
z_c(k^p_L) &= \frac{2\pi I^p}{L},\label{eq:rapidityRequirement1}\\
z_{c,0}(k^p) &=\frac{2\pi I^p}{L}, \label{eq:rapidityRequirement2}
\end{align}
with $k^p_L = k^p + \frac{\delta k^p}{L}$.
Expanding (\ref{eq:rapidityRequirement1}) in the deviation $\delta k^p$ and using (\ref{eq:rapidityRequirement2}) yields
\be
\delta k^p = -\frac{L}{2\pi \rho_{c,0}(k^p)} \left[ \sum_{\beta,\sigma} \Psi^{(\sigma)}_\beta(k^p) (X^\sigma_\beta - \sigma X^\beta)\right] - \frac{1}{2\pi \rho_{c,0}(k^p)} \int_{-A}^A \d \Lambda \rho_{s,1}(\Lambda) \theta\left( \frac{\Lambda - \sin k^p}{u} \right), \label{eq:deltakp}
\ee
where
\begin{align}
\Psi^{(\sigma)}_\beta(k) &= \sigma \rho_{s,0}(A) \theta\left( \frac{\sigma A - \sin k}{u} \right) \delta_{s,\beta} + \int_{-A}^A \d \Lambda\, r_{s,\beta}^{(\sigma)}(\Lambda) \theta \left( \frac{\Lambda - \sin k}{u} \right), \label{eq:psiDef}\\
r_{\alpha\beta}^{(\sigma)} &= \sigma \rho_{\beta,0}(X^\beta) K_{\alpha\beta}(z_\alpha,\sigma X^\beta) + K_{\alpha\gamma} * r_{\gamma\beta}^{(\sigma)}.
\end{align}
Using the results of Appendix \ref{app:fsm}, this can be shown to reduce to
\be
\delta k^p = \frac{1}{\rho_{c,0}(k^p)} \sum_{\alpha = c,s} \left(  N_\alpha^\imp D^{\phantom i}_\alpha + D_\alpha^\imp \Delta N^{\phantom i}_\alpha - D_\alpha^\imp N_\alpha^\imp \right). \label{eq:dkpSimple}
\ee
We then have that
\be
E = e_{GS}L + \varepsilon_c(k^p) + \varepsilon_c'(k^p) \frac{\delta k^p}{L} - \frac{\pi}{6L}(v_s+v_c) +\frac{2\pi}{L} \left[ \frac{1}{4} \Delta\widetilde{N}_\gamma (Z^\top)^{-1}_{\gamma\alpha} v_\alpha Z^{-1}_{\alpha\beta} \Delta \widetilde{N}_\beta + \widetilde{D}_\gamma Z_{\gamma\alpha} v_\alpha Z^\top_{\alpha\beta} \widetilde{D}_\beta \right], \label{eq:FSSchargeParticle}
\ee
with the form of $\widetilde{D}_\alpha$, $\Delta\widetilde{N}_\alpha$ and $Z_{\alpha\beta}$ given by (\ref{eq:delDt}), (\ref{eq:delNt}), (\ref{eq:dressedChargeDef}). 
\subsection{Simplification for $B\to 0$}

In the $B\to 0$ limit, the integral equations describing the impurity densities are given by
\be
\rho_{c,1}(k) = \cos k\, R(\sin k - \sin k^p) + \cos k \int_{-Q}^Q \d k'\, R(\sin k - \sin k')\rho_{c,1}(k'),
\ee
\be
N_c^\imp = \int_{-Q}^Q \d k \rho_{c,1}(k),\qquad
N_s^\imp = \frac{1}{2}(1+N_c^\imp),
\ee
\beA
2 D_c^\imp &= \int_Q^\pi \d k \left[ \rho_{c,1}(-k) - \rho_{c,1}(k) \right] + \frac{i}{\pi} \left\{ \ln \left[ \frac{ \Gamma\left( \frac{1}{2} - i \frac{\sin k^p}{4u} \right) \Gamma\left( 1 + i \frac{\sin k^p}{4u}\right)}{ \Gamma\left( \frac{1}{2} + i \frac{\sin k^p}{4u} \right) \Gamma\left( 1 - i \frac{\sin k^p}{4u}\right)}\right]\right\} \\
&+\frac{i}{\pi}\int_{-Q}^Q \d k \rho_{c,1}(k) \left\{ \ln \left[ \frac{ \Gamma\left( \frac{1}{2} - i \frac{\sin k}{4u} \right) \Gamma\left( 1 + i \frac{\sin k}{4u}\right)}{ \Gamma\left( \frac{1}{2} + i \frac{\sin k}{4u} \right) \Gamma\left( 1 - i \frac{\sin k}{4u}\right)}\right] \right\},
\eeA
\be
D_s^\imp = 0.
\ee
This gives the finite-size corrections to the energy as
\begin{align}
E &= e_{GS}L + \varepsilon_c(k^p) + \varepsilon'(k^p) \frac{\delta k^p}{L}  - \frac{\pi v_c}{6L} \nn
&+\frac{2\pi v_c}{L} \left[ \frac{ (\Delta N_c - N_c^\imp)^2}{4\xi^2} + \xi^2 \left( D_c - D_c^\imp  + \frac{D_s}{2} \right)^2 \right]\nn
&+\frac{2\pi v_s}{L} \left[ \frac{1}{2} \left( \Delta N_s - \frac{ \Delta N_c}{2} -\frac{1}{2} \right)^2 + \frac{D_s^2}{2} \right]
,
\end{align}
where $\xi = \xi(Q)$ and $\xi(k)$ obeys (\ref{eq:dressedChargeZeroField}).

\section{Bethe Ansatz results for two high-energy charge hole excitations}\label{app:holonPairFull}
We again start from (\ref{eq:takahashiStartingPoint}). 
Following similar steps to Appendices \ref{app:klBAfull} and \ref{app:chBAfull}, applying the Euler-Maclaurin summation formula (\ref{eq:eulerMaclaurin}) then allows us to write
\be
\rho_\alpha(z_\alpha) = \rho_{\alpha,0}(z_\alpha) + \frac{1}{L} \rho_{\alpha,1}(z_\alpha) + \frac{1}{24L^2} \sum_{\beta,\sigma} \frac{f_{\alpha\beta}^{(\sigma)}(z_\alpha)}{\rho_\beta(X_\sigma^\beta)}.
\ee
We can again split these linear integral equations into the form
(\ref{eq:split1}), (\ref{eq:split2}), (\ref{eq:split3}) 
where in this case 
\begin{align}
\rho_{\alpha,0}^{(0)} &= \frac{\delta_{\alpha,c}}{2\pi},\\
\rho_{\alpha,1}^{(0)} &= - \delta_{\alpha,s} \left[ a_1(\Lambda - \sin k^{h_1}) + a_1(\Lambda - \sin k^{h_2}) \right].
\end{align}
and the integral kernels are given by (\ref{eq:kernels}). 
We can now determine
\begin{align}
2D_c^\imp &= \int_{Q}^\pi \d k \left[ \rho_{c,1}(-k) - \rho_{c,1}(k) \right] - \frac{1}{\pi} \int_{-A}^A \d \Lambda\, \theta\left( \frac{\Lambda}{u} \right) \rho_{s,1}(\Lambda),\\
2D_s^\imp &= \int_A^\infty \d \Lambda\,\left[ \rho_{s,1}(-\Lambda) - \rho_{s,1}(\Lambda) \right].
\end{align}
We also have that
\be
z_c(k^{h_i}_L) = \frac{2\pi I^{h_i}}{L},\qquad 
z_{c,0}(k^{h_i}) = \frac{2\pi I^{h_i}}{L}, \label{eq:zckhi}
\ee
with $k^{h_i}_L = k^{h_i} + \frac{\delta k^{h_i}}{L}$, yielding
\be
\delta k^{h_i} = -\frac{L}{2\pi \rho_{c,0}(k^{h_i})} \left[ \sum_{\beta,\sigma} \Psi^{(\sigma)}_\beta(k^{h_i}) (X^\sigma_\beta - \sigma X^\beta)\right] - \frac{1}{2\pi \rho_{c,0}(k^{h_i})} \int_{-A}^A \d \Lambda \rho_{s,1}(\Lambda) \theta\left( \frac{\Lambda - \sin k^{h_i}}{u} \right), \label{eq:dkhi}
\ee
with $\Psi^{(\sigma)}(k)$ given by (\ref{eq:psiDef}).
We now have all of the quantities required to evaluate the finite-size spectrum in the presence of the two high-energy holons:
\begin{align}
E &= e_{GS}L - \varepsilon_c(k^{h_1}) - \varepsilon_c(k^{h_2}) -\varepsilon_c'(k^{h_1}) \frac{\delta k^{h_1}}{L}-\varepsilon_c'(k^{h_2}) \frac{\delta k^{h_2}}{L}  - \frac{\pi}{6L}(v_s+v_c)\nn
&+\frac{2\pi}{L} \left[ \frac{1}{4} \Delta\widetilde{N}_\gamma (Z^\top)^{-1}_{\gamma\alpha} v_\alpha Z^{-1}_{\alpha\beta} \Delta \widetilde{N}_\beta + \widetilde{D}_\gamma Z_{\gamma\alpha} v_\alpha Z^\top_{\alpha\beta} \widetilde{D}_\beta \right], 
\end{align}
with the form of $\widetilde{D}_\alpha$, $\Delta\widetilde{N}_\alpha$ and $Z_{\alpha\beta}$ given by (\ref{eq:delDt}), (\ref{eq:delNt}), (\ref{eq:dressedChargeDef}).

\subsection{Zero field}
In zero field, the integral equations for the functions $\rho_{c,1}$, $\rho_{s,1}$ simplify due to $A\to\infty$ allowing the use of a Fourier transform, specifically
\begin{align}
\rho_{c,1}(k) &= -\cos k \left[ R(\sin k - \sin k^{h_1}) + R(\sin k - \sin k^{h_2}) \right] + \cos k \int_{-Q}^Q \d k'\, R(\sin k - \sin k') \rho_{c,1}(k'), \\
\rho_{s,1} &= -s(\Lambda - \sin k^{h_1}) - s(\Lambda - \sin k^{h_2}) + \int_{-Q}^Q \d k\, s(\Lambda - \sin k) \rho_{c,1}(k).
\end{align}
We also have
\be
N_s^\imp = \frac{1}{2} N_c^\imp - 1.
\ee
The finite-size spectrum can then be written as
\be
\begin{aligned}
	E &= e_{GS} L -\varepsilon_c(k^{h_1}) - \varepsilon_c(k^{h_2}) - \frac{\delta k^{h_1}}{L} \varepsilon_c'(k^{h_1}) - \frac{\delta k^{h_2}}{L} \varepsilon_c'(k^{h_2})
	- \frac{\pi}{6L} (v_c + v_s)  \\
&+\frac{2\pi v_c}{L} \left[ \frac{ (\Delta N_c - N_c^\imp)^2}{4\xi^2} + \xi^2 \left( D_c - D_c^\imp  + \frac{D_s - D_s^\imp}{2} \right)^2 \right]  \\
&+\frac{2\pi v_s}{L} \left[ \frac{1}{2} \left( \Delta N_s - \frac{ \Delta N_c}{2} + 1 \right)^2 + \frac{(D_s - D_s^\imp)^2}{2} \right].
\end{aligned}
\ee

\section{Finite-size momentum spectrum} \label{app:fsm}
As well as the finite-size energies, it is also possible to  
match the finite-size momentum spectra.  
We consider here the simple case of a single high-energy 
charge excitation, but the reasoning is the same for other excitations.

\subsection{Mobile impurity model momentum spectrum} 
We bosonise the Hubbard chain at $U=0$,  
decomposing the fermionic annihilation operator as 
\be c_\sigma (x) = R_\sigma(x) e^{ik_F x} + L_\sigma(x) e^{-ik_F x}. \ee 
To identify the momentum operator, we consider it as the generator of translations by one site i.e.  
\be e^{-ia_0 P} c_\sigma(x) e^{ia_0 P} = c_\sigma(x+a_0). \ee
Which means that $R_\sigma(x) \to R_\sigma(x+a_0)e^{ik_Fa_0}$.
By utilising the refermionisation identities\cite{EsslerPereira}
\be
R_\up \sim \prod_{\alpha=c,s} e^{-\frac{i}{\sqrt{2}} \varphi_\alpha^* + \frac{i}{4\sqrt{2}} \Phi_\alpha^*},\qquad
L_\up \sim \prod_{\alpha=c,s} e^{\frac{i}{\sqrt{2}} \bar{\varphi}_\alpha^* - \frac{i}{4\sqrt{2}} \Phi_\alpha^*},
\ee
we can identify that, 
in terms of the mode expansion of the spin and charge modes, the momentum operator is given by
\bea
P &=& \frac{k_F}{\pi\sqrt{2}}\left( \bar{Q}^*_c -Q^*_c \right) +
\frac{1}{8\pi L} \left[  Q^*_c{}^2 - \bar{Q}^*_c{}^2 + Q^*_s{}^2 - \bar{Q}^*_s{}^2
  \right]\nn
&+& i \int \d x B^\dagger(x) \partial_x B(x) + \sum_{\alpha =
  c,s}\sum_{n=1}^\infty \frac{2\pi n}{L} \left(
c^\dagger_{\alpha,R,n} c^{\pd}_{\alpha,R,n}   -c^\dagger_{\alpha,L,n}
c^{\pd}_{\alpha,L,n}   \right). 
\eea 
Employing the unitary transformation, this can be written as 
\be
\begin{aligned}
	P &=\frac{k_F}{\pi\sqrt{2}} \left(  \bar{Q}^\circ_c - \bar{Q}^\circ_c - 4\pi\gamma_c + 4\pi\bar{\gamma}_c  \right) + \frac{1}{8\pi L} \left[ Q_c^\circ{}^2 - \bar{Q}_c^\circ{}^2 + Q_s^\circ{}^2 - \bar{Q}_s^\circ{}^2 \right] + i \int \d x \widetilde{B}^\dagger \partial_x \widetilde{B} \\ 
	&+\sum_{\alpha = c,s}\sum_{n=1}^\infty \frac{2\pi n}{L} \left(  c^\dagger_{\alpha,R,n} c^{\pd}_{\alpha,R,n}   -c^\dagger_{\alpha,L,n} c^{\pd}_{\alpha,L,n}    \right). 
\end{aligned}
\ee
Which therefore predicts a finite-size spectrum  of the form 
\be
\begin{aligned}
	P &=\frac{k_F}{\pi\sqrt{2}} \left(  \bar{q}_c - q_c \right) + P_{mimp}(k^p) + \frac{2\pi}{L} \left[ \left( \frac{q_c + \bar{q}_c}{4\pi} - \gamma_c +  \bar{\gamma}_c \right)\left( \frac{q_c - \bar{q}_c}{4\pi} - \gamma_c - \bar{\gamma}_c \right)+ \right.\\ 
	&\left.  \left( \frac{q_s + \bar{q}_s}{4\pi} - \gamma_s + \bar{\gamma}_s \right)\left( \frac{q_s - \bar{q}_s}{4\pi} - \gamma_s - \bar{\gamma}_s \right) \right] + \frac{2\pi}{L} \sum_{k=c,s} \left( N_k^+ - N_k^- \right), 
\end{aligned}
\ee
where the $N_k^\pm$ are non-negative integers enumerating the number of particle-hole pairs in the vicinity of the ``Fermi points''.
\subsection{Bethe Ansatz calculation: high-energy charge particle}
We wish to know the momentum contribution from the high-energy charge particle: there will be finite-size contributions to this from interactions with the low-energy sector.
As we know precisely the integers forming this state from (\ref{eq:heInts}), we can simply sum these integers to find the momentum. This approach, however, yields no information on which contributions come from the finite-size shift of the rapidity and which contributions come from interactions between the high-energy and low-energy degrees of freedom.
The solution is to explicitly include the finite-size shift of the rapidity and calculate the remaining corrections in terms of the quantites $N_\alpha^\imp$, $D_\alpha^\imp$, $N_\alpha$, $D_\alpha$, as we had for the finite-size energy.

\subsubsection{Basic integral equations}
The solution for $\rho_{\alpha,1}$  implicitly defined by (\ref{eq:rhoImp3}), can be formally written as
\be
\rho_{\alpha,1}(z_\alpha) = \left( K_{\alpha\beta} * (1-\hat{K})^{-1}_{\beta c} \right)(z_\alpha,k^p). \label{eq:rhoImp4}
\ee
We introduce the shift functions\cite{korepin}
\beA
F_{cc}^{(0)}(k,k') = 0,&\qquad F_{cs}^{(0)}(k,\Lambda) = \frac{1}{2\pi} \theta\left(  \frac{\sin k - \Lambda}{u} \right),\\
F_{sc}^{(0)}(\Lambda,k) = \frac{1}{2\pi} \theta\left( \frac{\Lambda - \sin k}{u} \right),&\qquad F_{ss}(\Lambda,\Lambda') = -\frac{1}{2\pi} \theta\left( \frac{\Lambda - \Lambda'}{2u} \right),
\eeA
and the ``dressed'' shift functions
\be
F_{\alpha\beta}(z_\alpha,z_\beta) = F_{\alpha\beta}^{(0)} (z_\alpha,z_\beta) + \left(F_{\alpha\gamma} * K_{\gamma\beta} \right) (z_\alpha,z_\beta)\label{eq:Fdef}.
\ee
It is useful to note that
\be
K_{\alpha\beta}(z_\alpha,z_\beta) = \partial_{z_\alpha} F^{(0)}_{\alpha\beta} \label{eq:FisderivK}.
\ee
Both the finite-size energy and momentum spectra involve the function
\be
\tilde{r}_{\alpha\beta}^{(\sigma)}(z_\alpha) = K_{\alpha\beta}(z_\alpha, \sigma X^\beta) + K_{\alpha\gamma} * \tilde{r}_{\gamma\beta}^{(\sigma)} \label{eq:trdef}.
\ee
\subsubsection{Finite-size momentum spectrum}
As for the energy of the system, the momentum can also be expanded as an asymptotic series in powers of $L^{-1}$. In the analysis of the finite-size energy calculation, when determining $\delta k^p$ as in (\ref{eq:deltakp}), one finds
\be
\begin{aligned}
z_c(k^p_L) &= z_{c,0}(k^p)  + z_{c,0}'(k^p) \frac{\delta k^p}{L} \\
&+\sum_{\sigma,\beta} \sigma \rho_{\beta,0}(X^\beta) \left[ \theta\left( \frac{\sin k^p - \sigma X^\beta}{u} \right) \delta_{\beta,s} + \int_{-A}^A \d \Lambda \theta\left(  \frac{\sin k^p - \Lambda}{u}\right) \tilde{r}_{s\beta}^{(\sigma)}(\Lambda)  \right] \left[ X_\sigma^\beta - \sigma X^\beta \right] \\
&+\frac{1}{L} \int_{-A}^A \d \Lambda\, \theta\left( \frac{\sin k^p - \Lambda}{u} \right) \rho_{s,1}(\Lambda) .
\end{aligned}
\ee
We will first look at the term in the sum multiplied by $X_\sigma^\beta - \sigma X^\beta$. 
(\ref{eq:Fdef}) and  (\ref{eq:trdef}) imply that 
\be
F_{\alpha\beta} = F^{(0)}_{\alpha\gamma} * (1-\hat{K})^{-1}_{\gamma\beta},\qquad 
\tilde{r}_{\alpha\beta}^{(\sigma)} = (1-\hat{K})^{-1}_{\alpha\gamma} * K_{\gamma\beta} (z_\alpha,\sigma X^\beta), \label{eq:F1-k}
\ee
allowing us to write
\be
F_{c \beta}^{(0)} (k^p, \sigma X^\beta) + F^{(0)}_{c\alpha} * \tilde{r}^{(\sigma)}_{\alpha\beta} (k^p) 
=
F_{c\beta}(k^p,\sigma X^\beta).
\ee
It can also be shown that
\be
\int_{-A}^A \d\Lambda\,\theta\left( \frac{\sin k -\Lambda}{u} \right) \rho_{s,1}(\Lambda) =2\pi F_{cc} (k,k^p) \label{eq:lastMomPart}.
\ee
The finite-size momentum can therefore be written in terms of the dressed shift functions as
\be
z_c(k^p_L) = z_{c_0}(k^p) + z_{c,0}'(k^p) \frac{\delta k^p}{L} + \sum_{\sigma,\beta}\sigma2\pi \rho_{\beta,0} (X^\beta) F_{c\beta}(k^p,\sigma X^\beta) \left[ X_\sigma^\beta - \sigma X^\beta \right] + \frac{2\pi}{L} F_{cc}(k^p,k^p). \label{eq:fsmsimplified1}
\ee
We now wish to relate the functions $F_{\alpha\beta}(z_\alpha,z_\beta)$ to the impurity densities $N_\alpha^\imp$, $D_\alpha^\imp$.
\subsubsection{Relating shift functions to impurity densities}
By using (\ref{eq:rhoImp4}) and (\ref{eq:FisderivK}) in (\ref{eq:naimpdef1}) and (\ref{eq:dcimpdef1}), it can be shown that
\begin{align}
2D_\alpha^\imp &= F_{\alpha c}(X^\alpha,k^p)  +F_{\alpha c}(-X^\alpha,k^p),\\
N_\alpha^\imp &=  F_{\alpha c}(X^\alpha, k^p) - F_{\alpha c}(-X_\alpha, k^p),
\end{align}
i.e.
\be
D_\alpha^\imp \pm \frac{N_\alpha^\imp}{2} = F_{\alpha c}(\pm X^\alpha, k^p). \label{eq:diniF}
\ee
\subsubsection{Determining boundary terms}
To express the finite-size momentum (\ref{eq:fsmsimplified1}) in terms of the $N_\alpha^\imp$, $D_\alpha^\imp$ (\ref{eq:diniF}), we need to relate $F_{\alpha c}(\sigma X^\alpha, k^p)$ to $F_{c\beta}(k^p,\sigma' X^\beta)$.
By considering the Neumann series of (\ref{eq:F1-k}) and integrating by parts, it can be shown that
\be
F_{\alpha\beta}(z_\alpha,z_\beta) + F_{\beta\alpha}(z_\beta,z_\alpha) = -\sum_{\gamma,\sigma} \sigma F_{\gamma\alpha}(\sigma X^\gamma, z_\alpha) F_{\gamma\beta}(\sigma X^\gamma, z_\beta) \label{eq:Fidentity} .
\ee
To establish the desired relationship, (\ref{eq:Fidentity}) implies that we require the values $F_{\alpha\beta}(\tau X^\alpha, \tau' X^\beta)$.
It is simple to show that
\be
F_{\alpha\beta}(X^\alpha, X^\beta) - F_{\alpha\beta}(-X^\alpha, X^\beta) = Z_{\alpha\beta} - \delta_{\alpha\beta} \label{eq:diffDressedCharge},
\ee
with $Z$ the dressed charge matrix as defined in (\ref{eq:dressedChargeDef}). (\ref{eq:Fidentity}) also implies that
\be
F_{\alpha\beta}(X^\alpha,X^\beta)+ F_{\beta\alpha}(X^\beta,X^\alpha) = - \sum_{\gamma} \left[
F_{\gamma\alpha}(X^\gamma, X^\alpha) F_{\gamma\beta}(X^\gamma,X^\beta)
-
F_{\gamma\alpha}(-X^\gamma, X^\alpha) F_{\gamma\beta}(-X^\gamma,X^\beta)
\right]. \label{eq:toSubIn}
\ee
Substituting (\ref{eq:diffDressedCharge}) into (\ref{eq:toSubIn}) and simplifying, if we define $F$ to be the matrix $F_{\alpha\beta}(X^\alpha,X^\beta)$, then it satisfies the equation
\be
Z^\top F + F^\top Z = (1-Z)^\top (1-Z) \label{eq:matEq1}. 
\ee
Considering
\be
F_{\beta\alpha}(-X^\beta,X^\alpha) - F_{\alpha\beta}(-X^\alpha,X^\beta) = \sum_{\sigma,\gamma} \sigma F_{\gamma\alpha}(\sigma X^\gamma,X^\alpha) F_{\gamma\beta}(-\sigma X^\gamma, X^\beta),
\ee
and using (\ref{eq:diffDressedCharge}) again, we find the similar equation 
\be
Z^\top F - F^\top Z = Z-Z^\top. \label{eq:matEq2}
\ee
(\ref{eq:matEq1}) and (\ref{eq:matEq2}) determine $F$ uniquely, giving 
\be
F_{\alpha\beta}(\tau X^\alpha, \tau' X^\beta) = \frac{\tau}{2} \left(  Z -1 \right)_{\alpha\beta} + \frac{\tau'}{2} \left({Z^{-1}}^\top  - 1\right)_{\alpha\beta}.
\ee
This therefore allows us to write down the dressed shift functions appearing in (\ref{eq:fsmsimplified1}) in terms of the known quantities $N_\alpha^\imp$, $D_\alpha^\imp$, viz.
\begin{align}
F_{cc}(k^p,k^p) &= - \sum_\gamma  D_\gamma^\imp N_\gamma^\imp,\\
F_{c\alpha}(k^p,\tau X^\alpha) &= -\sum_\gamma \left(  
\frac{\tau}{2} N_\gamma^\imp {Z^{-1}}^\top_{\gamma\alpha}
+D_\gamma^\imp Z_{\gamma\alpha}
\right).
\end{align}
Combining the previous results, we find
\be
z_c(k^p_L) = z_{c,0}(k^p) + z_{c,0}'(k^p) \frac{\delta k^p}{L} - \frac{2\pi}{L} \sum_\alpha \left[ 
N^\imp_\alpha D_\alpha
+D^\imp_\alpha \Delta N_\alpha
- D^\imp_\alpha N^\imp_\alpha
\right].
\ee
Using Eq.~(8.38) from Ref.~\onlinecite{book},
the full finite-size momentum spectrum in the presence of a high-energy charge particle is given by
\be
P = 2 D_c k_{F,\up} + 2(D_c + D_s) k_{F,\down} + z_{c,0}(k^p) + 2\pi \rho_{c,0}(k^p) \frac{\delta k^p}{L} +  \frac{2\pi}{L} \left( \Delta {\bf \widetilde{N}}^\top  \cdot \Delta {\bf \widetilde{D}} + \sum_{k\in\{c,s\}} (N_k^+ - N_k^- )\right) \label{eq:BAFSS} ,
\ee
where the $N_k^\pm$ are non-negative integers enumerating the number of particle-hole pairs in the vicinity of the Fermi points and $k_{F,\up(\down)} = \frac{1}{2} \left( \pi n_c \pm 2\pi m \right)$.
In the zero-field limit $m=0$ and therefore $k_{F,\up} = k_{F,\down} = k_F$, giving
\be
P = 2 k_F \left( 2D_c + D_s \right) + z_{c,0}(k^p) + 2\pi \rho_{c,0}(k^p) \frac{\delta k^p}{L} +  \frac{2\pi}{L} \left( \Delta {\bf \widetilde{N}}^\top  \cdot \Delta {\bf \widetilde{D}} + \sum_{k\in\{c,s\}} (N_k^+ - N_k^- )\right) \label{eq:BAFSS2} .
\ee

\section{Mobile impurity contributions to $\sigma^{(2)}(\omega)$} \label{app:sigma2MIM}
\subsection{Threshold of the ``particle-hole'' continuum in $\sigma^{(2)}(\omega)$}
\label{sec:antiHolSec}
Next we examine the thresholds in the second contribution
\fr{sigma_master} to the optical conductivity. The lowest threshold
arises in the ``particle-hole'' and ``two-particle'' excitations
considered in \ref{sec:shiftedAntiHolon} and \ref{descpp} respectively.
The threshold in both cases is given by
\bea
E^{\rm ph}_{\rm thres}(q) &=& \varepsilon_c(k(q)) - 2\mu,\nn
q &=& k + \int_{-\infty}^\infty \d \Lambda\, \theta\left( \frac{\sin k
  - \Lambda}{u} \right) \rho_{s,0}(\Lambda),
 \label{eq:fixMomentum}
\eea
where $\rho_{s,0}(\Lambda)$ is the ground state root density
(\ref{eq:gsrhos}). The threshold for the particle-hole (two-particle)
excitation is obtained by fixing the position of the hole (one of the
particles) in momentum space at one of the ``Fermi points'', so that it
contributes only at ${\cal   O}(L^{-1})$ to the excitation
energy. Hence the impurity degree of freedom corresponds to a particle
in both cases.

\subsubsection{Projection of the operator ${\cal O}_j$}
\label{sec:proj2}
We will use the representation \fr{eq:simplifiedSum} to determine the
contribution $C_{\rm JJ}^{(2)}(\ell,t)$ to the current-current
correlator. Hence we require the projection of the operator
${\cal O}_j$ defined in \fr{eq:ojDef} to the mobile impurity model
\fr{MIM}. This can be worked out by following
Ref. \onlinecite{EsslerPereira}. We start by taking the 
continuum limit of the lattice fermion operators
\be
c_{j,\sigma} \sim R_\sigma(x)e^{ik_Fx} + L_\sigma (x) e^{-ik_Fx}+\ldots,\quad x=ja_0
\label{cRL}
\ee
where $a_0$ is the lattice spacing. The continuum limit of $\OO_j$ then
takes the form
\be
\mathcal{O}(x) \sim e^{ix(2k_F-\pi)} R_c \partial_x R_c e^{i\pi
  \int_{-\infty}^x \d x'\, Q_c(x')} + \dots\, , 
\label{eq:oChargeFermionRep} 
\ee
with $Q_c(x) =  R_c^\dagger(x) R^\pd_c(x) + L_c^\dagger(x)L^\pd_c(x)$. 
Next we decompose the charge part into the low-energy and impurity
pieces
\be
R_c(x) \sim r_c + B^\dagger(x) e^{i(\pi-2k_F)x}+\ldots\, .
\label{eq:chargeFermionDecomposition}
\ee
Substituting this back into \fr{eq:oChargeFermionRep} and then
bosonising the low-energy degrees of freedom we obtain
\be
{\cal O}(x) \sim B^\dagger(x) e^{-\frac{i}{2\sqrt{2}} \Theta_c^*(x)}+\ldots,
\label{eq:lowEnergyOp}
\ee
where we have retained only the most relevant piece in the sector with
a single impurity.

\subsubsection{Finite-size excitation energy in the 
Mobile Impurity Model}
As the mobile impurity model is again given by (\ref{eq:MIM1}) to 
(\ref{MIM}), and the impurity again is located at a maximum of its
dispersion, we can follow through the same steps as in our analysis of
the $k$-$\Lambda$ string threshold. The finite-size spectrum is,
accordingly, of the same form as (\ref{eq:generalImpModelFSS}). The
values of $q^{(0)}_\alpha$ follow from the form of the Luttinger liquid
part of \fr{eq:lowEnergyOp} to be 
\be
q_c^{(0)} = \bar{q}_c^{(0)} = \pi \sqrt{2};\qquad q_s^{(0)} = \bar{q}_s^{(0)} = 0.
\ee
\subsubsection{Finite-size excitation energy from Bethe Ansatz}
We consider again the excitation described in Section
\ref{sec:shiftedAntiHolon}. The finite-size corrections to the
excitation energy are calculated in Appendix~\ref{app:chBAfull}. The
final result is of the form
\bea
E &=& e_{GS} L +\varepsilon_c(k^p) + \frac{\delta k^p}{L} \varepsilon_c'(k^p)	- \frac{\pi}{6L} (v_c + v_s)  \nn
&+& {\frac{2\pi v_c}{L} \left[ \frac{ (\Delta N_c -
      N_c^\imp)^2}{8K_c} + 2K_c \left( D_c - D_c^\imp  + \frac{D_s
      - D_s^\imp}{2} \right)^2 \right] } \nn
&+& \frac{2\pi v_s}{L} \left[ \frac{1}{2} \left( \Delta N_s - N_s^\imp
  - \frac{ \Delta N_c - N_c^\imp}{2} \right)^2 + \frac{(D_s -
    D_s^\imp)^2}{2} \right]. 
\label{eq:BAfinitespech} 
\eea
Here the ground state energy per site $e_{GS}$ and dressed energy
$\varepsilon_c(k)$ are given in \fr{eq:egs} and \fr{eq:ec}
respectively, while the velocities $v_{s,c}$ and the Luttinger
parameter $K_c$ are calculated in Appendix~\ref{app:vK}. The
thermodynamic value $k^p$ of the impurity rapidity and its finite-size
correction $\delta k^p$ are determined by 
(\ref{eq:rapidityRequirement2}) and (\ref{eq:dkpSimple}). 
Finally, we have
\bea
\Delta N_c &=& -3\ , \qquad \Delta N_s = -1\ ,\qquad D_c = 0\ ,
\qquad D_s = 0\ ,\nn
N_c^\imp &=& 2N_s^{\rm imp}-1=\int_{-Q}^Q \d k\,
\rho_{c,1}(k)\ ,\qquad D_s^{\rm imp}=0, \label{eq:nciDefault} \\
2 D_c^\imp &=& \int_Q^\pi \d k \left[ \rho_{c,1}(-k) - \rho_{c,1}(k) \right] + \frac{i}{\pi} \left\{ \ln \left[ \frac{ \Gamma\left( \frac{1}{2} - i \frac{\sin k^p}{4u} \right) \Gamma\left( 1 + i \frac{\sin k^p}{4u}\right)}{ \Gamma\left( \frac{1}{2} + i \frac{\sin k^p}{4u} \right) \Gamma\left( 1 - i \frac{\sin k^p}{4u}\right)}\right]\right\}\nn
&+&\frac{i}{\pi}\int_{-Q}^Q \d k\, \rho_{c,1}(k) \left\{ \ln \left[ \frac{ \Gamma\left( \frac{1}{2} - i \frac{\sin k}{4u} \right) \Gamma\left( 1 + i \frac{\sin k}{4u}\right)}{ \Gamma\left( \frac{1}{2} + i \frac{\sin k}{4u} \right) \Gamma\left( 1 - i \frac{\sin k}{4u}\right)}\right] \right\},
\eea
where $\rho_{c,1}(k)$ is the solution of the integral equation
\be
\rho_{c,1}(k) = \cos k\, R(\sin k - \sin k^p) + \cos k\, \int_{-Q}^Q \d k'\, R(\sin k - \sin k')\rho_{c,1}(k').
\ee
In order to fully specify our mobile impurity model we also require the
the curvature of the dispersion relations of $\varepsilon_c(k)$ at
$k=\pi$, which is given by
\be
-\frac{1}{m} = \frac{\d^2 \varepsilon_c(k(q))}{\d q^2} \Bigg\vert_{k=\pi} =  \frac{2 + \int_{-Q}^Q \d k\, R'(\sin k) \varepsilon_c'(k)}{(2\pi \rho_{c,0}(\pi) )^2}.
\ee
\subsubsection{Fixing the parameters $\gamma_\alpha$, $\bar\gamma_\alpha$}
\label{sec:fsComp2}
By matching the expressions \fr{eq:BAfinitespech} and
\fr{eq:generalImpModelFSS} for the finite-size energies we can fix the
parameters $\gamma_\alpha$, $\bar\gamma_\alpha$
\beA
\gamma_c &= \frac{1}{2\sqrt{2}} + \frac{1}{2\sqrt{2}} (\Delta N_c
- N_c^\imp)\ ,&&\quad\gamma_s=0\ ,\\
\bar{\gamma}_c &=  -\frac{1}{2\sqrt{2}} -
\frac{1}{2\sqrt{2}} (\Delta N_c - N_c^\imp)\ ,&&\quad \bar{\gamma}_s = 0.
\eeA
\subsubsection{Current-current correlator in the mobile impurity
model}
\begin{figure}[ht!]
	\includegraphics[width=0.6\columnwidth]{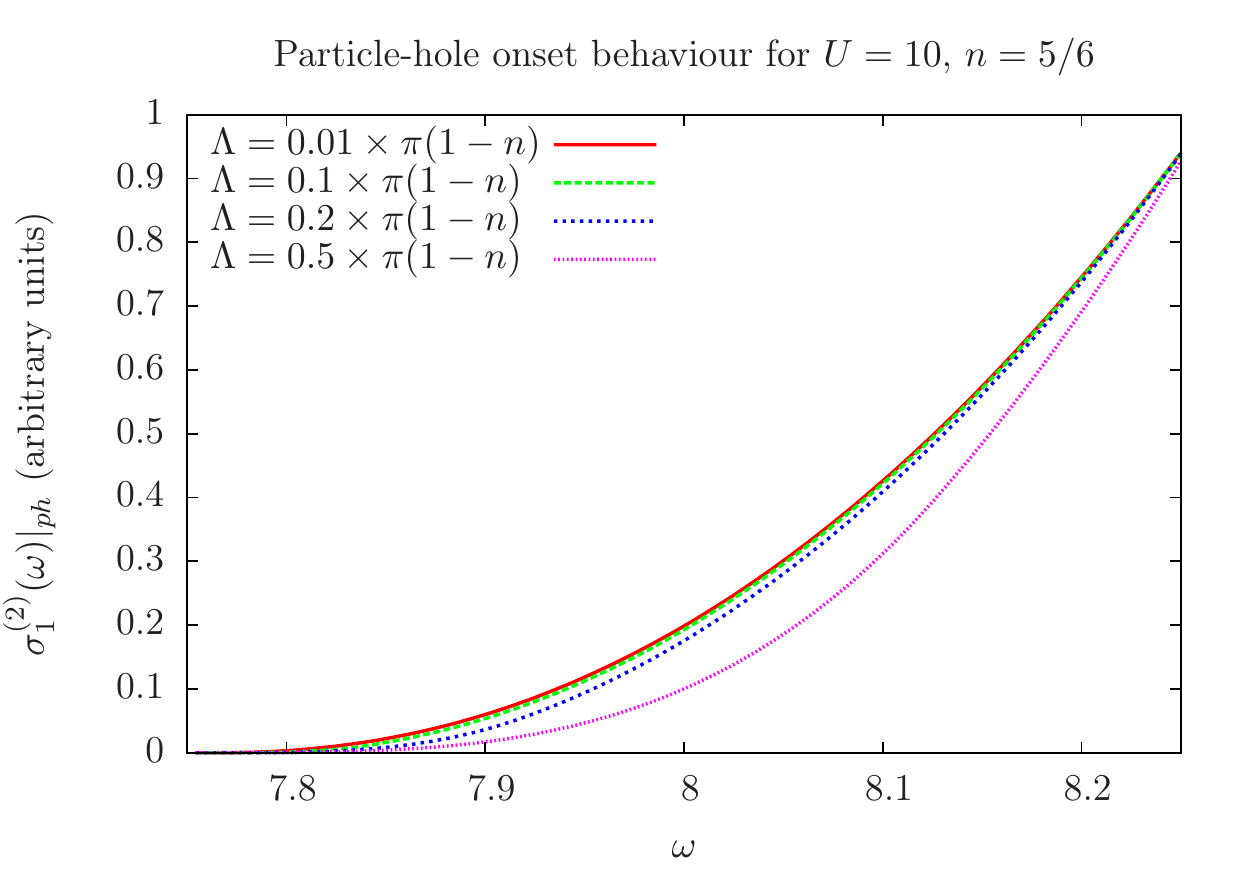}
	\caption{Contribution to onset of $\sigma_1^{(2)}(\omega)$
          from particle-hole excitation in (\ref{eq:sigma2ph}) for
          $U=10$, $n=5/6$}\label{fig:sigma2ph}
\end{figure}

Given the expression \fr{eq:lowEnergyOp} for the projection of the
operator $\mathcal{O}_j$, we have 
\bea
C^{(2)}_{\rm JJ}(\ell,t)\sim H(x,t)&=&\langle {\cal O}^\dagger(x,t)
{\cal O}(0,0)\rangle
\sim\langle  B(x,t) e^{\frac{i}{2\sqrt{2}} \Theta^*_c(x,t)}
B^\dagger (0,0) e^{-\frac{i}{2\sqrt{2}} \Theta^*_c(0,0)} \rangle . 
\label{odaggero}
\eea
This is readily calculated using the unitary transformation
\fr{eq:unitaryDef}. In the new basis the correlator factorises
\bea
H(x,t) &\sim& \langle
e^{\frac{i}{2\sqrt{2}} (\Delta N_c - N_c^\imp)
  \Theta_c^\circ(x,t)}  e^{- \frac{i}{2\sqrt{2}} (\Delta N_c -
  N_c^\imp) \Theta_c^\circ(0,0)} \rangle
\langle \widetilde{B}(x,t) \widetilde{B}^\dagger(0,0)\rangle\nn
&\sim&\frac{1}{(x^2 - v_c^2t^2)^{\eta}}
\int_{-\Lambda}^\Lambda \frac{\d p}{2\pi} e^{-ipx} e^{-i\varepsilon(p)t},
\eea
where in this case $\varepsilon(p)$ is given by $\varepsilon_c(\pi+p)$ and
\be
\eta = \frac{1}{2K_c} \left( \frac{3}{2} + \frac{N_c^\imp}{2}\right)^2.
\ee
Fourier transforming and using \fr{sigma_master} we arrive at
\be
\sigma_1^{(2)}(\omega)\Big\vert_{ph} \sim 
\frac{1}{\omega}\int_{-\Lambda}^\Lambda \d p\,
\widetilde{G}^c_{\eta,\eta}\big(\omega - \varepsilon(p),p\big)\ , \label{eq:sigma2ph}
\ee
where $\widetilde{G}^c_{\eta,\eta}(\omega,p)$ is given by
(\ref{eq:Gtildec}). The behaviour of (\ref{eq:sigma2ph}) is shown in
Fig.~\ref{fig:sigma2ph}. 
We see that the contribution vanishes smoothly at the threshold and
increases slowly above it.
\subsection{Threshold of the two-hole continuum in $\sigma_1^{(2)}(\omega)$}
\label{sec:halfFillSec}
Last but not least we wish to consider the threshold of the
contribution of the two-hole continuum to $\sigma_1^{(2)}(\omega)$.
This occurs at a higher energy than the threshold of the particle-hole
and particle-particle continua, but unlike the latter two persists
as we approach half-filling. The threshold is parametrised by
\be
E_{\rm thres}^{\rm hh}(q) = -2\varepsilon_c\left( \frac{k(q)}{2} \right) -2\mu, \label{eq:halfFillThresh}
\ee
where $k(q)$ is again fixed by (\ref{eq:fixMomentum}). The threshold
corresponds to having \emph{two} high-energy hole excitations with
momentum $q/2$ each. As we are now dealing with two impurities with
equal momenta, the appropriate mobile impurity model is of the form
\fr{MIM}, but we now have to retain impurity-impurity interactions
\bea
H_{\rm imp}&=& \int \d x\left[ 
B^\dagger(x)( \varepsilon -i u \partial_x - \frac{1}{2m} \partial_x^2)
B(x)+V\  B^\dagger(x)\partial_x B^\dagger(x)\ B(x)\partial_x B(x)\right].
\eea

\subsubsection{Projection of the operator ${\cal O}_j$}
\label{sec:proj3}
Next we require the projection of the operator ${\cal O}_j$ onto the
mobile impurity model. This proceeds as before,
cf. eqns \fr{cRL}, (\ref{eq:oChargeFermionRep}), but now we take
\be
R_c(x) \sim r_c(x) + B^\dagger (x) e^{-iqx/2} .
\ee
Substituting this into the expression \fr{eq:oChargeFermionRep} for
${\cal O}(x)$ we find
\be
\OO(x) \sim e^{i(2k_F - \pi - q)x} B^\dagger(x) \partial_x B^\dagger(x)
e^{\frac{i}{2\sqrt{2}} \Phi^*_c(x)}+\ldots\ ,
\label{eq:Oproj3} 
\ee
where we have retained only the most relevant term in the sector with
two impurities.

\subsubsection{Finite-size corrections to excitation energies in the
  mobile impurity model}
\label{sec:fieldTheoryThree}
The interactions between the mobile impurities and the Luttinger
liquid degrees of freedom can again be removed by the unitary
transformation \fr{eq:unitaryDef}. In the transformed basis
finite-size corrections to the excitation energies in the LL part of
the theory can then be calculated as before, and lead to the result
\fr{eq:generalImpModelFSS}.

The zero mode eigenvalues for the ``minimal'' excitation
(cf. \fr{q0s}) associated with $\OO(x)$ as defined in (\ref{eq:Oproj3})
are 
\be
q_c^{(0)} = -\pi\sqrt{2}\ ,\quad \bar{q}_c^{(0)} = \pi\sqrt{2}\ ,\quad
q_s^{(0)} = 0\ ,\quad \bar{q}_s^{(0)} = 0.
\ee

\subsubsection{Finite-size corrections to excitation energies from
the Bethe Ansatz} 
The two-hole excitation has been constructed in
\ref{sec:shiftedTwoHolon}, and the threshold of interest here occurs
when, in the thermodynamic limit, the two holes have equal momentum. The finite-size corrections to
the excitation energy can be calculated following Ref.~\onlinecite{FHLE},
details are given in Appendix \ref{app:holonPairFull}. The final result
in zero magnetic field is
\bea
E &=& e_{GS} L -\varepsilon_c(k^{h_1}) - \varepsilon_c(k^{h_2}) -
\frac{\delta k^{h_1}}{L} \varepsilon_c'(k^{h_1}) - \frac{\delta
k^{h_2}}{L} \varepsilon_c'(k^{h_2}) - \frac{\pi}{6L} (v_c + v_s)  \nn
&+&\frac{2\pi v_c}{L} \left[ \frac{ (\Delta N_c - N_c^\imp)^2}{8K_c}
  + 2K_c \left( D_c - D_c^\imp  + \frac{D_s - D_s^\imp}{2} \right)^2
  \right]  \nn
&+&\frac{2\pi v_s}{L} \left[ \frac{1}{2} \left( \Delta N_s - N_s^\imp
  - \frac{ \Delta N_c - N_c^\imp}{2} \right)^2 + \frac{(D_s -
    D_s^\imp)^2}{2} \right]. 
\label{eq:BAfinitespechh} 
\eea
Here the ground state energy per site $e_{GS}$ and dressed energy
$\varepsilon_c(k)$ are given in \fr{eq:egs} and \fr{eq:ec}
respectively, while the velocities $v_{s,c}$ and the Luttinger
parameter $K_c$ are calculated in Appendix~\ref{app:vK}. The
thermodynamic values $k^{h_i}$ of the impurity rapidities and the finite-size
corrections $\delta k^{h_i}$ are determined by (\ref{eq:zckhi}) and (\ref{eq:dkhi}). 
Finally, we have
\bea
\Delta N_c &=& 0\ , \qquad \Delta N_s = -1\ ,\qquad D_c = \frac{1}{2}\ ,
\qquad D_s = 0\ ,\nn
N_c^\imp &=& 2(N_s^{\rm imp}+1)=\int_{-Q}^Q \d k\,
\rho_{c,1}(k)\ ,\qquad D_s^{\rm imp}=0, \\
2 D_c^\imp &=& \int_Q^\pi \d k \left[ \rho_{c,1}(-k) - \rho_{c,1}(k) \right] - \sum_{j=1,2} \frac{i}{\pi} \left\{ \ln \left[ \frac{ \Gamma\left( \frac{1}{2} - i \frac{\sin k^{h_j}}{4u} \right) \Gamma\left( 1 + i \frac{\sin k^{h_j}}{4u}\right)}{ \Gamma\left( \frac{1}{2} + i \frac{\sin k^{h_j}}{4u} \right) \Gamma\left( 1 - i \frac{\sin k^{h_j}}{4u}\right)}\right]\right\}\nn
&+&\frac{i}{\pi}\int_{-Q}^Q \d k\, \rho_{c,1}(k) \left\{ \ln \left[ \frac{ \Gamma\left( \frac{1}{2} - i \frac{\sin k}{4u} \right) \Gamma\left( 1 + i \frac{\sin k}{4u}\right)}{ \Gamma\left( \frac{1}{2} + i \frac{\sin k}{4u} \right) \Gamma\left( 1 - i \frac{\sin k}{4u}\right)}\right] \right\},
\eea
where $\rho_{c,1}(k)$ is the solution of the integral equation
\be
\rho_{c,1}(k) = -\cos k\; \left[ R(\sin k - \sin k^{h_1}) + R(\sin k - \sin k^{h_2}) \right] + \cos k \int_{-Q}^Q dk'\ \rho_{c,1}(k') R(\sin k - \sin k').
\ee

\subsubsection{Fixing the parameters $\gamma_\alpha$, $\bar\gamma_\alpha$}
\label{sec:fsComp3}
By comparing the finite-size spectra calculated from the Bethe Ansatz
(\ref{eq:BAfinitespechh}) with those obtained from the mobile impurity 
model (\ref{eq:generalImpModelFSS}) we are again able to
determine the parameters $\gamma_\alpha$, $\bar\gamma_\alpha$. In the
case at hand we obtain
\be
\gamma_c+\bar{\gamma}_c = -\sqrt{2} D_c^\imp\ ,\quad
\gamma_c-\bar{\gamma}_c = -\frac{1}{\sqrt{2}} N_c^\imp\ ,\quad
\gamma_s = \bar{\gamma}_s = 0.
\ee

\subsubsection{Current-current correlator in the mobile impurity
model}
Given the expression \fr{eq:Oproj3} for the projection of the
operator $\mathcal{O}_j$, we have 
\bea
C^{(2)}_{\rm JJ}(\ell,t)&\sim& \langle {\cal O}^\dagger(x,t) {\cal
  O}(0,0)\rangle\nn
&\sim&\langle  \partial_x B(x,t) B(x,t) e^{-i \Phi_c^*(x,t)/2\sqrt{2}}  B^\dagger(0,0) \partial_x B^\dagger(0,0) e^{i \Phi_c^*(0,0)/2\sqrt{2}}\rangle
\equiv L(x,t).
\label{odaggerohh}
\eea
This is readily calculated using the unitary transformation
\fr{eq:unitaryDef}. In the new basis the correlator factorises
\bea
L(x,t) &=& \langle\partial_x\widetilde{B}(x,t) \widetilde{B}(x,t)
\widetilde{B}^\dagger(0,0)\partial_x\widetilde{B}^\dagger(0,0) \rangle\nn 
&&\times\big\langle 
e^{-i \frac{\frac{1}{2} - 2D_c^\imp}{\sqrt{2}} \Phi_c^\circ(x,t) +i
  \frac{N_c^\imp}{\sqrt{2}} \Theta_c^\circ(x,t)} e^{i
  \frac{\frac{1}{2} - 2D_c^\imp}{\sqrt{2}} \Phi_c^\circ(0,0) -i
  \frac{N_c^\imp}{\sqrt{2}} \Theta_c^\circ(0,0)}    
\big\rangle.
\eea
The Luttinger liquid part of the correlator is readily calculated
\be
L(x,t)=\langle\partial_x\widetilde{B}(x,t) \widetilde{B}(x,t)
\widetilde{B}^\dagger(0,0)\partial_x\widetilde{B}^\dagger(0,0) \rangle  
\big(x - v_ct\big)^{-\nu_+}
\big(x + v_ct\big)^{-\nu_-}
, \label{eq:simplifiedTwoImp}
\ee
where
\beA
\nu_\pm &= 2\left[ \sqrt{K_c} \left( \frac{1}{2} - 2D_c^\imp \right) \mp \frac{N_c^\imp}{2\sqrt{K_c}} \right]^2,\\
\nu &=\nu_++\nu_- =
4K_c\left(\frac{1}{2} - 2D_c^\imp\right)^2 + 
\frac{\left(N_c^\imp \right)^2}{K_c}.
\eeA
In the absence of interactions between our two high-energy impurities
($V=0$) the impurity part of the correlator is readily calculated as
\be 
\langle\partial_x\widetilde{B}(x,t) \widetilde{B}(x,t)
\widetilde{B}^\dagger(0,0)\partial_x\widetilde{B}^\dagger(0,0) \rangle
\sim \frac{1}{t^{3/2}} \delta(x-ut). 
\ee
In order to gain some insight in the importance of interactions, they
can be taken into account in a random phase approximation in the
impurity-impurity interaction. Summing up the RPA bubble diagrams does
not change the behaviour sufficiently close to the threshold.
Putting everything together we find
\be
L(x,t) \sim \frac{1}{(x - v_ct)^{\nu_+}} \frac{1}{(x + v_ct)^{\nu_-}}\frac{\delta(x-ut)}{t^{3/2}}+\ldots
\ee
The resulting contribution to $\sigma_1^{(2)}(\omega)$ for frequencies
close to $\omega_0 = -2\mu - 2\varepsilon_c\left( \frac{k(q))}{2}
\right)$ is thus
\be
\sigma_1^{(2)}(\omega)\Big\vert_{\rm two-hole} \sim \frac{1}{\omega} \left(\omega - \omega_0\right)^{\nu + \frac{1}{2}}\, \Theta(\omega - \omega_0).
\label{sigma2hh}
\ee

As we have pointed out before, the excitation with two high-energy
holes persists at half-filling.
Importantly, this contribution is no longer suppressed at
half-filling, and in fact gives rise to the square
root increase above the absolute threshold in the optical conductivity
in this limit\cite{JeckEssler,ControzziEsslerTsvelik,PereiraPenc}. 
Our result \fr{sigma2hh} is reconciled with this behaviour by
noting that the frequency range $\omega-\omega_0$ over which
\fr{sigma2hh} holds is related to the cutoff $\Lambda_c$ of the charge
sector of the Luttinger liquid degrees of freedom. As we approach 
half-filling this cutoff tends to zero i.e. the frequency window in
which \fr{sigma2hh} applies vanishes. At sufficiently high frequencies 
$\omega>\omega_0+\Lambda_c$ we expect on general grounds to recover
the square root behaviour observed at half-filling.



\begin{references}
\bibitem{endNote}
	In such a case, the difference between $\varepsilon(q)$ and
	$\tilde{\varepsilon}(q)$ is simply a constant shift of
	$\sum_\alpha \pi v_\alpha\left( 
	\frac{\left( \gamma_\alpha-\bar{\gamma}_\alpha \right)^2}{2K_\alpha} +
	2K_\alpha\left( \gamma_\alpha+\bar{\gamma}_\alpha \right)^2 \right)$.
\end{references}
\end{document}